\documentclass[aps,prl,reprint,groupedaddress]{revtex4-2}

\usepackage{graphicx}  
\usepackage{dcolumn}   
\usepackage{bm}        
\usepackage{amssymb}   
\usepackage{amsmath}
\usepackage{comment}
\usepackage[utf8]{inputenc}
\usepackage{float}
\usepackage[dvipsnames]{xcolor}
\usepackage{graphicx}
\usepackage{comment}
\usepackage{bigints}
\usepackage{tabularx}
\usepackage{pbox}
\usepackage{tabu}
\usepackage{ulem}

\newcommand{\Em}{\langle E \rangle}

\newcommand{\pltime}{\tau_{\rm{pl}}}

\newcommand{\yone}{y^{(1)}}
\newcommand{\ytwo}{y^{(2)}}

\newcommand{\tgamma}{\tilde{\gamma}_0}
\newcommand{\talpha}{\tilde{\alpha}}

\begin{document}


\title{Mean field theory of yielding under oscillatory shear}



\author{Jack T. Parley}
\email[Author to whom correspondence should be addressed: ]{jack.parley@uni-goettingen.de}
\affiliation{Institut f{\"u}r Theoretische Physik, University of G{\"o}ttingen,
Friedrich-Hund-Platz 1, 37077 G{\"o}ttingen, Germany}

\author{Srikanth Sastry}
\affiliation{Jawaharlal Nehru Centre for Advanced Scientific Research, Jakkar Campus, 560064 Bengaluru, India}

\author{Peter Sollich}
\affiliation{Institut f{\"u}r Theoretische Physik, University of G{\"o}ttingen,
Friedrich-Hund-Platz 1, 37077 G{\"o}ttingen, Germany}
\affiliation{Department of Mathematics, King’s College London, London WC2R 2LS, UK}


\date{\today}

\begin{abstract}
We study a mean field elastoplastic model, embedded within a disordered landscape of local yield barriers, to shed light on the behaviour of athermal amorphous solids subject to oscillatory shear. We show that the model presents a genuine dynamical transition between an elastic
and a yielded state, and qualitatively reproduces the dependence on the initial degree of annealing found in particle simulations. For initial conditions prepared below the analytically derived threshold energy, 
we observe a non-trivial, non-monotonic approach to the yielded state. The timescale diverges as one approaches the yielding point from above, which we identify with the fatigue limit. We finally discuss the connections to brittle yielding under uniform shear.

\end{abstract}


\maketitle

The behaviour of amorphous solids (characterised by the lack of any regular structure) under deformations is of great practical importance, and has long been an active topic because the disorder inherent in these systems poses a significant challenge to their understanding \cite{bonn_yield_2017,berthier_theoretical_2011,nicolas_deformation_2018}. These
materials typically show yielding behaviour: although they behave elastically at small deformation, plastic deformation eventually sets in, leading 
to a flowing state. Given the large variety of amorphous solids, ranging from hard metallic
glasses to soft colloidal gels or emulsions, so-called elastoplastic models~\cite{nicolas_deformation_2018} aim for a unified description from a statistical physics point of view.

A key aspect of yielding under uniform shear, which has received much attention recently \cite{shi_strain_2005,ozawa_random_2018,ozawa_rare_2021,ozawa_role_2020,singh_brittle_2020,barlow_ductile_2020,borja_da_rocha_rigidity-controlled_2020},
concerns the dependence on the initial degree of annealing --  quantified by potential energy -- of the amorphous solid (or ``glass'' for short) before deformation starts. Typically, it is found that poorly annealed glasses yield in a smooth, ductile manner, with plastic deformation appearing gradually, while well annealed glasses may yield in a brittle manner, accompanied by a macroscopic stress drop. Under startup of steady shear, although some features are still debated \cite{barlow_ductile_2020}, there is strong evidence that, at least in the brittle case, and under quasistatic loading, yielding corresponds to a discontinuous 
non-equilibrium transition, which in finite-dimensional systems is accompanied by the sudden appearance of a unique system-spanning shear band \cite{ozawa_random_2018,ozawa_rare_2021,ozawa_role_2020,singh_brittle_2020}.

Yielding under oscillatory shear
has until recently received less attention, although it may in some respects be a more informative protocol than the uniform case. One advantage is that one may probe directly the steady state after many cycles both below and above the yield point, whereas in the uniform case the states up to yielding are inherently transient. Furthermore, oscillatory strain allows one to relate  macroscopic yielding directly to a sharp absorbing-to-diffusive transition in the nature of the microscopic trajectories~\cite{fiocco_oscillatory_2013,priezjev_heterogeneous_2013,regev_onset_2013,kawasaki_macroscopic_2016,leishangthem_yielding_2017}
and shear jamming~\cite{pine_chaos_2005,corte_random_2008,nagasawa_classification_2019,das_unified_2020,babu_dilatancy_2021}.



Behaviour under oscillatory shear also shows intriguing dependencies on the initial degree of annealing. Atomistic simulations of model glasses \cite{yeh_glass_2020,bhaumik_role_2021,bhaumik_yielding_2021} reveal the appearance of a threshold initial energy.
Samples prepared above this threshold show mechanical annealing up to a common strain amplitude, the yield point, where the energy achieves the threshold value irrespectively of the initial condition. On approaching the yield point in strain, the timescale to anneal to the threshold energy
appears to diverge
\cite{kawasaki_macroscopic_2016,regev_onset_2013,fiocco_oscillatory_2013,leishangthem_yielding_2017}. 
On the other hand, samples prepared below the threshold are insensitive to shear up to an initial condition-dependent critical strain {\it above the common yield point}, where they then yield abruptly. 

Recent attempts to tackle this problem 
include energy landscape based \cite{szulc_forced_2020,sastry_models_2021}
and $2$D lattice elastoplastic models \cite{liu_fate_2021,khirallah_yielding_2021}.
While the elastoplastic models~\cite{liu_fate_2021,khirallah_yielding_2021} defy analytical progress as they implement the full spatial interaction kernel, the approach of Ref.~\cite{sastry_models_2021} ignores interactions between elements, and indeed does not display a genuine yielding transition as the steady state after sufficiently 
many cycles is always elastic.
An Ehrenfest-type model has also been proposed along these lines \cite{mungan_metastability_2021}. This incorporates a simplified form of mechanical noise but does not explicitly 
represent oscillatory shear.


Here, we consider a model with a similar single element description to \cite{liu_fate_2021,sastry_models_2021}
that accounts for the disordered energy landscape, while including the elastic interactions in a mean field manner that allows for analytical progress. Importantly, we find a steady state yielding transition 
and are able to reproduce qualitatively all the main features of yielding under oscillatory shear described above. 


\textit{Disordered HL model.}---The H\'{e}braud-Lequeux (HL) model for the rheology of amorphous solids \cite{hebraud_mode-coupling_1998} is a mean field mesoscopic elastoplastic model, which despite its many idealisations has had remarkable successes and been widely studied \cite{agoritsas_non-trivial_2017,liu_mean-field_2018,agoritsas_relevance_2015,puosi_probing_2015,bouchaud_spontaneous_2016,ekeh_power_2021,sollich_aging_2017}. The material is conceptually divided into mesoscopic elements, large enough to carry a local elastic strain $l$ and stress $\sigma = k~l$; these are related by an elastic modulus $k$ that is considered to be uniform throughout the system for simplicity. In the elastoplastic approach, the dynamics of the elements is 
described as consisting of periods of elastic loading interrupted by plastic events that are accompanied by a local stress drop. In a mean field fashion, the effect of stress propagation from other yield events is considered as a mechanical noise \cite{nicolas_deformation_2018,agoritsas_relevance_2015}, leading to a diffusive dynamics in the local strain $l$ (or equivalently the stress).



In the original HL model~\cite{hebraud_mode-coupling_1998}, all elements have a common strain threshold related to the common yield energy $E$ as $l_c=\sqrt{2E/k}$. However, due to the over-simplification of considering one single energy barrier throughout the system, this model is unable to capture the rich phenomenology under oscillatory shear found in particle simulations, for which it is essential to take into account the full energy landscape each mesoscopic element has access to.

An extension of the HL model to include this energy landscape, following previous approaches such as the SGR model \cite{sollich_rheology_1997}, was introduced in \cite{agoritsas_relevance_2015}. The essential ingredient is the disorder in the depth $E$ of the energy minima relative to a common reference energy, 
characterised by a distribution $\rho(E)$. Each time an element yields, it occupies a new local minimum with a depth extracted from this distribution. The depth $E$ of the current local energy minimum is thus promoted to a stochastic variable, and the system is described by a joint distribution $P(E,l)$ evolving as
\begin{eqnarray}\label{ME}
    \partial_t P(E,l,t)=-\dot{\gamma}\partial_l P+D(t)k^{-2}\partial_l^2 P \nonumber\\{}+Y(t)\rho(E)\delta(l)-\pltime^{-1}\theta(|l|-\sqrt{{2E}/{k}})P
\end{eqnarray}
with
\begin{equation}\label{Y}
    Y(t)=\frac{1}{\pltime}\int_0^{\infty} \mathrm{d}E\int_{-\infty}^{\infty} \mathrm{d}l \  P(E,l,t)  \ \theta\left(|l|-\sqrt{\frac{2E}{k}}\right)
\end{equation}
where $\theta$ and $\delta$ denote the Heaviside and delta functions, respectively, and $\dot{\gamma}$ is the applied shear rate.
%
$\pltime^{-1}$, the plastic rate, is the rate at which a plastic event occurs
once an element is strained beyond its yield threshold. We fix energy and time units by setting $k=1$ and $\pltime=1$.
The quantity $Y(t)$ in  (\ref{Y}) is the \textit{yield rate}, i.e.\ the fraction of elements that yield per unit time.
The key feature of the model is the \textit{closure relation} relating the yield rate to the diffusion constant $D(t)$.
We adopt the simple proportionality $D(t)=\alpha Y(t)$ \cite{hebraud_mode-coupling_1998}. The coupling constant $\alpha$ effectively sets the strength of the interactions, and under certain assumptions can be directly related to the elastic stress propagator~\cite{bocquet_kinetic_2009,agoritsas_relevance_2015}.
In the SM \cite{sm} we check that a more general closure relation, which reflects the fact that yield events contribute differently to the noise depending on their local barrier,
leaves the theory essentially unchanged, with only slight quantitative changes in the transient behaviour.

After its introduction in~\cite{agoritsas_relevance_2015}, the approach described by Eqs.~(\ref{ME},\ref{Y}) has not been developed further as it is somewhat unwieldy to tackle analytically; in particular it has not been used to study oscillatory shear. Our first contribution 
will be to determine a dynamical transition in Eq.~(\ref{ME}) under oscillatory shear, separating a frozen elastically-deforming solid state from a yielded state.


\begin{figure}
\includegraphics[width=0.35\textwidth]{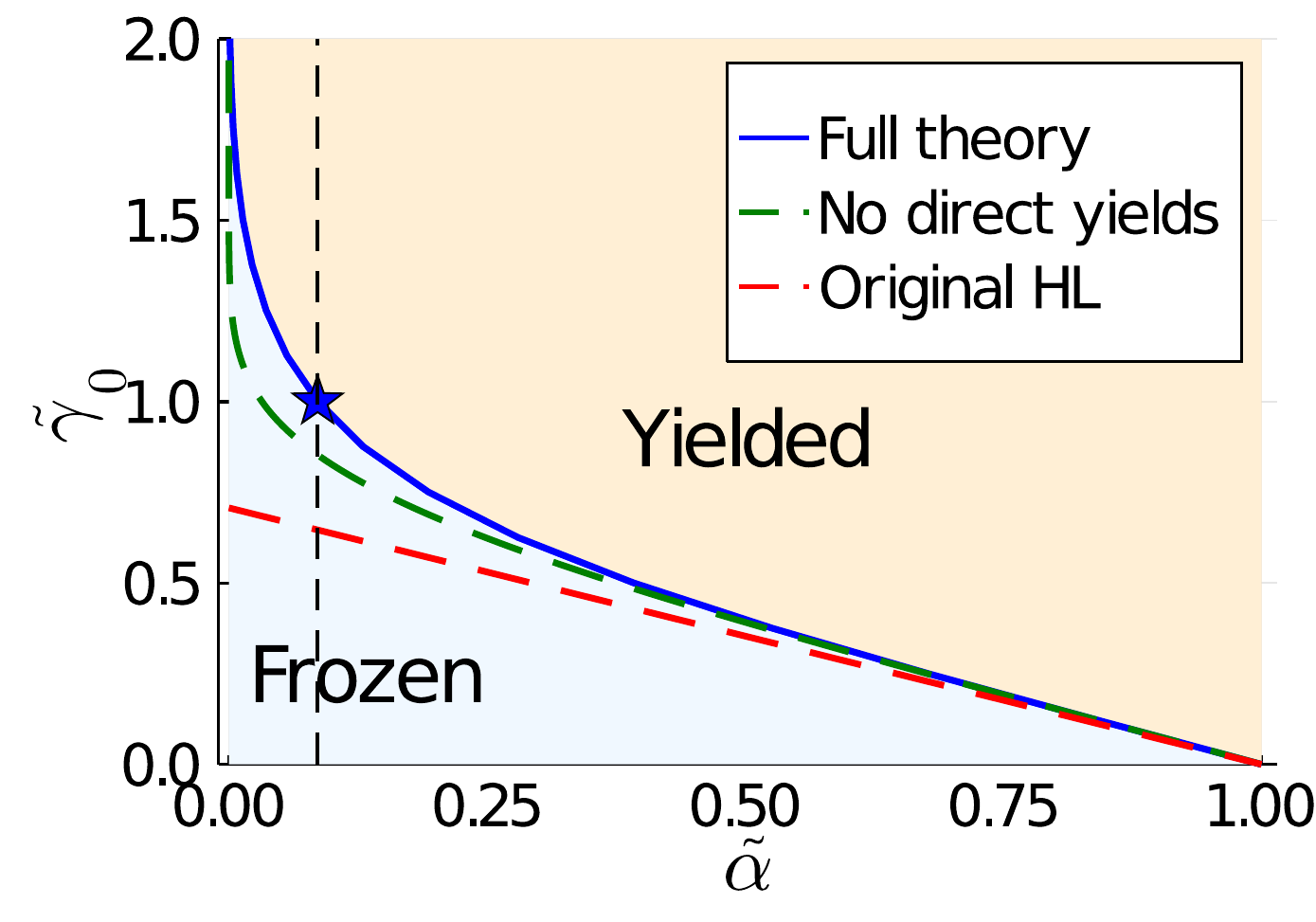}
\caption{\label{fig:phase} 
Phase diagram of the model in the $\talpha-\tgamma$ plane for a Gaussian $\rho(E)$. Vertical dashed lines indicate the fixed coupling value $\talpha=0.086$, where ${\tgamma}^{*}=1$ (see star), chosen for studying the initial annealing dependence.}
\end{figure}

\textit{Transition line.}---We consider applying oscillatory shear $\gamma(t)=\gamma_0 \sin{\left(\omega t\right)}$ in (\ref{ME}), with a fixed low frequency $
\omega 
\ll 1$ so that we are in the quasistatic regime. 
The two control parameters are thus the strain amplitude $\gamma_0$ and the coupling constant $\alpha$. For convenience, we introduce the rescaled versions $\tgamma=\gamma_0/\sqrt{\Em}$ and $\talpha=\alpha /\Em$, $\Em$ being the average over the disorder distribution $\rho(E)$. From \cite{agoritsas_relevance_2015}, the physically relevant parameter regime of the disordered HL model is known to be $\talpha<1$, where the system is jammed in the absence of shear. Within this jammed regime, we now calculate the transition line $\tgamma^{*}(\talpha)$ above which there exists a yielded steady state.


We proceed as follows. At a fixed $\tilde{\alpha}$, suppose $\tgamma$ is large enough so that (\ref{ME}) has a yielded steady state.
Rescaling time by the period $T$ so that $\tau=t/T=\omega t/(2\pi)$, $\tau \in [0,1] $, this steady state is characterised by a non-zero period-averaged yield rate $\overline{Y}=\int_0^{1}Y(\tau)\mathrm{d}\tau$. As $\tgamma$ is decreased towards $\tgamma^{*}$, we take $\overline{Y}$  to vanish smoothly -- an assumption we show to be self-consistent in the end -- with the rescaled yield rate $Y(\tau)/\overline{Y}=y(\tau)$ approaching a limiting form. 
In this limit, the key observation
from the dynamical equations (\ref{ME},\ref{Y}) is that the local yielding events can be classified into two distinct groups.



Suppose an element yields at a time $\tau' \in [0,1]$ within the period, and is assigned a new energy depth $E$. Neglecting strain diffusion, its local strain will subsequently evolve as $l(\tau)=\gamma(\tau)-\gamma(\tau')$. If $\gamma_0+|\gamma(\tau')|<\sqrt{2E}$, this element will therefore not be able to yield again in the next cycle; its strain will have to change diffusively (due to mechanical noise) during a large number of ensuing cycles until it comes close enough to the threshold $\sqrt{2E}$ to be swept across it by the external shear. In the limit of vanishing strain diffusion,
this will occur precisely at either the strain maximum or minimum within the cycle ($\tau=1/4$ or $3/4$).


The second group of events are the \textit{direct yields}. If $\gamma_0+|\gamma(\tau')|\geq \sqrt{2E}$, the element will yield within the ensuing cycle. It will do so at a time $\tau_y$ during the cycle that will depend on the previous yield time $\tau'$ and the corresponding shear strain
$\gamma(\tau')=\gamma_0 \sin(2\pi \tau')$, as well as on $E$ and $\gamma_0$.

Overall, one can therefore separate the limiting yield rate into two contributions as $y(\tau)=\yone(\tau)+\ytwo (\tau)$, corresponding to indirect and direct yields respectively. Conservation of probability then implies the following pair of self-consistent equations:
\begin{widetext}
\begin{eqnarray}\label{eq:pair}
    \yone(\tau)&=&
\frac{1}{2}\left[\delta\left(\tau-{1}/{4}\right)+\delta \left(\tau-{3}/{4}\right)\right]
    \int_{\gamma_0^2/2}^{\infty}\mathrm{d}E \ \rho(E)\int_0^{1}\mathrm{d}\tau' \ y(\tau') \ \theta\left(\sqrt{2E}-\gamma_0-|\gamma_0 \sin(2\pi\tau')|\right) \nonumber \\
    \ytwo(\tau)&=&\int_{0}^{\infty}\mathrm{d}E \  \rho(E) \ \int_{0}^{1}\mathrm{d}\tau' \ y(\tau') \ \delta\left(\tau-\tau_y(\tau',\gamma_0 \sin(2\pi \tau'),\gamma_0,E)\right)
\end{eqnarray}
\end{widetext}
which can be solved numerically in an iterative way~\cite{sm}.

Once the limiting form of the yield rate $y(\tau)$ is known, the full steady state distribution at the transition $P^{*}(E,l)$
can be obtained straightforwardly by applying the diffusion propagator with absorbing boundary conditions at the local yield thresholds. 
The critical coupling $\talpha^{*}(\tgamma)$ is then found by imposing normalisation of this distribution, and arises from the interplay between the disordered landscape and the timescale set by the mechanical noise.
A key property of $P^{*}(E,l)$ is that it is nonzero only for values $(E,l)$ from which all yields are indirect. 
The steady state probability of other elements vanishes as $\overline{Y}/\omega$ at the transition, but they still contribute to the total yield rate as they have yield rates $\sim\omega$.


Fig. \ref{fig:phase} shows the transition line 
for the specific case of a Gaussian yield energy distribution $\rho(E)\sim e^{-E^2/(2\sigma^2)}$. This is the form for $\rho(E)$ we will adopt in the rest of the work~\footnote{In the SM~\cite{sm} we show that an exponential $\rho(E)$ does not qualitatively change the form of the transition line, and argue more generally that our main results are qualitatively robust to the precise form of $\rho(E)$.}, to match the results of earlier numerical studies \cite{sastry_relationship_2001}. In Fig.~\ref{fig:phase} we also show the approximate solution obtained if one neglects direct yields; this is exact for $\talpha\rightarrow 1$. This approximation is useful to derive an exact bound \cite{sm} proving in general that in the presence of disorder the phase boundary lies above the original HL model: 
the inclusion of disorder (which entails deep traps where elements may get stuck) always tends to extend the size of the frozen region.


\begin{figure}
\includegraphics[width=0.4\textwidth]{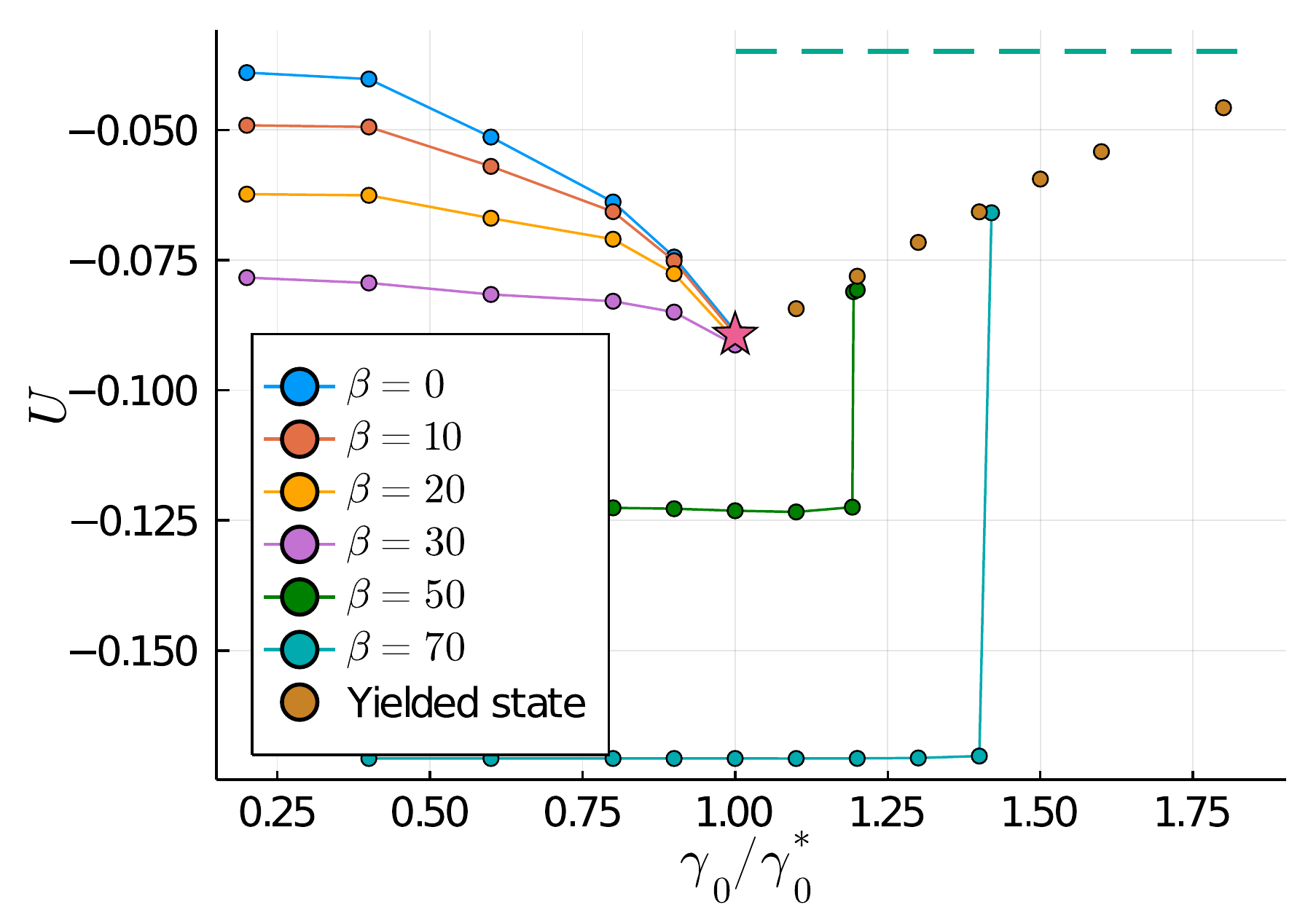}
\caption{\label{fig:energies}Stroboscopic energy in the steady state after application of many cycles of shear with amplitude $\gamma_0$. Star indicates analytically calculated threshold energy $U^{*}$; dashed line corresponds to the steady shear limit reached as $\gamma_0 \rightarrow \infty$, known from~\cite{agoritsas_relevance_2015}. Steady state energy values for $\gamma_0=\gamma_0^{*}$ and $ 0.9\gamma_0^{*}$ are obtained from a power-law extrapolation of the slow relaxation \cite{sm}.}
\end{figure}

\textit{Dependence on initial degree of annealing}---Although we have proven that in the yielded region of Fig.~\ref{fig:phase} a fluid steady state \textit{exists}, 
whether this ergodic state is reached depends crucially on the initial condition. We now study the master equation (\ref{ME}) numerically, while fixing $\talpha=0.086$ (where $\tgamma^{*}\approx 1$, see star in Fig.~\ref{fig:phase}), and setting the variance of the Gaussian to $\sigma=0.05$ as in \cite{sastry_models_2021}. Numerical solutions entail choosing a discrete set of energy levels $\{E_i\}$, and solving a PDE in the strain variable for each \cite{sm}. As a proxy for different degrees of \textit{thermal annealing} of the initial glass, we generate initial conditions of the form $P(E,t=0)\sim \rho(E)e^{\beta E}$, introducing an inverse temperature $\beta$. Physically, increasing $\beta$ can be interpreted as decreasing the density of weak zones in the system, here represented by the shallow energy levels. As regards the initial local strains, we consider them to be well-relaxed (narrowly-distributed) within each energy level, with standard deviation in strain $l_c(E)/6$. 

In Fig.~\ref{fig:energies} we show the stroboscopic ($\gamma=0$) energy in the steady state after application of many cycles of shear at a given amplitude $\gamma_0$. On the solid side, this corresponds to a frozen state with $\overline{Y}=0$; on the yielded side, this is the ergodic state with $\overline{Y}
>0$. 
The total energy is measured within the model as $U=\int \mathrm{d}E \int \mathrm{d}l \ \left(-E + l^2/2\right) P(E,l)$, i.e.\ energy at the bottom of each minimum plus elastic energy, while the macroscopic stress (see below) is $\Sigma=\int \mathrm{d}E \int \mathrm{d}l \ l \ P(E,l)$. 
The main features found in MD simulations \cite{yeh_glass_2020,bhaumik_role_2021,bhaumik_yielding_2021} are reproduced in Fig.~\ref{fig:energies}. Within the precision and range of our numerics, yielding for poorly annealed samples appears as a cusp in $U$ at the common yield point $\gamma_0^{*}$, 
while well-annealed samples are insensitive to shear up to a critical strain $\gamma_c(\beta)>\gamma_0^{*}$. The threshold energy (and corresponding $\beta^{*}$) separating the two types of yielding simply arise as the lower limit of the ergodic state. 
The corresponding data for the macroscopic stress amplitude in the steady state \cite{sm} 
also qualitatively reproduce the behaviour in \cite{yeh_glass_2020,bhaumik_role_2021,leishangthem_yielding_2017}, with a finite drop in steady state macroscopic stress appearing for samples prepared below the threshold. 

We note that the MD studies of \cite{bhaumik_role_2021,leishangthem_yielding_2017,parmar_strain_2019} report a small (essentially invisible on the scale of Fig.~\ref{fig:energies}) jump in energy and macroscopic stress amplitude at $\gamma_0^{*}$. The origin of this effect, which appears to survive for large system size, is unclear. We expect that in our mean field model both energy and stress remain continuous on approaching from the solid side, and our numerics are consistent with this. From the fluid side, our theory predicts that $\overline{Y}$ vanishes continuously, reminiscent of e.g.\ the second order transition scenario of \cite{ness_absorbing-state_2020}. Closer inspection reveals that samples initialized above the threshold energy display critical behaviour at $\gamma_0^{*}$, where the yield rate decays as $\overline{Y}(t)\sim t^{-b}$, with an $\talpha$-dependent exponent $b\leq 1$. This implies a diverging number of events for long time, allowing the system to lose memory of its initial condition. The critical power law decay of $\overline{Y}(t)$ also means that relaxation timescales must diverge on the approach from either side of the transition.

\begin{figure}
\includegraphics[width=0.44\textwidth]{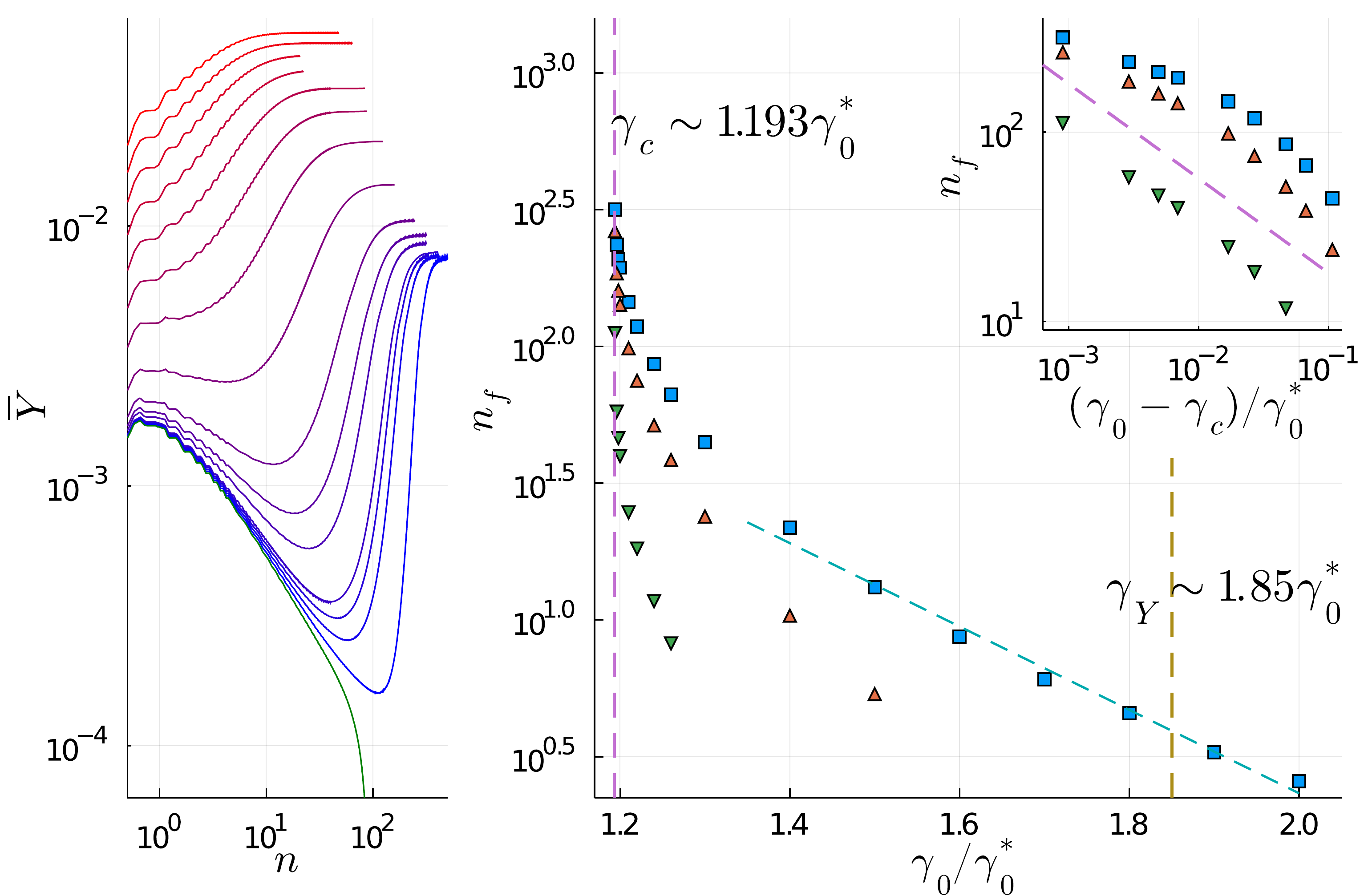}
\caption{\label{fig:well_annealed}\textit{Fatigue} behaviour for well-annealed sample ($\beta=50$). Left: non-monotonic behaviour of period-averaged yield rate $\overline{Y}$ against number of cycles. Strain amplitudes $\gamma_0$ range from $1.194\gamma_0^{*}$ (blue) to $2\gamma_0^{*}$ (red) (see SM \cite{sm} for precise values); also shown is $1.192\gamma_0^{*}$ (green), below the fatigue limit. Right: for the same strain amplitudes, three measures of the fluidisation time as described in the text, $75\%$ of final yield rate (squares), inflection point (up-triangles) and cycles to reach the minimum (down-triangles). Inset: divergence of timescales above $\gamma_c (\beta)$, consistent with an inverse square root (dashed line) for the minimum.}
\end{figure}

\textit{Fatigue.}---Turning to samples with initial energy below the threshold energy, which yield at $\gamma_c(\beta)>\gamma_0^{*}$, we find very interesting transient behaviour. 
As shown in Fig.~\ref{fig:well_annealed} for $\beta=50$, close to $\gamma_c(\beta)$ the yield rate $\overline{Y}$ displays strongly non-monotonic behaviour.
Although in our mean field model the yielded state is reached smoothly, one generally finds that in finite dimensional systems, once the plastic activity starts to increase, an instability develops leading to shear banding or even material failure \cite{fielding_triggers_2016}.
As a proxy for the time to failure (expressed in number of cycles, $n_{\rm f}$~\footnote{For the number of cycles we simply take the time divided by the period, so that it takes continuous values in Fig.~\ref{fig:well_annealed}}), we take the inflection point of $\overline{Y}(t)$ as done in \cite{liu_mean-field_2018} for creep (where it is associated to banding~\cite{divoux_stress-induced_2011}), as well as the point at which $\overline{Y}$ reaches $75 \%$ of its steady state value, which allows us to analyse larger $\gamma_0$ where an inflection is not present. We additionally consider the number of cycles at which the minimum of $\overline{Y}$ is reached. We find (Fig.~\ref{fig:well_annealed}) that the number of cycles $n_{\rm f}$ decreases rapidly (consistent with an exponential) towards unity as $\gamma_0$ is increased towards $\gamma_Y$, the yield point determined for $\beta=50$ under startup of steady shear~\cite{sm}. This is very reminiscent of \textit{fatigue failure} \cite{carmona_computer_2007,pradhan_failure_2010,bhowmik_fatigue_2021,kun_fatigue_2007,sha_cyclic_2015} found e.g.\ in metallic glasses. Close to $\gamma_c(\beta)$, the timescale associated to the minimum shows a clear power-law divergence, consistent with an inverse square root. The similarity of this mean field fatigue behaviour with creep flow suggests the intriguing possibility that this divergence may be understood from a Landau-type scaling argument as recently proposed in \cite{popovic_scaling_2021} for creep.

A closer look at the dynamics of the mean field model near $\gamma_c$ reveals that, during the initial cycles, the plastic activity is dominated by direct yielding of rare shallow elements~\cite{sm}, which we recall may be thought of as weak zones in the material. 
At intermediate times, the energy distribution $P(E)$ then almost settles down to a frozen fixed point. However, eventually the accumulated strain diffusion is enough to trigger yield events across the entire energy spectrum (including deep levels where the bulk of the population lies), driving the system away 
towards the yielded steady state.


\textit{Concluding remarks.}---In this Letter, we have presented a mean field mesoscopic elastoplastic approach to study yielding behaviour under oscillatory shear. Our first contribution was to demonstrate the existence of a dynamical yielding transition and to characterize it analytically. Secondly, despite its relative simplicity, we have shown that the model reproduces the key phenomena related to initial annealing dependence found in MD studies \cite{yeh_glass_2020,bhaumik_role_2021,fiocco_oscillatory_2013}. Thirdly, we showed that the dynamics of well annealed samples exhibits characteristics of fatigue failure so that the model also contributes to the understanding of this phenomenon.

We comment finally on the contrast to brittle yielding under uniform shear, and the relative importance of shear banding. There are important differences between the two shear protocols, and these are also reflected in the mean field model. On the one hand, under oscillatory shear, both the existence of a sharp yield point and the initial annealing dependence (Fig.~\ref{fig:energies}), which we recall concern \textit{steady state} quantities, are largely unaffected by the shear rate $\dot{\gamma}_0$ (equivalent to frequency via $\dot{\gamma}_0=\gamma_0 \omega$), and are independent of the presence of banding. This was found in MD studies~\cite{yeh_glass_2020} and is also supported here, where numerical results with finite frequency ($\omega=0.1$) largely agree with the $\omega \rightarrow 0$ theory. Under uniform shear, on the other hand, brittle yielding can only strictly be defined for $\dot{\gamma}\rightarrow 0$, where a macroscopic stress drop is caused by the formation of a system-spanning shear band~\cite{singh_brittle_2020}. Indeed, under uniform shear, the disordered HL model shows no sign of brittle yielding even in the $\dot{\gamma}\rightarrow 0$ limit, reflecting the absence of banding in mean field~\footnote{In the SM~\cite{sm} we provide more discussion and contrast this to the mean field results of \cite{ozawa_random_2018,popovic_elastoplastic_2018}}. Regarding the \textit{transient} dynamics under oscillatory shear, we expect the mean field model to become more accurate away from the quasistatic limit, as should the approximation of Gaussian mechanical noise~\cite{liu_driving_2016}.

As avenues of future research, one could improve upon the diffusive approximation and study the model with power-law mechanical noise~\cite{parley_aging_2020,lin_mean-field_2016}.
It would also be interesting
to include thermal activation over barriers (as in SGR \cite{sollich_rheology_1997},\cite{mungan_metastability_2021}) within the elastoplastic model, following \cite{popovic_thermally_2021,ferrero_yielding_2021}. Fascinating questions arise, starting with the phase diagram: would the existence of activation always lead to a yielded state? How will temperature influence the fatigue behaviour?

\begin{acknowledgements}
The authors thank Suzanne Fielding, James Cochran and Muhittin Mungan for helpful discussions.
SS acknowledges support through the JC Bose Fellowship  (JBR/2020/000015) SERB, DST (India).
\end{acknowledgements}

\bibliography{ms.bib}

\begin{thebibliography}{59}%
\makeatletter
\providecommand \@ifxundefined [1]{%
 \@ifx{#1\undefined}
}%
\providecommand \@ifnum [1]{%
 \ifnum #1\expandafter \@firstoftwo
 \else \expandafter \@secondoftwo
 \fi
}%
\providecommand \@ifx [1]{%
 \ifx #1\expandafter \@firstoftwo
 \else \expandafter \@secondoftwo
 \fi
}%
\providecommand \natexlab [1]{#1}%
\providecommand \enquote  [1]{``#1''}%
\providecommand \bibnamefont  [1]{#1}%
\providecommand \bibfnamefont [1]{#1}%
\providecommand \citenamefont [1]{#1}%
\providecommand \href@noop [0]{\@secondoftwo}%
\providecommand \href [0]{\begingroup \@sanitize@url \@href}%
\providecommand \@href[1]{\@@startlink{#1}\@@href}%
\providecommand \@@href[1]{\endgroup#1\@@endlink}%
\providecommand \@sanitize@url [0]{\catcode `\\12\catcode `\$12\catcode
  `\&12\catcode `\#12\catcode `\^12\catcode `\_12\catcode `\%12\relax}%
\providecommand \@@startlink[1]{}%
\providecommand \@@endlink[0]{}%
\providecommand \url  [0]{\begingroup\@sanitize@url \@url }%
\providecommand \@url [1]{\endgroup\@href {#1}{\urlprefix }}%
\providecommand \urlprefix  [0]{URL }%
\providecommand \Eprint [0]{\href }%
\providecommand \doibase [0]{https://doi.org/}%
\providecommand \selectlanguage [0]{\@gobble}%
\providecommand \bibinfo  [0]{\@secondoftwo}%
\providecommand \bibfield  [0]{\@secondoftwo}%
\providecommand \translation [1]{[#1]}%
\providecommand \BibitemOpen [0]{}%
\providecommand \bibitemStop [0]{}%
\providecommand \bibitemNoStop [0]{.\EOS\space}%
\providecommand \EOS [0]{\spacefactor3000\relax}%
\providecommand \BibitemShut  [1]{\csname bibitem#1\endcsname}%
\let\auto@bib@innerbib\@empty
\bibitem [{\citenamefont {Bonn}\ \emph {et~al.}(2017)\citenamefont {Bonn},
  \citenamefont {Denn}, \citenamefont {Berthier}, \citenamefont {Divoux},\ and\
  \citenamefont {Manneville}}]{bonn_yield_2017}%
  \BibitemOpen
  \bibfield  {author} {\bibinfo {author} {\bibfnamefont {D.}~\bibnamefont
  {Bonn}}, \bibinfo {author} {\bibfnamefont {M.~M.}\ \bibnamefont {Denn}},
  \bibinfo {author} {\bibfnamefont {L.}~\bibnamefont {Berthier}}, \bibinfo
  {author} {\bibfnamefont {T.}~\bibnamefont {Divoux}},\ and\ \bibinfo {author}
  {\bibfnamefont {S.}~\bibnamefont {Manneville}},\ }\bibfield  {title}
  {\bibinfo {title} {Yield stress materials in soft condensed matter},\ }\href
  {https://doi.org/10.1103/RevModPhys.89.035005} {\bibfield  {journal}
  {\bibinfo  {journal} {Rev. Mod. Phys.}\ }\textbf {\bibinfo {volume} {89}},\
  \bibinfo {pages} {035005} (\bibinfo {year} {2017})}\BibitemShut {NoStop}%
\bibitem [{\citenamefont {Berthier}\ and\ \citenamefont
  {Biroli}(2011)}]{berthier_theoretical_2011}%
  \BibitemOpen
  \bibfield  {author} {\bibinfo {author} {\bibfnamefont {L.}~\bibnamefont
  {Berthier}}\ and\ \bibinfo {author} {\bibfnamefont {G.}~\bibnamefont
  {Biroli}},\ }\bibfield  {title} {\bibinfo {title} {Theoretical perspective on
  the glass transition and amorphous materials},\ }\href
  {https://doi.org/10.1103/RevModPhys.83.587} {\bibfield  {journal} {\bibinfo
  {journal} {Rev. Mod. Phys.}\ }\textbf {\bibinfo {volume} {83}},\ \bibinfo
  {pages} {587} (\bibinfo {year} {2011})}\BibitemShut {NoStop}%
\bibitem [{\citenamefont {Nicolas}\ \emph {et~al.}(2018)\citenamefont
  {Nicolas}, \citenamefont {Ferrero}, \citenamefont {Martens},\ and\
  \citenamefont {Barrat}}]{nicolas_deformation_2018}%
  \BibitemOpen
  \bibfield  {author} {\bibinfo {author} {\bibfnamefont {A.}~\bibnamefont
  {Nicolas}}, \bibinfo {author} {\bibfnamefont {E.~E.}\ \bibnamefont
  {Ferrero}}, \bibinfo {author} {\bibfnamefont {K.}~\bibnamefont {Martens}},\
  and\ \bibinfo {author} {\bibfnamefont {J.-L.}\ \bibnamefont {Barrat}},\
  }\bibfield  {title} {\bibinfo {title} {Deformation and flow of amorphous
  solids: {Insights} from elastoplastic models},\ }\href
  {https://doi.org/10.1103/RevModPhys.90.045006} {\bibfield  {journal}
  {\bibinfo  {journal} {Rev. Mod. Phys.}\ }\textbf {\bibinfo {volume} {90}},\
  \bibinfo {pages} {045006} (\bibinfo {year} {2018})}\BibitemShut {NoStop}%
\bibitem [{\citenamefont {Shi}\ and\ \citenamefont
  {Falk}(2005)}]{shi_strain_2005}%
  \BibitemOpen
  \bibfield  {author} {\bibinfo {author} {\bibfnamefont {Y.}~\bibnamefont
  {Shi}}\ and\ \bibinfo {author} {\bibfnamefont {M.~L.}\ \bibnamefont {Falk}},\
  }\bibfield  {title} {\bibinfo {title} {Strain {Localization} and
  {Percolation} of {Stable} {Structure} in {Amorphous} {Solids}},\ }\href
  {https://doi.org/10.1103/PhysRevLett.95.095502} {\bibfield  {journal}
  {\bibinfo  {journal} {Phys. Rev. Lett.}\ }\textbf {\bibinfo {volume} {95}},\
  \bibinfo {pages} {095502} (\bibinfo {year} {2005})}\BibitemShut {NoStop}%
\bibitem [{\citenamefont {Ozawa}\ \emph {et~al.}(2018)\citenamefont {Ozawa},
  \citenamefont {Berthier}, \citenamefont {Biroli}, \citenamefont {Rosso},\
  and\ \citenamefont {Tarjus}}]{ozawa_random_2018}%
  \BibitemOpen
  \bibfield  {author} {\bibinfo {author} {\bibfnamefont {M.}~\bibnamefont
  {Ozawa}}, \bibinfo {author} {\bibfnamefont {L.}~\bibnamefont {Berthier}},
  \bibinfo {author} {\bibfnamefont {G.}~\bibnamefont {Biroli}}, \bibinfo
  {author} {\bibfnamefont {A.}~\bibnamefont {Rosso}},\ and\ \bibinfo {author}
  {\bibfnamefont {G.}~\bibnamefont {Tarjus}},\ }\bibfield  {title} {\bibinfo
  {title} {A random critical point separates brittle and ductile yielding
  transitions in amorphous materials},\ }\href
  {https://doi.org/10.1073/pnas.1806156115} {\bibfield  {journal} {\bibinfo
  {journal} {Proceedings of the National Academy of Sciences}\ }\textbf
  {\bibinfo {volume} {115}},\ \bibinfo {pages} {6656} (\bibinfo {year}
  {2018})}\BibitemShut {NoStop}%
\bibitem [{\citenamefont {Ozawa}\ \emph {et~al.}(2021)\citenamefont {Ozawa},
  \citenamefont {Berthier}, \citenamefont {Biroli},\ and\ \citenamefont
  {Tarjus}}]{ozawa_rare_2021}%
  \BibitemOpen
  \bibfield  {author} {\bibinfo {author} {\bibfnamefont {M.}~\bibnamefont
  {Ozawa}}, \bibinfo {author} {\bibfnamefont {L.}~\bibnamefont {Berthier}},
  \bibinfo {author} {\bibfnamefont {G.}~\bibnamefont {Biroli}},\ and\ \bibinfo
  {author} {\bibfnamefont {G.}~\bibnamefont {Tarjus}},\ }\bibfield  {title}
  {\bibinfo {title} {Rare events and disorder control the brittle yielding of
  amorphous solids},\ }\href {http://arxiv.org/abs/2102.05846} {\bibfield
  {journal} {\bibinfo  {journal} {arXiv:2102.05846 [cond-mat]}\ } (\bibinfo
  {year} {2021})}\BibitemShut {NoStop}%
\bibitem [{\citenamefont {Ozawa}\ \emph {et~al.}(2020)\citenamefont {Ozawa},
  \citenamefont {Berthier}, \citenamefont {Biroli},\ and\ \citenamefont
  {Tarjus}}]{ozawa_role_2020}%
  \BibitemOpen
  \bibfield  {author} {\bibinfo {author} {\bibfnamefont {M.}~\bibnamefont
  {Ozawa}}, \bibinfo {author} {\bibfnamefont {L.}~\bibnamefont {Berthier}},
  \bibinfo {author} {\bibfnamefont {G.}~\bibnamefont {Biroli}},\ and\ \bibinfo
  {author} {\bibfnamefont {G.}~\bibnamefont {Tarjus}},\ }\bibfield  {title}
  {\bibinfo {title} {Role of fluctuations in the yielding transition of
  two-dimensional glasses},\ }\href
  {https://doi.org/10.1103/PhysRevResearch.2.023203} {\bibfield  {journal}
  {\bibinfo  {journal} {Phys. Rev. Research}\ }\textbf {\bibinfo {volume}
  {2}},\ \bibinfo {pages} {023203} (\bibinfo {year} {2020})}\BibitemShut
  {NoStop}%
\bibitem [{\citenamefont {Singh}\ \emph {et~al.}(2020)\citenamefont {Singh},
  \citenamefont {Ozawa},\ and\ \citenamefont {Berthier}}]{singh_brittle_2020}%
  \BibitemOpen
  \bibfield  {author} {\bibinfo {author} {\bibfnamefont {M.}~\bibnamefont
  {Singh}}, \bibinfo {author} {\bibfnamefont {M.}~\bibnamefont {Ozawa}},\ and\
  \bibinfo {author} {\bibfnamefont {L.}~\bibnamefont {Berthier}},\ }\bibfield
  {title} {\bibinfo {title} {Brittle yielding of amorphous solids at finite
  shear rates},\ }\href {https://doi.org/10.1103/PhysRevMaterials.4.025603}
  {\bibfield  {journal} {\bibinfo  {journal} {Phys. Rev. Materials}\ }\textbf
  {\bibinfo {volume} {4}},\ \bibinfo {pages} {025603} (\bibinfo {year}
  {2020})}\BibitemShut {NoStop}%
\bibitem [{\citenamefont {Barlow}\ \emph {et~al.}(2020)\citenamefont {Barlow},
  \citenamefont {Cochran},\ and\ \citenamefont
  {Fielding}}]{barlow_ductile_2020}%
  \BibitemOpen
  \bibfield  {author} {\bibinfo {author} {\bibfnamefont {H.~J.}\ \bibnamefont
  {Barlow}}, \bibinfo {author} {\bibfnamefont {J.~O.}\ \bibnamefont
  {Cochran}},\ and\ \bibinfo {author} {\bibfnamefont {S.~M.}\ \bibnamefont
  {Fielding}},\ }\bibfield  {title} {\bibinfo {title} {Ductile and brittle
  yielding in thermal and athermal amorphous materials},\ }\href
  {https://doi.org/10.1103/PhysRevLett.125.168003} {\bibfield  {journal}
  {\bibinfo  {journal} {Phys. Rev. Lett.}\ }\textbf {\bibinfo {volume} {125}},\
  \bibinfo {pages} {168003} (\bibinfo {year} {2020})}\BibitemShut {NoStop}%
\bibitem [{\citenamefont {Borja~da Rocha}\ and\ \citenamefont
  {Truskinovsky}(2020)}]{borja_da_rocha_rigidity-controlled_2020}%
  \BibitemOpen
  \bibfield  {author} {\bibinfo {author} {\bibfnamefont {H.}~\bibnamefont
  {Borja~da Rocha}}\ and\ \bibinfo {author} {\bibfnamefont {L.}~\bibnamefont
  {Truskinovsky}},\ }\bibfield  {title} {\bibinfo {title}
  {Rigidity-{Controlled} {Crossover}: {From} {Spinodal} to {Critical}
  {Failure}},\ }\href {https://doi.org/10.1103/PhysRevLett.124.015501}
  {\bibfield  {journal} {\bibinfo  {journal} {Phys. Rev. Lett.}\ }\textbf
  {\bibinfo {volume} {124}},\ \bibinfo {pages} {015501} (\bibinfo {year}
  {2020})}\BibitemShut {NoStop}%
\bibitem [{\citenamefont {Fiocco}\ \emph {et~al.}(2013)\citenamefont {Fiocco},
  \citenamefont {Foffi},\ and\ \citenamefont
  {Sastry}}]{fiocco_oscillatory_2013}%
  \BibitemOpen
  \bibfield  {author} {\bibinfo {author} {\bibfnamefont {D.}~\bibnamefont
  {Fiocco}}, \bibinfo {author} {\bibfnamefont {G.}~\bibnamefont {Foffi}},\ and\
  \bibinfo {author} {\bibfnamefont {S.}~\bibnamefont {Sastry}},\ }\bibfield
  {title} {\bibinfo {title} {Oscillatory athermal quasistatic deformation of a
  model glass},\ }\href {https://doi.org/10.1103/PhysRevE.88.020301} {\bibfield
   {journal} {\bibinfo  {journal} {Phys. Rev. E}\ }\textbf {\bibinfo {volume}
  {88}},\ \bibinfo {pages} {020301} (\bibinfo {year} {2013})}\BibitemShut
  {NoStop}%
\bibitem [{\citenamefont {Priezjev}(2013)}]{priezjev_heterogeneous_2013}%
  \BibitemOpen
  \bibfield  {author} {\bibinfo {author} {\bibfnamefont {N.~V.}\ \bibnamefont
  {Priezjev}},\ }\bibfield  {title} {\bibinfo {title} {Heterogeneous relaxation
  dynamics in amorphous materials under cyclic loading},\ }\href
  {https://doi.org/10.1103/PhysRevE.87.052302} {\bibfield  {journal} {\bibinfo
  {journal} {Phys. Rev. E}\ }\textbf {\bibinfo {volume} {87}},\ \bibinfo
  {pages} {052302} (\bibinfo {year} {2013})}\BibitemShut {NoStop}%
\bibitem [{\citenamefont {Regev}\ \emph {et~al.}(2013)\citenamefont {Regev},
  \citenamefont {Lookman},\ and\ \citenamefont
  {Reichhardt}}]{regev_onset_2013}%
  \BibitemOpen
  \bibfield  {author} {\bibinfo {author} {\bibfnamefont {I.}~\bibnamefont
  {Regev}}, \bibinfo {author} {\bibfnamefont {T.}~\bibnamefont {Lookman}},\
  and\ \bibinfo {author} {\bibfnamefont {C.}~\bibnamefont {Reichhardt}},\
  }\bibfield  {title} {\bibinfo {title} {Onset of irreversibility and chaos in
  amorphous solids under periodic shear},\ }\href
  {https://doi.org/10.1103/PhysRevE.88.062401} {\bibfield  {journal} {\bibinfo
  {journal} {Phys. Rev. E}\ }\textbf {\bibinfo {volume} {88}},\ \bibinfo
  {pages} {062401} (\bibinfo {year} {2013})}\BibitemShut {NoStop}%
\bibitem [{\citenamefont {Kawasaki}\ and\ \citenamefont
  {Berthier}(2016)}]{kawasaki_macroscopic_2016}%
  \BibitemOpen
  \bibfield  {author} {\bibinfo {author} {\bibfnamefont {T.}~\bibnamefont
  {Kawasaki}}\ and\ \bibinfo {author} {\bibfnamefont {L.}~\bibnamefont
  {Berthier}},\ }\bibfield  {title} {\bibinfo {title} {Macroscopic yielding in
  jammed solids is accompanied by a nonequilibrium first-order transition in
  particle trajectories},\ }\href {https://doi.org/10.1103/PhysRevE.94.022615}
  {\bibfield  {journal} {\bibinfo  {journal} {Phys. Rev. E}\ }\textbf {\bibinfo
  {volume} {94}},\ \bibinfo {pages} {022615} (\bibinfo {year}
  {2016})}\BibitemShut {NoStop}%
\bibitem [{\citenamefont {Leishangthem}\ \emph {et~al.}(2017)\citenamefont
  {Leishangthem}, \citenamefont {Parmar},\ and\ \citenamefont
  {Sastry}}]{leishangthem_yielding_2017}%
  \BibitemOpen
  \bibfield  {author} {\bibinfo {author} {\bibfnamefont {P.}~\bibnamefont
  {Leishangthem}}, \bibinfo {author} {\bibfnamefont {A.~D.~S.}\ \bibnamefont
  {Parmar}},\ and\ \bibinfo {author} {\bibfnamefont {S.}~\bibnamefont
  {Sastry}},\ }\bibfield  {title} {\bibinfo {title} {The yielding transition in
  amorphous solids under oscillatory shear deformation},\ }\href
  {https://doi.org/10.1038/ncomms14653} {\bibfield  {journal} {\bibinfo
  {journal} {Nat Commun}\ }\textbf {\bibinfo {volume} {8}},\ \bibinfo {pages}
  {14653} (\bibinfo {year} {2017})}\BibitemShut {NoStop}%
\bibitem [{\citenamefont {Pine}\ \emph {et~al.}(2005)\citenamefont {Pine},
  \citenamefont {Gollub}, \citenamefont {Brady},\ and\ \citenamefont
  {Leshansky}}]{pine_chaos_2005}%
  \BibitemOpen
  \bibfield  {author} {\bibinfo {author} {\bibfnamefont {D.~J.}\ \bibnamefont
  {Pine}}, \bibinfo {author} {\bibfnamefont {J.~P.}\ \bibnamefont {Gollub}},
  \bibinfo {author} {\bibfnamefont {J.~F.}\ \bibnamefont {Brady}},\ and\
  \bibinfo {author} {\bibfnamefont {A.~M.}\ \bibnamefont {Leshansky}},\
  }\bibfield  {title} {\bibinfo {title} {Chaos and threshold for
  irreversibility in sheared suspensions},\ }\href
  {https://doi.org/10.1038/nature04380} {\bibfield  {journal} {\bibinfo
  {journal} {Nature}\ }\textbf {\bibinfo {volume} {438}},\ \bibinfo {pages}
  {997} (\bibinfo {year} {2005})}\BibitemShut {NoStop}%
\bibitem [{\citenamefont {Corté}\ \emph {et~al.}(2008)\citenamefont {Corté},
  \citenamefont {Chaikin}, \citenamefont {Gollub},\ and\ \citenamefont
  {Pine}}]{corte_random_2008}%
  \BibitemOpen
  \bibfield  {author} {\bibinfo {author} {\bibfnamefont {L.}~\bibnamefont
  {Corté}}, \bibinfo {author} {\bibfnamefont {P.~M.}\ \bibnamefont {Chaikin}},
  \bibinfo {author} {\bibfnamefont {J.~P.}\ \bibnamefont {Gollub}},\ and\
  \bibinfo {author} {\bibfnamefont {D.~J.}\ \bibnamefont {Pine}},\ }\bibfield
  {title} {\bibinfo {title} {Random organization in periodically driven
  systems},\ }\href {https://doi.org/10.1038/nphys891} {\bibfield  {journal}
  {\bibinfo  {journal} {Nature Phys}\ }\textbf {\bibinfo {volume} {4}},\
  \bibinfo {pages} {420} (\bibinfo {year} {2008})}\BibitemShut {NoStop}%
\bibitem [{\citenamefont {Nagasawa}\ \emph {et~al.}(2019)\citenamefont
  {Nagasawa}, \citenamefont {Miyazaki},\ and\ \citenamefont
  {Kawasaki}}]{nagasawa_classification_2019}%
  \BibitemOpen
  \bibfield  {author} {\bibinfo {author} {\bibfnamefont {K.}~\bibnamefont
  {Nagasawa}}, \bibinfo {author} {\bibfnamefont {K.}~\bibnamefont {Miyazaki}},\
  and\ \bibinfo {author} {\bibfnamefont {T.}~\bibnamefont {Kawasaki}},\
  }\bibfield  {title} {\bibinfo {title} {Classification of the
  reversible–irreversible transitions in particle trajectories across the
  jamming transition point},\ }\href {https://doi.org/10.1039/C9SM01488H}
  {\bibfield  {journal} {\bibinfo  {journal} {Soft Matter}\ }\textbf {\bibinfo
  {volume} {15}},\ \bibinfo {pages} {7557} (\bibinfo {year}
  {2019})}\BibitemShut {NoStop}%
\bibitem [{\citenamefont {Das}\ \emph {et~al.}(2020)\citenamefont {Das},
  \citenamefont {Vinutha},\ and\ \citenamefont {Sastry}}]{das_unified_2020}%
  \BibitemOpen
  \bibfield  {author} {\bibinfo {author} {\bibfnamefont {P.}~\bibnamefont
  {Das}}, \bibinfo {author} {\bibfnamefont {H.~A.}\ \bibnamefont {Vinutha}},\
  and\ \bibinfo {author} {\bibfnamefont {S.}~\bibnamefont {Sastry}},\
  }\bibfield  {title} {\bibinfo {title} {Unified phase diagram of
  reversible–irreversible, jamming, and yielding transitions in cyclically
  sheared soft-sphere packings},\ }\href
  {https://doi.org/10.1073/pnas.1912482117} {\bibfield  {journal} {\bibinfo
  {journal} {PNAS}\ }\textbf {\bibinfo {volume} {117}},\ \bibinfo {pages}
  {10203} (\bibinfo {year} {2020})}\BibitemShut {NoStop}%
\bibitem [{\citenamefont {Babu}\ \emph {et~al.}(2021)\citenamefont {Babu},
  \citenamefont {Pan}, \citenamefont {Jin}, \citenamefont {Chakraborty},\ and\
  \citenamefont {Sastry}}]{babu_dilatancy_2021}%
  \BibitemOpen
  \bibfield  {author} {\bibinfo {author} {\bibfnamefont {V.}~\bibnamefont
  {Babu}}, \bibinfo {author} {\bibfnamefont {D.}~\bibnamefont {Pan}}, \bibinfo
  {author} {\bibfnamefont {Y.}~\bibnamefont {Jin}}, \bibinfo {author}
  {\bibfnamefont {B.}~\bibnamefont {Chakraborty}},\ and\ \bibinfo {author}
  {\bibfnamefont {S.}~\bibnamefont {Sastry}},\ }\bibfield  {title} {\bibinfo
  {title} {Dilatancy, shear jamming, and a generalized jamming phase diagram of
  frictionless sphere packings},\ }\href {https://doi.org/10.1039/D0SM02186E}
  {\bibfield  {journal} {\bibinfo  {journal} {Soft Matter}\ }\textbf {\bibinfo
  {volume} {17}},\ \bibinfo {pages} {3121} (\bibinfo {year}
  {2021})}\BibitemShut {NoStop}%
\bibitem [{\citenamefont {Yeh}\ \emph {et~al.}(2020)\citenamefont {Yeh},
  \citenamefont {Ozawa}, \citenamefont {Miyazaki}, \citenamefont {Kawasaki},\
  and\ \citenamefont {Berthier}}]{yeh_glass_2020}%
  \BibitemOpen
  \bibfield  {author} {\bibinfo {author} {\bibfnamefont {W.-T.}\ \bibnamefont
  {Yeh}}, \bibinfo {author} {\bibfnamefont {M.}~\bibnamefont {Ozawa}}, \bibinfo
  {author} {\bibfnamefont {K.}~\bibnamefont {Miyazaki}}, \bibinfo {author}
  {\bibfnamefont {T.}~\bibnamefont {Kawasaki}},\ and\ \bibinfo {author}
  {\bibfnamefont {L.}~\bibnamefont {Berthier}},\ }\bibfield  {title} {\bibinfo
  {title} {Glass {Stability} {Changes} the {Nature} of {Yielding} under
  {Oscillatory} {Shear}},\ }\href@noop {} {\bibfield  {journal} {\bibinfo
  {journal} {Phys. Rev. Lett.}\ }\textbf {\bibinfo {volume} {124}},\ \bibinfo
  {pages} {225502} (\bibinfo {year} {2020})}\BibitemShut {NoStop}%
\bibitem [{\citenamefont {Bhaumik}\ \emph
  {et~al.}(2021{\natexlab{a}})\citenamefont {Bhaumik}, \citenamefont {Foffi},\
  and\ \citenamefont {Sastry}}]{bhaumik_role_2021}%
  \BibitemOpen
  \bibfield  {author} {\bibinfo {author} {\bibfnamefont {H.}~\bibnamefont
  {Bhaumik}}, \bibinfo {author} {\bibfnamefont {G.}~\bibnamefont {Foffi}},\
  and\ \bibinfo {author} {\bibfnamefont {S.}~\bibnamefont {Sastry}},\
  }\bibfield  {title} {\bibinfo {title} {The role of annealing in determining
  the yielding behavior of glasses under cyclic shear deformation},\ }\href
  {https://www.pnas.org/content/118/16/e2100227118} {\bibfield  {journal}
  {\bibinfo  {journal} {PNAS}\ }\textbf {\bibinfo {volume} {118}},\ \bibinfo
  {pages} {e2100227118} (\bibinfo {year} {2021}{\natexlab{a}})}\BibitemShut
  {NoStop}%
\bibitem [{\citenamefont {Bhaumik}\ \emph
  {et~al.}(2021{\natexlab{b}})\citenamefont {Bhaumik}, \citenamefont {Foffi},\
  and\ \citenamefont {Sastry}}]{bhaumik_yielding_2021}%
  \BibitemOpen
  \bibfield  {author} {\bibinfo {author} {\bibfnamefont {H.}~\bibnamefont
  {Bhaumik}}, \bibinfo {author} {\bibfnamefont {G.}~\bibnamefont {Foffi}},\
  and\ \bibinfo {author} {\bibfnamefont {S.}~\bibnamefont {Sastry}},\
  }\bibfield  {title} {\bibinfo {title} {Yielding transition of a two
  dimensional glass former under athermal cyclic shear deformation},\ }\href
  {http://arxiv.org/abs/2108.07497} {\bibfield  {journal} {\bibinfo  {journal}
  {arXiv:2108.07497 [cond-mat]}\ } (\bibinfo {year}
  {2021}{\natexlab{b}})}\BibitemShut {NoStop}%
\bibitem [{\citenamefont {Szulc}\ \emph {et~al.}(2020)\citenamefont {Szulc},
  \citenamefont {Gat},\ and\ \citenamefont {Regev}}]{szulc_forced_2020}%
  \BibitemOpen
  \bibfield  {author} {\bibinfo {author} {\bibfnamefont {A.}~\bibnamefont
  {Szulc}}, \bibinfo {author} {\bibfnamefont {O.}~\bibnamefont {Gat}},\ and\
  \bibinfo {author} {\bibfnamefont {I.}~\bibnamefont {Regev}},\ }\bibfield
  {title} {\bibinfo {title} {Forced deterministic dynamics on a random energy
  landscape: {Implications} for the physics of amorphous solids},\ }\href
  {https://doi.org/10.1103/PhysRevE.101.052616} {\bibfield  {journal} {\bibinfo
   {journal} {Phys. Rev. E}\ }\textbf {\bibinfo {volume} {101}},\ \bibinfo
  {pages} {052616} (\bibinfo {year} {2020})}\BibitemShut {NoStop}%
\bibitem [{\citenamefont {Sastry}(2021)}]{sastry_models_2021}%
  \BibitemOpen
  \bibfield  {author} {\bibinfo {author} {\bibfnamefont {S.}~\bibnamefont
  {Sastry}},\ }\bibfield  {title} {\bibinfo {title} {Models for the {Yielding}
  {Behavior} of {Amorphous} {Solids}},\ }\href
  {https://doi.org/10.1103/PhysRevLett.126.255501} {\bibfield  {journal}
  {\bibinfo  {journal} {Phys. Rev. Lett.}\ }\textbf {\bibinfo {volume} {126}},\
  \bibinfo {pages} {255501} (\bibinfo {year} {2021})}\BibitemShut {NoStop}%
\bibitem [{\citenamefont {Liu}\ \emph {et~al.}(2021)\citenamefont {Liu},
  \citenamefont {Ferrero}, \citenamefont {Jagla}, \citenamefont {Martens},
  \citenamefont {Rosso},\ and\ \citenamefont {Talon}}]{liu_fate_2021}%
  \BibitemOpen
  \bibfield  {author} {\bibinfo {author} {\bibfnamefont {C.}~\bibnamefont
  {Liu}}, \bibinfo {author} {\bibfnamefont {E.~E.}\ \bibnamefont {Ferrero}},
  \bibinfo {author} {\bibfnamefont {E.~A.}\ \bibnamefont {Jagla}}, \bibinfo
  {author} {\bibfnamefont {K.}~\bibnamefont {Martens}}, \bibinfo {author}
  {\bibfnamefont {A.}~\bibnamefont {Rosso}},\ and\ \bibinfo {author}
  {\bibfnamefont {L.}~\bibnamefont {Talon}},\ }\bibfield  {title} {\bibinfo
  {title} {The {Fate} of {Shear}-{Oscillated} {Amorphous} {Solids}},\ }\href
  {http://arxiv.org/abs/2012.15310} {\bibfield  {journal} {\bibinfo  {journal}
  {arXiv:2012.15310 [cond-mat]}\ } (\bibinfo {year} {2021})}\BibitemShut
  {NoStop}%
\bibitem [{\citenamefont {Khirallah}\ \emph {et~al.}(2021)\citenamefont
  {Khirallah}, \citenamefont {Tyukodi}, \citenamefont {Vandembroucq},\ and\
  \citenamefont {Maloney}}]{khirallah_yielding_2021}%
  \BibitemOpen
  \bibfield  {author} {\bibinfo {author} {\bibfnamefont {K.}~\bibnamefont
  {Khirallah}}, \bibinfo {author} {\bibfnamefont {B.}~\bibnamefont {Tyukodi}},
  \bibinfo {author} {\bibfnamefont {D.}~\bibnamefont {Vandembroucq}},\ and\
  \bibinfo {author} {\bibfnamefont {C.~E.}\ \bibnamefont {Maloney}},\
  }\bibfield  {title} {\bibinfo {title} {Yielding in an {Integer} {Automaton}
  {Model} for {Amorphous} {Solids} under {Cyclic} {Shear}},\ }\href
  {https://doi.org/10.1103/PhysRevLett.126.218005} {\bibfield  {journal}
  {\bibinfo  {journal} {Phys. Rev. Lett.}\ }\textbf {\bibinfo {volume} {126}},\
  \bibinfo {pages} {218005} (\bibinfo {year} {2021})}\BibitemShut {NoStop}%
\bibitem [{\citenamefont {Mungan}\ and\ \citenamefont
  {Sastry}(2021)}]{mungan_metastability_2021}%
  \BibitemOpen
  \bibfield  {author} {\bibinfo {author} {\bibfnamefont {M.}~\bibnamefont
  {Mungan}}\ and\ \bibinfo {author} {\bibfnamefont {S.}~\bibnamefont
  {Sastry}},\ }\bibfield  {title} {\bibinfo {title} {Metastability as a
  {Mechanism} for {Yielding} in {Amorphous} {Solids} under {Cyclic} {Shear}},\
  }\href {https://doi.org/10.1103/PhysRevLett.127.248002} {\bibfield  {journal}
  {\bibinfo  {journal} {Phys. Rev. Lett.}\ }\textbf {\bibinfo {volume} {127}},\
  \bibinfo {pages} {248002} (\bibinfo {year} {2021})}\BibitemShut {NoStop}%
\bibitem [{\citenamefont {Hébraud}\ and\ \citenamefont
  {Lequeux}(1998)}]{hebraud_mode-coupling_1998}%
  \BibitemOpen
  \bibfield  {author} {\bibinfo {author} {\bibfnamefont {P.}~\bibnamefont
  {Hébraud}}\ and\ \bibinfo {author} {\bibfnamefont {F.}~\bibnamefont
  {Lequeux}},\ }\bibfield  {title} {\bibinfo {title} {Mode-{Coupling} {Theory}
  for the {Pasty} {Rheology} of {Soft} {Glassy} {Materials}},\ }\href
  {https://doi.org/10.1103/PhysRevLett.81.2934} {\bibfield  {journal} {\bibinfo
   {journal} {Physical Review Letters}\ }\textbf {\bibinfo {volume} {81}},\
  \bibinfo {pages} {2934} (\bibinfo {year} {1998})}\BibitemShut {NoStop}%
\bibitem [{\citenamefont {Agoritsas}\ and\ \citenamefont
  {Martens}(2017)}]{agoritsas_non-trivial_2017}%
  \BibitemOpen
  \bibfield  {author} {\bibinfo {author} {\bibfnamefont {E.}~\bibnamefont
  {Agoritsas}}\ and\ \bibinfo {author} {\bibfnamefont {K.}~\bibnamefont
  {Martens}},\ }\bibfield  {title} {\bibinfo {title} {Non-trivial rheological
  exponents in sheared yield stress fluids},\ }\href
  {https://doi.org/10.1039/C6SM02702D} {\bibfield  {journal} {\bibinfo
  {journal} {Soft Matter}\ }\textbf {\bibinfo {volume} {13}},\ \bibinfo {pages}
  {4653} (\bibinfo {year} {2017})}\BibitemShut {NoStop}%
\bibitem [{\citenamefont {Liu}\ \emph {et~al.}(2018)\citenamefont {Liu},
  \citenamefont {Martens},\ and\ \citenamefont {Barrat}}]{liu_mean-field_2018}%
  \BibitemOpen
  \bibfield  {author} {\bibinfo {author} {\bibfnamefont {C.}~\bibnamefont
  {Liu}}, \bibinfo {author} {\bibfnamefont {K.}~\bibnamefont {Martens}},\ and\
  \bibinfo {author} {\bibfnamefont {J.-L.}\ \bibnamefont {Barrat}},\ }\bibfield
   {title} {\bibinfo {title} {Mean-{Field} {Scenario} for the {Athermal}
  {Creep} {Dynamics} of {Yield}-{Stress} {Fluids}},\ }\href
  {https://doi.org/10.1103/PhysRevLett.120.028004} {\bibfield  {journal}
  {\bibinfo  {journal} {Phys. Rev. Lett.}\ }\textbf {\bibinfo {volume} {120}},\
  \bibinfo {pages} {028004} (\bibinfo {year} {2018})}\BibitemShut {NoStop}%
\bibitem [{\citenamefont {Agoritsas}\ \emph {et~al.}(2015)\citenamefont
  {Agoritsas}, \citenamefont {Bertin}, \citenamefont {Martens},\ and\
  \citenamefont {Barrat}}]{agoritsas_relevance_2015}%
  \BibitemOpen
  \bibfield  {author} {\bibinfo {author} {\bibfnamefont {E.}~\bibnamefont
  {Agoritsas}}, \bibinfo {author} {\bibfnamefont {E.}~\bibnamefont {Bertin}},
  \bibinfo {author} {\bibfnamefont {K.}~\bibnamefont {Martens}},\ and\ \bibinfo
  {author} {\bibfnamefont {J.-L.}\ \bibnamefont {Barrat}},\ }\bibfield  {title}
  {\bibinfo {title} {On the relevance of disorder in athermal amorphous
  materials under shear},\ }\href {https://doi.org/10.1140/epje/i2015-15071-x}
  {\bibfield  {journal} {\bibinfo  {journal} {Eur. Phys. J. E}\ }\textbf
  {\bibinfo {volume} {38}},\ \bibinfo {pages} {71} (\bibinfo {year}
  {2015})}\BibitemShut {NoStop}%
\bibitem [{\citenamefont {Puosi}\ \emph {et~al.}(2015)\citenamefont {Puosi},
  \citenamefont {Olivier},\ and\ \citenamefont {Martens}}]{puosi_probing_2015}%
  \BibitemOpen
  \bibfield  {author} {\bibinfo {author} {\bibfnamefont {F.}~\bibnamefont
  {Puosi}}, \bibinfo {author} {\bibfnamefont {J.}~\bibnamefont {Olivier}},\
  and\ \bibinfo {author} {\bibfnamefont {K.}~\bibnamefont {Martens}},\
  }\bibfield  {title} {\bibinfo {title} {Probing relevant ingredients in
  mean-field approaches for the athermal rheology of yield stress materials},\
  }\href {https://doi.org/10.1039/C5SM01694K} {\bibfield  {journal} {\bibinfo
  {journal} {Soft Matter}\ }\textbf {\bibinfo {volume} {11}},\ \bibinfo {pages}
  {7639} (\bibinfo {year} {2015})}\BibitemShut {NoStop}%
\bibitem [{\citenamefont {Bouchaud}\ \emph {et~al.}(2016)\citenamefont
  {Bouchaud}, \citenamefont {Gualdi}, \citenamefont {Tarzia},\ and\
  \citenamefont {Zamponi}}]{bouchaud_spontaneous_2016}%
  \BibitemOpen
  \bibfield  {author} {\bibinfo {author} {\bibfnamefont {J.-P.}\ \bibnamefont
  {Bouchaud}}, \bibinfo {author} {\bibfnamefont {S.}~\bibnamefont {Gualdi}},
  \bibinfo {author} {\bibfnamefont {M.}~\bibnamefont {Tarzia}},\ and\ \bibinfo
  {author} {\bibfnamefont {F.}~\bibnamefont {Zamponi}},\ }\bibfield  {title}
  {\bibinfo {title} {Spontaneous instabilities and stick-slip motion in a
  generalized {Hébraud}–{Lequeux} model},\ }\href
  {https://doi.org/10.1039/C5SM02216A} {\bibfield  {journal} {\bibinfo
  {journal} {Soft Matter}\ }\textbf {\bibinfo {volume} {12}},\ \bibinfo {pages}
  {1230} (\bibinfo {year} {2016})}\BibitemShut {NoStop}%
\bibitem [{\citenamefont {Ekeh}\ \emph {et~al.}(2021)\citenamefont {Ekeh},
  \citenamefont {Fodor}, \citenamefont {Fielding},\ and\ \citenamefont
  {Cates}}]{ekeh_power_2021}%
  \BibitemOpen
  \bibfield  {author} {\bibinfo {author} {\bibfnamefont {T.}~\bibnamefont
  {Ekeh}}, \bibinfo {author} {\bibfnamefont {E.}~\bibnamefont {Fodor}},
  \bibinfo {author} {\bibfnamefont {S.~M.}\ \bibnamefont {Fielding}},\ and\
  \bibinfo {author} {\bibfnamefont {M.~E.}\ \bibnamefont {Cates}},\ }\bibfield
  {title} {\bibinfo {title} {Power fluctuations in sheared amorphous materials:
  {A} minimal model},\ }\href {http://arxiv.org/abs/2106.12962} {\bibfield
  {journal} {\bibinfo  {journal} {arXiv:2106.12962 [cond-mat]}\ } (\bibinfo
  {year} {2021})}\BibitemShut {NoStop}%
\bibitem [{\citenamefont {Sollich}\ \emph {et~al.}(2017)\citenamefont
  {Sollich}, \citenamefont {Olivier},\ and\ \citenamefont
  {Bresch}}]{sollich_aging_2017}%
  \BibitemOpen
  \bibfield  {author} {\bibinfo {author} {\bibfnamefont {P.}~\bibnamefont
  {Sollich}}, \bibinfo {author} {\bibfnamefont {J.}~\bibnamefont {Olivier}},\
  and\ \bibinfo {author} {\bibfnamefont {D.}~\bibnamefont {Bresch}},\
  }\bibfield  {title} {\bibinfo {title} {Aging and linear response in the
  {Hébraud}–{Lequeux} model for amorphous rheology},\ }\href
  {https://doi.org/10.1088/1751-8121/aa6261} {\bibfield  {journal} {\bibinfo
  {journal} {J. Phys. A: Math. Theor.}\ }\textbf {\bibinfo {volume} {50}},\
  \bibinfo {pages} {165002} (\bibinfo {year} {2017})}\BibitemShut {NoStop}%
\bibitem [{\citenamefont {Sollich}\ \emph {et~al.}(1997)\citenamefont
  {Sollich}, \citenamefont {Lequeux}, \citenamefont {Hébraud},\ and\
  \citenamefont {Cates}}]{sollich_rheology_1997}%
  \BibitemOpen
  \bibfield  {author} {\bibinfo {author} {\bibfnamefont {P.}~\bibnamefont
  {Sollich}}, \bibinfo {author} {\bibfnamefont {F.}~\bibnamefont {Lequeux}},
  \bibinfo {author} {\bibfnamefont {P.}~\bibnamefont {Hébraud}},\ and\
  \bibinfo {author} {\bibfnamefont {M.~E.}\ \bibnamefont {Cates}},\ }\bibfield
  {title} {\bibinfo {title} {Rheology of {Soft} {Glassy} {Materials}},\ }\href
  {https://doi.org/10.1103/PhysRevLett.78.2020} {\bibfield  {journal} {\bibinfo
   {journal} {Phys. Rev. Lett.}\ }\textbf {\bibinfo {volume} {78}},\ \bibinfo
  {pages} {2020} (\bibinfo {year} {1997})}\BibitemShut {NoStop}%
\bibitem [{\citenamefont {Bocquet}\ \emph {et~al.}(2009)\citenamefont
  {Bocquet}, \citenamefont {Colin},\ and\ \citenamefont
  {Ajdari}}]{bocquet_kinetic_2009}%
  \BibitemOpen
  \bibfield  {author} {\bibinfo {author} {\bibfnamefont {L.}~\bibnamefont
  {Bocquet}}, \bibinfo {author} {\bibfnamefont {A.}~\bibnamefont {Colin}},\
  and\ \bibinfo {author} {\bibfnamefont {A.}~\bibnamefont {Ajdari}},\
  }\bibfield  {title} {\bibinfo {title} {Kinetic {Theory} of {Plastic} {Flow}
  in {Soft} {Glassy} {Materials}},\ }\href
  {https://doi.org/10.1103/PhysRevLett.103.036001} {\bibfield  {journal}
  {\bibinfo  {journal} {Phys. Rev. Lett.}\ }\textbf {\bibinfo {volume} {103}},\
  \bibinfo {pages} {036001} (\bibinfo {year} {2009})}\BibitemShut {NoStop}%
\bibitem [{sm()}]{sm}%
  \BibitemOpen
  \href@noop {} {\bibinfo  {journal} {See Supplemental Material at [URL will be
  inserted by publisher] for details on analytical derivations and numerical
  solutions}\ }\BibitemShut {NoStop}%
\bibitem [{Note1()}]{Note1}%
  \BibitemOpen
\bibfield  {journal} {  }\bibinfo {note} {In the SM~\cite {sm} we show that an
  exponential $\rho (E)$ does not qualitatively change the form of the
  transition line, and argue more generally that our main results are
  qualitatively robust to the precise form of $\rho (E)$.}\BibitemShut {Stop}%
\bibitem [{\citenamefont {Sastry}(2001)}]{sastry_relationship_2001}%
  \BibitemOpen
  \bibfield  {author} {\bibinfo {author} {\bibfnamefont {S.}~\bibnamefont
  {Sastry}},\ }\bibfield  {title} {\bibinfo {title} {The relationship between
  fragility, configurational entropy and the potential energy landscape of
  glass-forming liquids},\ }\href {https://doi.org/10.1038/35051524} {\bibfield
   {journal} {\bibinfo  {journal} {Nature}\ }\textbf {\bibinfo {volume}
  {409}},\ \bibinfo {pages} {164} (\bibinfo {year} {2001})}\BibitemShut
  {NoStop}%
\bibitem [{\citenamefont {Parmar}\ \emph {et~al.}(2019)\citenamefont {Parmar},
  \citenamefont {Kumar},\ and\ \citenamefont {Sastry}}]{parmar_strain_2019}%
  \BibitemOpen
  \bibfield  {author} {\bibinfo {author} {\bibfnamefont {A.~D.}\ \bibnamefont
  {Parmar}}, \bibinfo {author} {\bibfnamefont {S.}~\bibnamefont {Kumar}},\ and\
  \bibinfo {author} {\bibfnamefont {S.}~\bibnamefont {Sastry}},\ }\bibfield
  {title} {\bibinfo {title} {Strain {Localization} {Above} the {Yielding}
  {Point} in {Cyclically} {Deformed} {Glasses}},\ }\href
  {https://doi.org/10.1103/PhysRevX.9.021018} {\bibfield  {journal} {\bibinfo
  {journal} {Phys. Rev. X}\ }\textbf {\bibinfo {volume} {9}},\ \bibinfo {pages}
  {021018} (\bibinfo {year} {2019})}\BibitemShut {NoStop}%
\bibitem [{\citenamefont {Ness}\ and\ \citenamefont
  {Cates}(2020)}]{ness_absorbing-state_2020}%
  \BibitemOpen
  \bibfield  {author} {\bibinfo {author} {\bibfnamefont {C.}~\bibnamefont
  {Ness}}\ and\ \bibinfo {author} {\bibfnamefont {M.~E.}\ \bibnamefont
  {Cates}},\ }\bibfield  {title} {\bibinfo {title} {Absorbing-{State}
  {Transitions} in {Granular} {Materials} {Close} to {Jamming}},\ }\href
  {https://doi.org/10.1103/PhysRevLett.124.088004} {\bibfield  {journal}
  {\bibinfo  {journal} {Phys. Rev. Lett.}\ }\textbf {\bibinfo {volume} {124}},\
  \bibinfo {pages} {088004} (\bibinfo {year} {2020})}\BibitemShut {NoStop}%
\bibitem [{\citenamefont {Fielding}(2016)}]{fielding_triggers_2016}%
  \BibitemOpen
  \bibfield  {author} {\bibinfo {author} {\bibfnamefont {S.~M.}\ \bibnamefont
  {Fielding}},\ }\bibfield  {title} {\bibinfo {title} {Triggers and signatures
  of shear banding in steady and time-dependent flows},\ }\href
  {https://doi.org/10.1122/1.4961480} {\bibfield  {journal} {\bibinfo
  {journal} {Journal of Rheology}\ }\textbf {\bibinfo {volume} {60}},\ \bibinfo
  {pages} {821} (\bibinfo {year} {2016})}\BibitemShut {NoStop}%
\bibitem [{Note2()}]{Note2}%
  \BibitemOpen
  \bibinfo {note} {For the number of cycles we simply take the time divided by
  the period, so that it takes continuous values in Fig.~\ref
  {fig:well_annealed}}\BibitemShut {NoStop}%
\bibitem [{\citenamefont {Divoux}\ \emph {et~al.}(2011)\citenamefont {Divoux},
  \citenamefont {Barentin},\ and\ \citenamefont
  {Manneville}}]{divoux_stress-induced_2011}%
  \BibitemOpen
  \bibfield  {author} {\bibinfo {author} {\bibfnamefont {T.}~\bibnamefont
  {Divoux}}, \bibinfo {author} {\bibfnamefont {C.}~\bibnamefont {Barentin}},\
  and\ \bibinfo {author} {\bibfnamefont {S.}~\bibnamefont {Manneville}},\
  }\bibfield  {title} {\bibinfo {title} {From stress-induced fluidization
  processes to {Herschel}-{Bulkley} behaviour in simple yield stress fluids},\
  }\href {https://doi.org/10.1039/c1sm05607g} {\bibfield  {journal} {\bibinfo
  {journal} {Soft Matter}\ }\textbf {\bibinfo {volume} {7}},\ \bibinfo {pages}
  {8409} (\bibinfo {year} {2011})}\BibitemShut {NoStop}%
\bibitem [{\citenamefont {Carmona}\ \emph {et~al.}(2007)\citenamefont
  {Carmona}, \citenamefont {Kun}, \citenamefont {Andrade},\ and\ \citenamefont
  {Herrmann}}]{carmona_computer_2007}%
  \BibitemOpen
  \bibfield  {author} {\bibinfo {author} {\bibfnamefont {H.~A.}\ \bibnamefont
  {Carmona}}, \bibinfo {author} {\bibfnamefont {F.}~\bibnamefont {Kun}},
  \bibinfo {author} {\bibfnamefont {J.~S.}\ \bibnamefont {Andrade}},\ and\
  \bibinfo {author} {\bibfnamefont {H.~J.}\ \bibnamefont {Herrmann}},\
  }\bibfield  {title} {\bibinfo {title} {Computer simulation of fatigue under
  diametrical compression},\ }\href
  {https://doi.org/10.1103/PhysRevE.75.046115} {\bibfield  {journal} {\bibinfo
  {journal} {Phys. Rev. E}\ }\textbf {\bibinfo {volume} {75}},\ \bibinfo
  {pages} {046115} (\bibinfo {year} {2007})}\BibitemShut {NoStop}%
\bibitem [{\citenamefont {Pradhan}\ \emph {et~al.}(2010)\citenamefont
  {Pradhan}, \citenamefont {Hansen},\ and\ \citenamefont
  {Chakrabarti}}]{pradhan_failure_2010}%
  \BibitemOpen
  \bibfield  {author} {\bibinfo {author} {\bibfnamefont {S.}~\bibnamefont
  {Pradhan}}, \bibinfo {author} {\bibfnamefont {A.}~\bibnamefont {Hansen}},\
  and\ \bibinfo {author} {\bibfnamefont {B.~K.}\ \bibnamefont {Chakrabarti}},\
  }\bibfield  {title} {\bibinfo {title} {Failure processes in elastic fiber
  bundles},\ }\href {https://doi.org/10.1103/RevModPhys.82.499} {\bibfield
  {journal} {\bibinfo  {journal} {Rev. Mod. Phys.}\ }\textbf {\bibinfo {volume}
  {82}},\ \bibinfo {pages} {499} (\bibinfo {year} {2010})}\BibitemShut
  {NoStop}%
\bibitem [{\citenamefont {Bhowmik}\ \emph {et~al.}(2021)\citenamefont
  {Bhowmik}, \citenamefont {Hentschel},\ and\ \citenamefont
  {Procaccia}}]{bhowmik_fatigue_2021}%
  \BibitemOpen
  \bibfield  {author} {\bibinfo {author} {\bibfnamefont {B.~P.}\ \bibnamefont
  {Bhowmik}}, \bibinfo {author} {\bibfnamefont {H.~G.~E.}\ \bibnamefont
  {Hentschel}},\ and\ \bibinfo {author} {\bibfnamefont {I.}~\bibnamefont
  {Procaccia}},\ }\bibfield  {title} {\bibinfo {title} {Fatigue and {Collapse}
  of {Cyclically} {Bent} {Strip} of {Amorphous} {Solid}},\ }\href
  {http://arxiv.org/abs/2103.03040} {\bibfield  {journal} {\bibinfo  {journal}
  {arXiv:2103.03040 [cond-mat]}\ } (\bibinfo {year} {2021})}\BibitemShut
  {NoStop}%
\bibitem [{\citenamefont {Kun}\ \emph {et~al.}(2007)\citenamefont {Kun},
  \citenamefont {Costa}, \citenamefont {Filho}, \citenamefont {Andrade},
  \citenamefont {Soares}, \citenamefont {Zapperi},\ and\ \citenamefont
  {Herrmann}}]{kun_fatigue_2007}%
  \BibitemOpen
  \bibfield  {author} {\bibinfo {author} {\bibfnamefont {F.}~\bibnamefont
  {Kun}}, \bibinfo {author} {\bibfnamefont {M.~H.}\ \bibnamefont {Costa}},
  \bibinfo {author} {\bibfnamefont {R.~N.~C.}\ \bibnamefont {Filho}}, \bibinfo
  {author} {\bibfnamefont {J.~S.}\ \bibnamefont {Andrade}}, \bibinfo {author}
  {\bibfnamefont {J.~B.}\ \bibnamefont {Soares}}, \bibinfo {author}
  {\bibfnamefont {S.}~\bibnamefont {Zapperi}},\ and\ \bibinfo {author}
  {\bibfnamefont {H.~J.}\ \bibnamefont {Herrmann}},\ }\bibfield  {title}
  {\bibinfo {title} {Fatigue failure of disordered materials},\ }\href
  {https://doi.org/10.1088/1742-5468/2007/02/P02003} {\bibfield  {journal}
  {\bibinfo  {journal} {J. Stat. Mech.}\ }\textbf {\bibinfo {volume} {2007}},\
  \bibinfo {pages} {P02003} (\bibinfo {year} {2007})}\BibitemShut {NoStop}%
\bibitem [{\citenamefont {Sha}\ \emph {et~al.}(2015)\citenamefont {Sha},
  \citenamefont {Qu}, \citenamefont {Liu}, \citenamefont {Wang},\ and\
  \citenamefont {Gao}}]{sha_cyclic_2015}%
  \BibitemOpen
  \bibfield  {author} {\bibinfo {author} {\bibfnamefont {Z.~D.}\ \bibnamefont
  {Sha}}, \bibinfo {author} {\bibfnamefont {S.~X.}\ \bibnamefont {Qu}},
  \bibinfo {author} {\bibfnamefont {Z.~S.}\ \bibnamefont {Liu}}, \bibinfo
  {author} {\bibfnamefont {T.~J.}\ \bibnamefont {Wang}},\ and\ \bibinfo
  {author} {\bibfnamefont {H.}~\bibnamefont {Gao}},\ }\bibfield  {title}
  {\bibinfo {title} {Cyclic {Deformation} in {Metallic} {Glasses}},\ }\href
  {https://doi.org/10.1021/acs.nanolett.5b03045} {\bibfield  {journal}
  {\bibinfo  {journal} {Nano Lett.}\ }\textbf {\bibinfo {volume} {15}},\
  \bibinfo {pages} {7010} (\bibinfo {year} {2015})}\BibitemShut {NoStop}%
\bibitem [{\citenamefont {Popović}\ \emph
  {et~al.}(2021{\natexlab{a}})\citenamefont {Popović}, \citenamefont
  {de~Geus}, \citenamefont {Ji}, \citenamefont {Rosso},\ and\ \citenamefont
  {Wyart}}]{popovic_scaling_2021}%
  \BibitemOpen
  \bibfield  {author} {\bibinfo {author} {\bibfnamefont {M.}~\bibnamefont
  {Popović}}, \bibinfo {author} {\bibfnamefont {T.~W.~J.}\ \bibnamefont
  {de~Geus}}, \bibinfo {author} {\bibfnamefont {W.}~\bibnamefont {Ji}},
  \bibinfo {author} {\bibfnamefont {A.}~\bibnamefont {Rosso}},\ and\ \bibinfo
  {author} {\bibfnamefont {M.}~\bibnamefont {Wyart}},\ }\bibfield  {title}
  {\bibinfo {title} {Scaling description of creep flow in amorphous solids},\
  }\href {http://arxiv.org/abs/2111.04061} {\bibfield  {journal} {\bibinfo
  {journal} {arXiv:2111.04061 [cond-mat]}\ } (\bibinfo {year}
  {2021}{\natexlab{a}})}\BibitemShut {NoStop}%
\bibitem [{Note3()}]{Note3}%
  \BibitemOpen
  \bibinfo {note} {In the SM~\cite {sm} we provide more discussion and contrast
  this to the mean field results of \cite
  {ozawa_random_2018,popovic_elastoplastic_2018}}\BibitemShut {NoStop}%
\bibitem [{\citenamefont {Liu}\ \emph {et~al.}(2016)\citenamefont {Liu},
  \citenamefont {Ferrero}, \citenamefont {Puosi}, \citenamefont {Barrat},\ and\
  \citenamefont {Martens}}]{liu_driving_2016}%
  \BibitemOpen
  \bibfield  {author} {\bibinfo {author} {\bibfnamefont {C.}~\bibnamefont
  {Liu}}, \bibinfo {author} {\bibfnamefont {E.~E.}\ \bibnamefont {Ferrero}},
  \bibinfo {author} {\bibfnamefont {F.}~\bibnamefont {Puosi}}, \bibinfo
  {author} {\bibfnamefont {J.-L.}\ \bibnamefont {Barrat}},\ and\ \bibinfo
  {author} {\bibfnamefont {K.}~\bibnamefont {Martens}},\ }\bibfield  {title}
  {\bibinfo {title} {Driving {Rate} {Dependence} of {Avalanche} {Statistics}
  and {Shapes} at the {Yielding} {Transition}},\ }\href
  {https://link.aps.org/doi/10.1103/PhysRevLett.116.065501} {\bibfield
  {journal} {\bibinfo  {journal} {Phys. Rev. Lett.}\ }\textbf {\bibinfo
  {volume} {116}},\ \bibinfo {pages} {065501} (\bibinfo {year}
  {2016})}\BibitemShut {NoStop}%
\bibitem [{\citenamefont {Parley}\ \emph {et~al.}(2020)\citenamefont {Parley},
  \citenamefont {Fielding},\ and\ \citenamefont {Sollich}}]{parley_aging_2020}%
  \BibitemOpen
  \bibfield  {author} {\bibinfo {author} {\bibfnamefont {J.~T.}\ \bibnamefont
  {Parley}}, \bibinfo {author} {\bibfnamefont {S.~M.}\ \bibnamefont
  {Fielding}},\ and\ \bibinfo {author} {\bibfnamefont {P.}~\bibnamefont
  {Sollich}},\ }\bibfield  {title} {\bibinfo {title} {Aging in a mean field
  elastoplastic model of amorphous solids},\ }\href
  {https://doi.org/10.1063/5.0033196} {\bibfield  {journal} {\bibinfo
  {journal} {Physics of Fluids}\ }\textbf {\bibinfo {volume} {32}},\ \bibinfo
  {pages} {127104} (\bibinfo {year} {2020})}\BibitemShut {NoStop}%
\bibitem [{\citenamefont {Lin}\ and\ \citenamefont
  {Wyart}(2016)}]{lin_mean-field_2016}%
  \BibitemOpen
  \bibfield  {author} {\bibinfo {author} {\bibfnamefont {J.}~\bibnamefont
  {Lin}}\ and\ \bibinfo {author} {\bibfnamefont {M.}~\bibnamefont {Wyart}},\
  }\bibfield  {title} {\bibinfo {title} {Mean-{Field} {Description} of
  {Plastic} {Flow} in {Amorphous} {Solids}},\ }\href
  {https://doi.org/10.1103/PhysRevX.6.011005} {\bibfield  {journal} {\bibinfo
  {journal} {Phys. Rev. X}\ }\textbf {\bibinfo {volume} {6}},\ \bibinfo {pages}
  {011005} (\bibinfo {year} {2016})}\BibitemShut {NoStop}%
\bibitem [{\citenamefont {Popović}\ \emph
  {et~al.}(2021{\natexlab{b}})\citenamefont {Popović}, \citenamefont
  {de~Geus}, \citenamefont {Ji},\ and\ \citenamefont
  {Wyart}}]{popovic_thermally_2021}%
  \BibitemOpen
  \bibfield  {author} {\bibinfo {author} {\bibfnamefont {M.}~\bibnamefont
  {Popović}}, \bibinfo {author} {\bibfnamefont {T.~W.~J.}\ \bibnamefont
  {de~Geus}}, \bibinfo {author} {\bibfnamefont {W.}~\bibnamefont {Ji}},\ and\
  \bibinfo {author} {\bibfnamefont {M.}~\bibnamefont {Wyart}},\ }\bibfield
  {title} {\bibinfo {title} {Thermally activated flow in models of amorphous
  solids},\ }\href {https://doi.org/10.1103/PhysRevE.104.025010} {\bibfield
  {journal} {\bibinfo  {journal} {Phys. Rev. E}\ }\textbf {\bibinfo {volume}
  {104}},\ \bibinfo {pages} {025010} (\bibinfo {year}
  {2021}{\natexlab{b}})}\BibitemShut {NoStop}%
\bibitem [{\citenamefont {Ferrero}\ \emph {et~al.}(2021)\citenamefont
  {Ferrero}, \citenamefont {Kolton},\ and\ \citenamefont
  {Jagla}}]{ferrero_yielding_2021}%
  \BibitemOpen
  \bibfield  {author} {\bibinfo {author} {\bibfnamefont {E.~E.}\ \bibnamefont
  {Ferrero}}, \bibinfo {author} {\bibfnamefont {A.~B.}\ \bibnamefont
  {Kolton}},\ and\ \bibinfo {author} {\bibfnamefont {E.~A.}\ \bibnamefont
  {Jagla}},\ }\bibfield  {title} {\bibinfo {title} {Yielding of amorphous
  solids at finite temperatures},\ }\href
  {https://doi.org/10.1103/PhysRevMaterials.5.115602} {\bibfield  {journal}
  {\bibinfo  {journal} {Phys. Rev. Materials}\ }\textbf {\bibinfo {volume}
  {5}},\ \bibinfo {pages} {115602} (\bibinfo {year} {2021})}\BibitemShut
  {NoStop}%
\bibitem [{\citenamefont {Popović}\ \emph {et~al.}(2018)\citenamefont
  {Popović}, \citenamefont {de~Geus},\ and\ \citenamefont
  {Wyart}}]{popovic_elastoplastic_2018}%
  \BibitemOpen
  \bibfield  {author} {\bibinfo {author} {\bibfnamefont {M.}~\bibnamefont
  {Popović}}, \bibinfo {author} {\bibfnamefont {T.~W.~J.}\ \bibnamefont
  {de~Geus}},\ and\ \bibinfo {author} {\bibfnamefont {M.}~\bibnamefont
  {Wyart}},\ }\bibfield  {title} {\bibinfo {title} {Elastoplastic description
  of sudden failure in athermal amorphous materials during quasistatic
  loading},\ }\href {https://doi.org/10.1103/PhysRevE.98.040901} {\bibfield
  {journal} {\bibinfo  {journal} {Phys. Rev. E}\ }\textbf {\bibinfo {volume}
  {98}},\ \bibinfo {pages} {040901} (\bibinfo {year} {2018})}\BibitemShut
  {NoStop}%
\end{thebibliography}%


\begin{thebibliography}{15}%
\makeatletter
\providecommand \@ifxundefined [1]{%
 \@ifx{#1\undefined}
}%
\providecommand \@ifnum [1]{%
 \ifnum #1\expandafter \@firstoftwo
 \else \expandafter \@secondoftwo
 \fi
}%
\providecommand \@ifx [1]{%
 \ifx #1\expandafter \@firstoftwo
 \else \expandafter \@secondoftwo
 \fi
}%
\providecommand \natexlab [1]{#1}%
\providecommand \enquote  [1]{``#1''}%
\providecommand \bibnamefont  [1]{#1}%
\providecommand \bibfnamefont [1]{#1}%
\providecommand \citenamefont [1]{#1}%
\providecommand \href@noop [0]{\@secondoftwo}%
\providecommand \href [0]{\begingroup \@sanitize@url \@href}%
\providecommand \@href[1]{\@@startlink{#1}\@@href}%
\providecommand \@@href[1]{\endgroup#1\@@endlink}%
\providecommand \@sanitize@url [0]{\catcode `\\12\catcode `\$12\catcode
  `\&12\catcode `\#12\catcode `\^12\catcode `\_12\catcode `\%12\relax}%
\providecommand \@@startlink[1]{}%
\providecommand \@@endlink[0]{}%
\providecommand \url  [0]{\begingroup\@sanitize@url \@url }%
\providecommand \@url [1]{\endgroup\@href {#1}{\urlprefix }}%
\providecommand \urlprefix  [0]{URL }%
\providecommand \Eprint [0]{\href }%
\providecommand \doibase [0]{https://doi.org/}%
\providecommand \selectlanguage [0]{\@gobble}%
\providecommand \bibinfo  [0]{\@secondoftwo}%
\providecommand \bibfield  [0]{\@secondoftwo}%
\providecommand \translation [1]{[#1]}%
\providecommand \BibitemOpen [0]{}%
\providecommand \bibitemStop [0]{}%
\providecommand \bibitemNoStop [0]{.\EOS\space}%
\providecommand \EOS [0]{\spacefactor3000\relax}%
\providecommand \BibitemShut  [1]{\csname bibitem#1\endcsname}%
\let\auto@bib@innerbib\@empty
\bibitem [{Note1()}]{Note1}%
  \BibitemOpen
  \bibinfo {note} {Strictly speaking (\ref {bound}) only proves that the
  transition line for the model with disorder starts above the original model
  as $\protect \tilde {\alpha }$ is decreased from $1$, but our numerics (for
  both exponential and Gaussian $\rho (E)$) confirm that this holds down to
  $\protect \tilde {\alpha }=0$.}\BibitemShut {Stop}%
\bibitem [{\citenamefont {Sastry}(2001)}]{sastry_relationship_2001}%
  \BibitemOpen
  \bibfield  {author} {\bibinfo {author} {\bibfnamefont {S.}~\bibnamefont
  {Sastry}},\ }\bibfield  {title} {\bibinfo {title} {The relationship between
  fragility, configurational entropy and the potential energy landscape of
  glass-forming liquids},\ }\href {https://doi.org/10.1038/35051524} {\bibfield
   {journal} {\bibinfo  {journal} {Nature}\ }\textbf {\bibinfo {volume}
  {409}},\ \bibinfo {pages} {164} (\bibinfo {year} {2001})}\BibitemShut
  {NoStop}%
\bibitem [{\citenamefont {Sastry}(2021)}]{sastry_models_2021}%
  \BibitemOpen
  \bibfield  {author} {\bibinfo {author} {\bibfnamefont {S.}~\bibnamefont
  {Sastry}},\ }\bibfield  {title} {\bibinfo {title} {Models for the {Yielding}
  {Behavior} of {Amorphous} {Solids}},\ }\href
  {https://doi.org/10.1103/PhysRevLett.126.255501} {\bibfield  {journal}
  {\bibinfo  {journal} {Phys. Rev. Lett.}\ }\textbf {\bibinfo {volume} {126}},\
  \bibinfo {pages} {255501} (\bibinfo {year} {2021})}\BibitemShut {NoStop}%
\bibitem [{\citenamefont {Parley}\ \emph {et~al.}(2020)\citenamefont {Parley},
  \citenamefont {Fielding},\ and\ \citenamefont {Sollich}}]{parley_aging_2020}%
  \BibitemOpen
  \bibfield  {author} {\bibinfo {author} {\bibfnamefont {J.~T.}\ \bibnamefont
  {Parley}}, \bibinfo {author} {\bibfnamefont {S.~M.}\ \bibnamefont
  {Fielding}},\ and\ \bibinfo {author} {\bibfnamefont {P.}~\bibnamefont
  {Sollich}},\ }\bibfield  {title} {\bibinfo {title} {Aging in a mean field
  elastoplastic model of amorphous solids},\ }\href
  {https://doi.org/10.1063/5.0033196} {\bibfield  {journal} {\bibinfo
  {journal} {Physics of Fluids}\ }\textbf {\bibinfo {volume} {32}},\ \bibinfo
  {pages} {127104} (\bibinfo {year} {2020})}\BibitemShut {NoStop}%
\bibitem [{Note2()}]{Note2}%
  \BibitemOpen
  \bibinfo {note} {As commented on below, we expect the actual yield point for
  this frequency to be slightly above $\gamma ^{*}_0$, although we are not able
  to determine this precisely within our numerics. It is possible therefore
  that the stress amplitude in the yielded state monotonically decreases above
  the yield point. Another possibility is that the non-monotonicity disappears
  for $\omega \rightarrow 0$, but we have not studied this in
  detail.}\BibitemShut {Stop}%
\bibitem [{\citenamefont {Bhaumik}\ \emph {et~al.}(2021)\citenamefont
  {Bhaumik}, \citenamefont {Foffi},\ and\ \citenamefont
  {Sastry}}]{bhaumik_role_2021}%
  \BibitemOpen
  \bibfield  {author} {\bibinfo {author} {\bibfnamefont {H.}~\bibnamefont
  {Bhaumik}}, \bibinfo {author} {\bibfnamefont {G.}~\bibnamefont {Foffi}},\
  and\ \bibinfo {author} {\bibfnamefont {S.}~\bibnamefont {Sastry}},\
  }\bibfield  {title} {\bibinfo {title} {The role of annealing in determining
  the yielding behavior of glasses under cyclic shear deformation},\ }\href
  {https://www.pnas.org/content/118/16/e2100227118} {\bibfield  {journal}
  {\bibinfo  {journal} {PNAS}\ }\textbf {\bibinfo {volume} {118}},\ \bibinfo
  {pages} {e2100227118} (\bibinfo {year} {2021})}\BibitemShut {NoStop}%
\bibitem [{\citenamefont {Agoritsas}\ \emph {et~al.}(2015)\citenamefont
  {Agoritsas}, \citenamefont {Bertin}, \citenamefont {Martens},\ and\
  \citenamefont {Barrat}}]{agoritsas_relevance_2015}%
  \BibitemOpen
  \bibfield  {author} {\bibinfo {author} {\bibfnamefont {E.}~\bibnamefont
  {Agoritsas}}, \bibinfo {author} {\bibfnamefont {E.}~\bibnamefont {Bertin}},
  \bibinfo {author} {\bibfnamefont {K.}~\bibnamefont {Martens}},\ and\ \bibinfo
  {author} {\bibfnamefont {J.-L.}\ \bibnamefont {Barrat}},\ }\bibfield  {title}
  {\bibinfo {title} {On the relevance of disorder in athermal amorphous
  materials under shear},\ }\href {https://doi.org/10.1140/epje/i2015-15071-x}
  {\bibfield  {journal} {\bibinfo  {journal} {Eur. Phys. J. E}\ }\textbf
  {\bibinfo {volume} {38}},\ \bibinfo {pages} {71} (\bibinfo {year}
  {2015})}\BibitemShut {NoStop}%
\bibitem [{\citenamefont {Yeh}\ \emph {et~al.}(2020)\citenamefont {Yeh},
  \citenamefont {Ozawa}, \citenamefont {Miyazaki}, \citenamefont {Kawasaki},\
  and\ \citenamefont {Berthier}}]{yeh_glass_2020}%
  \BibitemOpen
  \bibfield  {author} {\bibinfo {author} {\bibfnamefont {W.-T.}\ \bibnamefont
  {Yeh}}, \bibinfo {author} {\bibfnamefont {M.}~\bibnamefont {Ozawa}}, \bibinfo
  {author} {\bibfnamefont {K.}~\bibnamefont {Miyazaki}}, \bibinfo {author}
  {\bibfnamefont {T.}~\bibnamefont {Kawasaki}},\ and\ \bibinfo {author}
  {\bibfnamefont {L.}~\bibnamefont {Berthier}},\ }\bibfield  {title} {\bibinfo
  {title} {Glass {Stability} {Changes} the {Nature} of {Yielding} under
  {Oscillatory} {Shear}},\ }\href@noop {} {\bibfield  {journal} {\bibinfo
  {journal} {Phys. Rev. Lett.}\ }\textbf {\bibinfo {volume} {124}},\ \bibinfo
  {pages} {225502} (\bibinfo {year} {2020})}\BibitemShut {NoStop}%
\bibitem [{Note3()}]{Note3}%
  \BibitemOpen
  \bibinfo {note} {We note that these transient effects due to the energy
  redistribution could slightly modify the numerically measured value of the
  exponent $b$.}\BibitemShut {Stop}%
\bibitem [{\citenamefont {Sollich}\ \emph {et~al.}(2017)\citenamefont
  {Sollich}, \citenamefont {Olivier},\ and\ \citenamefont
  {Bresch}}]{sollich_aging_2017}%
  \BibitemOpen
  \bibfield  {author} {\bibinfo {author} {\bibfnamefont {P.}~\bibnamefont
  {Sollich}}, \bibinfo {author} {\bibfnamefont {J.}~\bibnamefont {Olivier}},\
  and\ \bibinfo {author} {\bibfnamefont {D.}~\bibnamefont {Bresch}},\
  }\bibfield  {title} {\bibinfo {title} {Aging and linear response in the
  {Hébraud}–{Lequeux} model for amorphous rheology},\ }\href
  {https://doi.org/10.1088/1751-8121/aa6261} {\bibfield  {journal} {\bibinfo
  {journal} {J. Phys. A: Math. Theor.}\ }\textbf {\bibinfo {volume} {50}},\
  \bibinfo {pages} {165002} (\bibinfo {year} {2017})}\BibitemShut {NoStop}%
\bibitem [{\citenamefont {Leishangthem}\ \emph {et~al.}(2017)\citenamefont
  {Leishangthem}, \citenamefont {Parmar},\ and\ \citenamefont
  {Sastry}}]{leishangthem_yielding_2017}%
  \BibitemOpen
  \bibfield  {author} {\bibinfo {author} {\bibfnamefont {P.}~\bibnamefont
  {Leishangthem}}, \bibinfo {author} {\bibfnamefont {A.~D.~S.}\ \bibnamefont
  {Parmar}},\ and\ \bibinfo {author} {\bibfnamefont {S.}~\bibnamefont
  {Sastry}},\ }\bibfield  {title} {\bibinfo {title} {The yielding transition in
  amorphous solids under oscillatory shear deformation},\ }\href
  {https://doi.org/10.1038/ncomms14653} {\bibfield  {journal} {\bibinfo
  {journal} {Nat Commun}\ }\textbf {\bibinfo {volume} {8}},\ \bibinfo {pages}
  {14653} (\bibinfo {year} {2017})}\BibitemShut {NoStop}%
\bibitem [{\citenamefont {Bocquet}\ \emph {et~al.}(2009)\citenamefont
  {Bocquet}, \citenamefont {Colin},\ and\ \citenamefont
  {Ajdari}}]{bocquet_kinetic_2009}%
  \BibitemOpen
  \bibfield  {author} {\bibinfo {author} {\bibfnamefont {L.}~\bibnamefont
  {Bocquet}}, \bibinfo {author} {\bibfnamefont {A.}~\bibnamefont {Colin}},\
  and\ \bibinfo {author} {\bibfnamefont {A.}~\bibnamefont {Ajdari}},\
  }\bibfield  {title} {\bibinfo {title} {Kinetic {Theory} of {Plastic} {Flow}
  in {Soft} {Glassy} {Materials}},\ }\href
  {https://doi.org/10.1103/PhysRevLett.103.036001} {\bibfield  {journal}
  {\bibinfo  {journal} {Phys. Rev. Lett.}\ }\textbf {\bibinfo {volume} {103}},\
  \bibinfo {pages} {036001} (\bibinfo {year} {2009})}\BibitemShut {NoStop}%
\bibitem [{\citenamefont {Ozawa}\ \emph {et~al.}(2018)\citenamefont {Ozawa},
  \citenamefont {Berthier}, \citenamefont {Biroli}, \citenamefont {Rosso},\
  and\ \citenamefont {Tarjus}}]{ozawa_random_2018}%
  \BibitemOpen
  \bibfield  {author} {\bibinfo {author} {\bibfnamefont {M.}~\bibnamefont
  {Ozawa}}, \bibinfo {author} {\bibfnamefont {L.}~\bibnamefont {Berthier}},
  \bibinfo {author} {\bibfnamefont {G.}~\bibnamefont {Biroli}}, \bibinfo
  {author} {\bibfnamefont {A.}~\bibnamefont {Rosso}},\ and\ \bibinfo {author}
  {\bibfnamefont {G.}~\bibnamefont {Tarjus}},\ }\bibfield  {title} {\bibinfo
  {title} {A random critical point separates brittle and ductile yielding
  transitions in amorphous materials},\ }\href
  {https://doi.org/10.1073/pnas.1806156115} {\bibfield  {journal} {\bibinfo
  {journal} {Proceedings of the National Academy of Sciences}\ }\textbf
  {\bibinfo {volume} {115}},\ \bibinfo {pages} {6656} (\bibinfo {year}
  {2018})}\BibitemShut {NoStop}%
\bibitem [{\citenamefont {Popović}\ \emph {et~al.}(2018)\citenamefont
  {Popović}, \citenamefont {de~Geus},\ and\ \citenamefont
  {Wyart}}]{popovic_elastoplastic_2018}%
  \BibitemOpen
  \bibfield  {author} {\bibinfo {author} {\bibfnamefont {M.}~\bibnamefont
  {Popović}}, \bibinfo {author} {\bibfnamefont {T.~W.~J.}\ \bibnamefont
  {de~Geus}},\ and\ \bibinfo {author} {\bibfnamefont {M.}~\bibnamefont
  {Wyart}},\ }\bibfield  {title} {\bibinfo {title} {Elastoplastic description
  of sudden failure in athermal amorphous materials during quasistatic
  loading},\ }\href {https://doi.org/10.1103/PhysRevE.98.040901} {\bibfield
  {journal} {\bibinfo  {journal} {Phys. Rev. E}\ }\textbf {\bibinfo {volume}
  {98}},\ \bibinfo {pages} {040901} (\bibinfo {year} {2018})}\BibitemShut
  {NoStop}%
\bibitem [{Note4()}]{Note4}%
  \BibitemOpen
  \bibinfo {note} {Note that Eq. (\ref {plastic_aqs}) with absorbing boundary
  conditions only strictly conserves normalisation if the derivatives at the
  boundaries satisfy $\partial _l P(E,l_c)-\partial _l P(E,-l_c)=-\rho
  (E)/\alpha $ (valid in steady state). This severely restricts the class of
  initial conditions that can be studied.}\BibitemShut {Stop}%
\end{thebibliography}%

\end{document}



\title{Supplemental Material to ``Mean field theory of yielding under oscillatory shear''}



\author{Jack T. Parley}
\email[Author to whom correspondence should be addressed: ]{jack.parley@uni-goettingen.de}
\affiliation{Institut f{\"u}r Theoretische Physik, University of G{\"o}ttingen,
Friedrich-Hund-Platz 1, 37077 G{\"o}ttingen, Germany}

\author{Srikanth Sastry}
\affiliation{Jawaharlal Nehru Centre for Advanced Scientific Research, Jakkar Campus, 560064 Bengaluru, India}

\author{Peter Sollich}
\affiliation{Institut f{\"u}r Theoretische Physik, University of G{\"o}ttingen,
Friedrich-Hund-Platz 1, 37077 G{\"o}ttingen, Germany}
\affiliation{Department of Mathematics, King’s College London, London WC2R 2LS, UK}


\date{\today}

\maketitle

The supplemental material below contains (a) an elaboration of the calculations outlined in the main text, which is structured in the same order as the main paper, and (b) additional results that further support the statements and conclusions discussed.
We start by giving more details on the analytical calculations required to obtain the transition line and the threshold energy. Next, we present the numerical methods employed to study the initial annealing dependence. We then give more information on this dependence. In addition to the data on the energy in Fig.~2 of the paper,  this includes data for the stress, the critical behaviour for samples above the threshold, and the behaviour in the yielded state. We then move on to the fatigue behaviour of well-annealed samples, showing data for the energy as well as substantiating some of the observations pointed out in the paper. Finally, we show results for the model under uniform shear.
\section{Transition line}
\subsection{Limiting yield rate}

We give here details regarding the pair of self-consistent equations determining the limiting form of the yield rate as the transition is approached. We focus firstly on the contribution from the indirect yields $\yone(\tau)$, which we recall is determined by
\begin{equation}\label{deltas}
    \yone(\tau)=c[y]\frac{1}{2}\left(\delta \left(\tau-\frac{1}{4}\right)+\delta \left(\tau-\frac{3}{4}\right)\right)
\end{equation}
As explained in the main text, the prefactor (given explicitly in the main text and below) is simply set by the fraction of elements which (after the previous yield event at $\taup$) now satisfy the condition $\gamma_0+|\gamma(\tau')|<\sqrt{2E}$, so that they cannot yield again in the next cycle. This gives
\begin{eqnarray}
   c[y]&=&\int_{\gamma_0^2/2}^{\infty}\mathrm{d}E \ \rho(E)\int_0^{1}\mathrm{d}\tau' \ y(\tau') \ \theta\left(\sqrt{2E}-\gamma_0-|\gamma_0 \sin(2\pi\tau')|\right) \nonumber\\
   &=&\int_{0}^{1}\mathrm{d}\tau' \  y(\tau') \ \Frho^0\!\left(\frac{\gamma_0^2}{2}\left(1+|\sin(2\pi \tau')|\right)^2\right)
\end{eqnarray}
where in the second line we have performed the integral over $E$, and introduced the following notation for the cumulative integrals over $\rho(E)$:
\begin{equation}
    \Frho^n(E)=\int_{E}^{\infty}\ (E')^{n}\rho(E')\mathrm{d}E'
\end{equation}
For the numerics it is inconvenient to work with the delta peaks in (\ref{deltas}), and we instead write the self-consistency equation directly in terms of the scalar quantity $c[y]$. Performing the integral over $\yone$, this becomes a functional of $\ytwo$ only, which obeys
\begin{equation}\label{eq_c}
    c=c\Frho^0(2\gamma_0^2)+\int_{0}^{1}\mathrm{d}\tau' \  \ytwo(\tau') \ \Frho^0\!\left(\frac{\gamma_0^2}{2}\left(1+|\sin(2\pi \tau')|\right)^2\right)
\end{equation}

\begin{figure}\center{\includegraphics[width=0.5\textwidth]{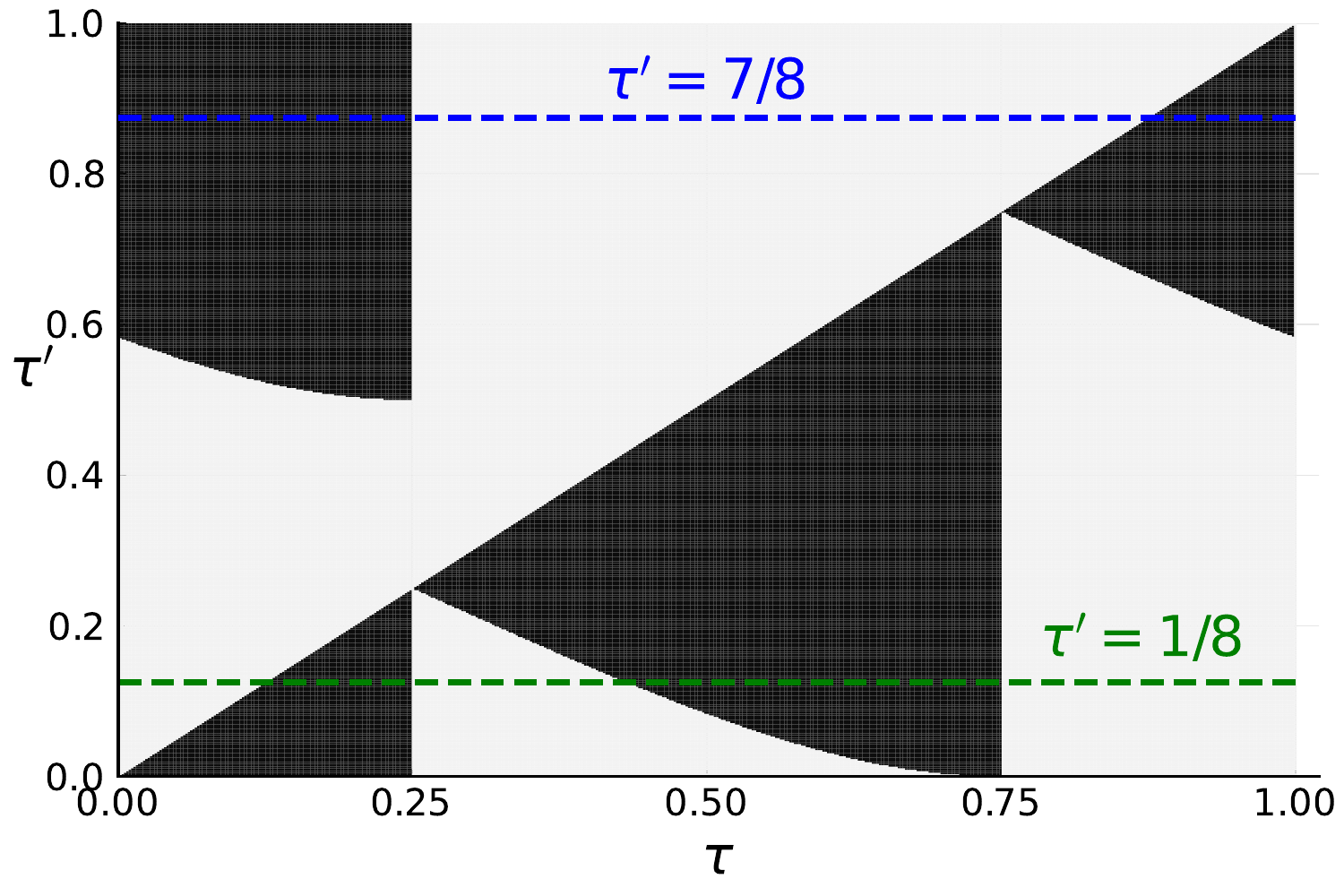}}
\caption{\label{fig:chi}
The function $\chi$ in the $\tau, \ \taup$ plane, where it takes the values $1$ (black) or $0$ (white). Note that the condition for the first next yield is not simply $\tau>\taup$, and also that the next event may occur in the following period (upper left section). Horizontal dashed lines mark the two horizontal slices of $\chi$ exemplified in Fig.~\ref{fig:first}.}
\end{figure}

We now consider the contribution from the direct yields, denoted by $\ytwo(\tau)$. In the main text we wrote this as 
\begin{equation}
\ytwo(\tau)=\int_{0}^{\infty}\mathrm{d}E \  \rho(E) \ \int_{0}^{1}\mathrm{d}\tau' \ y(\tau') \ \delta\left(\tau-\tau_y(\tau',\gamma_0 \sin(2\pi \tau'),\gamma_0,E)\right)
\end{equation}
Here the function 
$\tau_y(\tau',\gamma_0 \sin(2\pi \tau'),\gamma_0,E)$ gives, for each combination of previous yield time $\tau'$ and energy level $E$, the time $\tau_y \in (0,1)$ of the next direct yield event. In practice, it is not necessary to know this function, and we instead change variables in the delta function, writing the condition now in terms of energy so that $\delta (\tau-\tau_y(\tau',E))\rightarrow\delta(E-\Estar(\tau,\tau'))$ (along with the Jacobian associated with the transformation, see below). In other words, a combination of previous yield time $\tau'$ and a subsequent direct yield event at $\tau$ fixes the value of the depth $\Estar(\tau,\tau')$ of the energy minimum that the element must be in. Explicitly, one has simply
\begin{equation}
    \Estar(\tau,\tau')=\frac{1}{2}\left\{\gamma_0^2[\sin(2\pi\tau)-\sin(2\pi\tau')]\right\}^2
\end{equation}
because the expression in square brackets on the right (times $\gamma_0$) is the strain the element has acquired between times $\tau'$ and $\tau$. 
Expressing the delta function in terms of $E-\Estar$ now gives explicitly 
\begin{equation}
    \int_0^{\infty}\mathrm{d}E \ \rho(E) \ \delta\left(\tau-\tau_y(\tau',\gamma_0 \sin(2\pi \tau'),\gamma_0,E)\right)
    = \rho(\Estar(\tau,\tau'))\left|\frac{\mathrm{d}E}{\mathrm{d}\tau}\right|\chi(\tau,\tau')
    \label{chi_eqn}
\end{equation}
where
\begin{equation}
    \left|\frac{\mathrm{d}E}{\mathrm{d}\tau}\right|=2\pi\gamma_0^2\left|\sin(2\pi\tau)-\sin(2\pi\tau')\right|\left|\cos(2 \pi \tau)\right|
\end{equation}
and we have carried out the integration over $E$. The final factor
$\chi(\tau,\tau')$ in (\ref{chi_eqn})
is an indicator function ($\chi=1$ or 0) enforcing the condition that $\tau$ corresponds to the \textit{first} direct yield event in the cycle after a previous yield event at $\tau'$. From trigonometric arguments, this condition is satisfied, and hence  $\chi=1$, in the following four cases: 
\begin{enumerate}
    \item Case $(0<\taup<1/4)$: \\
    $(\taup<\tau<1/4)$ $\cup$ $(\tau_1^{*}<\tau<3/4)$, with $\tau_1^{*}=1/(2 \pi)\left(\pi+\arcsin(1-2\sin(2\pi\taup))\right)$.
    \item Case $(1/4<\taup<1/2)$: \\
    $(\taup<\tau<3/4)$.
    \item Case ($1/2<\taup<3/4$): \\
    Define $\tau_2^{*}=1/(2\pi)\left(\arcsin(1+2\sin(2\pi\taup))+2\pi\right)$.\\If $\tau_2^{*}<1$, $(\taup<\tau<3/4)$ $\cup$ $(\tau_2^{*}<\tau<1)$ $\cup$ $(0<\tau<1/4)$.\\
    If $\tau_2^{*}>1$, $(\taup<\tau<3/4)$ $\cup$ $(\tau_2^{*}-1<\tau<1/4)$.
    \item Case $(3/4<\taup<1)$:\\ 
    $(\taup<\tau<1)$ $\cup$ $(0<\tau<1/4)$. 
\end{enumerate}
Otherwise, $\chi=0$. A plot of $\chi(\tau,\tau')$ is shown in Fig.~\ref{fig:chi}. 
Note that values of $\tau$ that would be in the next period ($\tau>1$) are mapped back to $\tau\in[0,1]$ in determining $\chi$.

\begin{figure}\center{\includegraphics[width=0.5\textwidth]{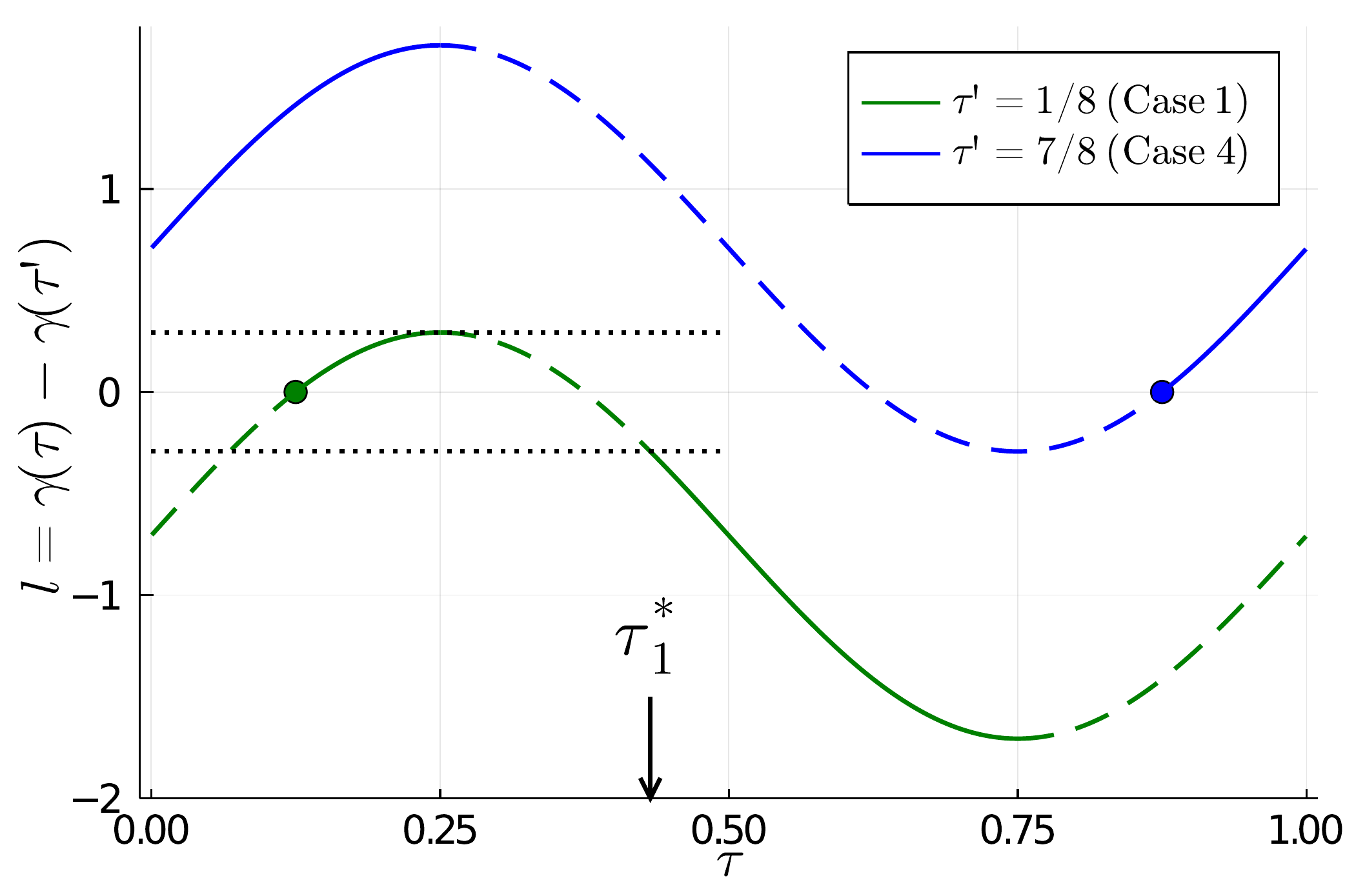}}
\caption{\label{fig:first}Plot of the local strain $l$ against $\tau$, with $\gamma_0=1$, illustrating two examples of how the $\chi$ function is obtained. The initial strains at $\tau'=1/8$ and $\tau'=7/8$ are marked with circles. Full lines indicate where $\chi(\tau,\taup)=1$, dashed lines where $\chi(\tau,\taup)=0$. Dotted lines mark $l=\pm (\gamma_0-\gamma(\tau'=1/8))$, which is the maximum absolute value of strain attained in the initial sweep for $\tau'=1/8$, and determines the point $\tau_1^{*}(\tau'=1/8)$ (see text for definition) beyond which the first yields with negative $l$ can occur. Note that $\gamma_0$ just sets the vertical scale and does not feature in the $\chi$ function.   
}
\end{figure}

For clarity we illustrate two example cases in Fig.~\ref{fig:first}, $\tau'=1/8$ and $\tau'=7/8$, which belong respectively to case 1 and case 4 above. Plotted is the strain variable $l$, starting from $l=0$ at the previous yield time $\tau'$. Dashed lines indicate where $\chi(\tau,\taup)=0$: intuitively, these are simply segments of the period where a yield event would have already happened, given that $|l|$ has previously reached a higher or equal value. Note that for $\tau'=1/8$, yields can occur both at positive and at negative $l$, while for $\tau'=7/8$ yields always happen first for positive $l$. 

Putting all this together, the second self-consistency equation becomes
\begin{equation}{\label{y2_final}}
    \ytwo(\tau)=\int_0^{1}\mathrm{d}\tau' \ y(\tau')\rho(\Estar(\tau,\tau'))\left|\frac{\mathrm{d}E}{\mathrm{d}\tau}\right|\chi(\tau,\tau')
\end{equation}
Eqs (\ref{eq_c}) and (\ref{y2_final}) constitute a pair of self-consistent equations which completely determine the yield rate through $c$ and $\ytwo(\tau)$. To solve for $\ytwo(\tau)$ and $c$, one can start from a normalised arbitrary initial condition $y_0(\tau)$ and apply them iteratively until convergence. The iterative scheme indeed conserves the norm $\int_0^{1} \mathrm{d}\tau\, y(\tau)=c+\int_0^{1} \mathrm{d}\tau\, \ytwo(\tau)=1$. This condition reads explicitly
\begin{equation}
    \int_0^{1}\mathrm{d}\tau' \ \rho(\Estar(\tau,\tau'))\left|\frac{\mathrm{d}E}{\mathrm{d}\tau}\right|\chi(\tau,\tau')+\Frho^0\left(\frac{\gamma_0^2}{2}\left(1+|\sin(2\pi \tau')|\right)^2\right)=1
\end{equation}
and can be verified to hold for $\forall \tau' \in [0,1]$ as it must. Intuitively, the normalization condition simply expresses the fact that an element with previous yield time $\tau'$ must either subsequently yield directly (first term) or fall into a deep energy level where it then yields diffusively (second term).

\begin{figure}[H]\center{\includegraphics[width=0.5\textwidth]{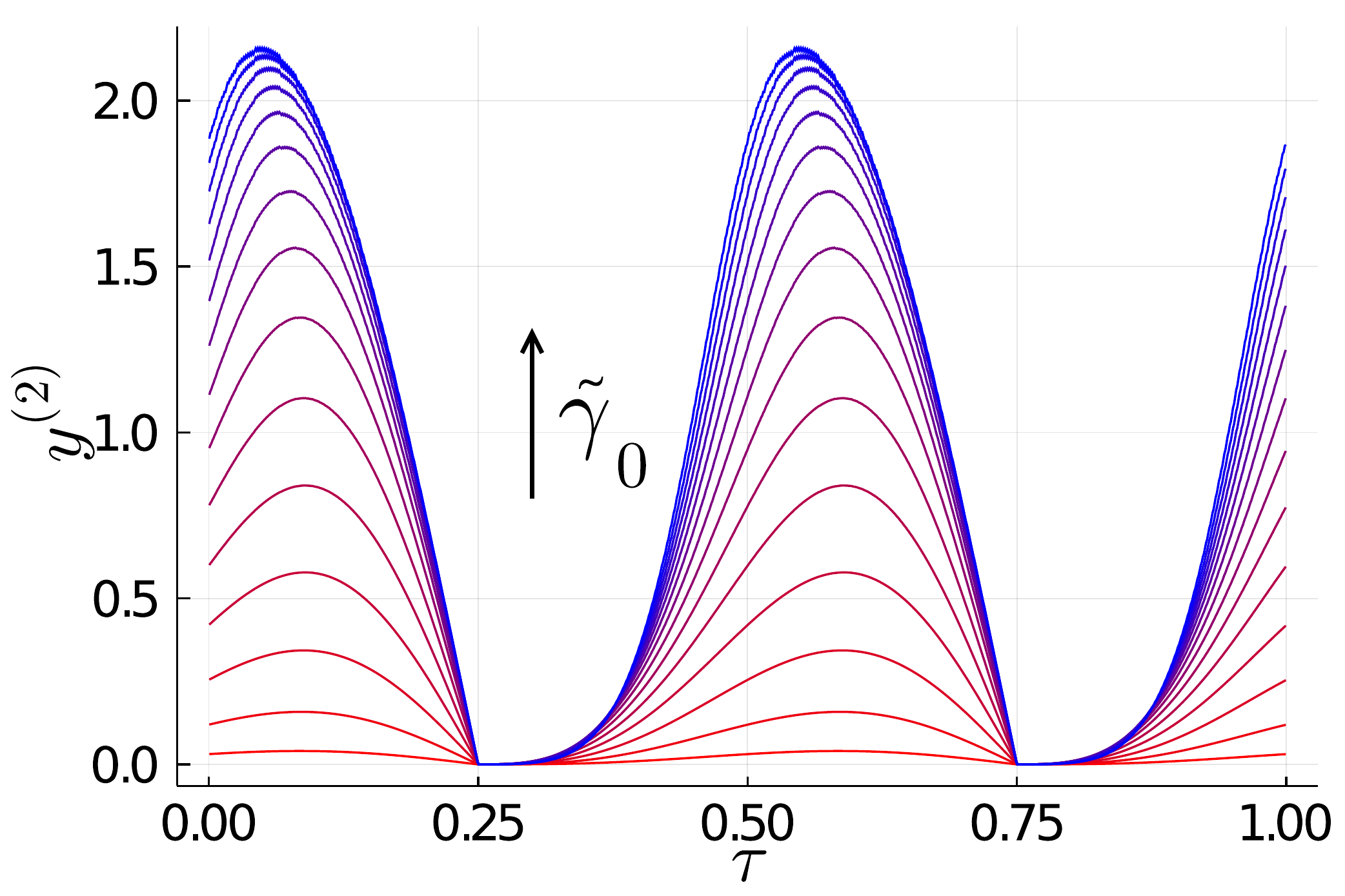}}
\caption{\label{fig:yends}$\ytwo$ obtained from iteratively solving (\ref{eq_c}) and (\ref{y2_final}), for a Gaussian $\rho(E)$. Results are shown for 15 values of $\tgamma=\gamma_0/\sqrt{\Em}$ that  are linearly spaced between 0.125 (red) and 1.875 (blue). The $\tau-$axis has been discretised into $N=1000$ points.}
\end{figure}

We show the converged $\ytwo(\tau)$ in Fig.~\ref{fig:yends}, for a Gaussian $\rho(E)$ and a range of shear amplitudes. As initial condition of the iteration we simply use in each case $c=1$, $\ytwo=0$. We note that as expected the rate of direct yields rate is periodic with double the frequency of the oscillatory shear, and increases slowly from zero at $\tau=1/4$ and $3/4$, where the modulus of the strain reaches a maximum given that $\gamma(t)=\gamma_0 \sin (2\pi \tau)$. 
Notice also that as $\gamma_0$ is increased the peaks in the direct yield rate move towards $\tau=0$ and $1/2$, where the shear {\it rate} is maximal.

\subsection{Limiting distribution}
Once the limiting yield rate is known we can work out the limiting distribution $P(E,l)$ at the transition, the normalisation of which determines the transition line $\gamma_0^*(\alpha)$. The calculation is in fact done the other way around, by fixing a strain amplitude $\gamma_0$ and working out the $\alpha^*$ where the yielded state ceases to exist.

For the calculation it is useful to switch to a ``co-moving'' frame, defined by $u(\tau)=l-\gamma(\tau)$. In this frame of reference, yielded elements are reinserted at $u=-\gamma(\tau)$, corresponding to $l=0$. We furthermore rescale the time-independent pdf at the transition by the prior distribution of energies, so that we work with
\begin{equation}
    \tilde{P}(E,u)=\frac{P(E,u)}{\rho(E)}
\end{equation}
At the transition, where the timescale of diffusion diverges, only elements undergoing indirect yields have nonzero steady state probability. The direct yields can effectively be considered as occurring instantaneously (on a timescale $\sim 1/\omega \ll 1/\overline{Y}$) 
so are negligible in the steady state distribution, but they do affect the behaviour via their contribution to $y(\tau)$. The indirect yielding processes, on the other hand, amount in the same limit to absorbing boundary conditions at $\pm u_c(E)=\sqrt{2E}-\gamma_0$. With these boundary conditions, the steady state master equation (Eq. (1) in the main text) reduces to
\begin{equation}
    0=\alpha Y(t)\partial_u^2 \tilde{P}(E,u)+Y(t)\delta\left(u+\gamma(t)\right)
\end{equation}
Averaging this over a period, and defining the rate of yield events producing yielded elements with some given $u$ as
\begin{equation}\label{R(u)}
    R(u)=\int_0^1\mathrm{d}\tau\, y(\tau)\delta \left(u+\gamma(\tau)\right)
\end{equation}
we obtain a simple equation for $\tilde{P}(E,u)$
\begin{equation}\label{eq_Ru}
    0=\alpha \partial_u^2\tilde{P}(E,u)+R(u)
\end{equation}
with the boundary conditions $\tilde{P}(E,u)=0$, $|u|>u_c(E)$.

Eq. (\ref{eq_Ru}) can be solved analytically, as the diffusive propagator with absorbing boundary conditions is well known. The solution, which depends on $E$ only via $u_c(E)$, is
\begin{equation}\label{p_e_u}
    \tilde{P}(E,u)=\frac{1}{2 \alpha u_c}\int_{-u_c}^{u_c}R(u_0)\bigg((u_c+u_0)(u_c-u)\theta(u-u_0)+(u_c+u)(u_c-u_0)\theta(u_0-u)\bigg)\theta(u_c-|u|) \mathrm{d}u_0
\end{equation}
Reverting to $P(E,u)$ and performing the integral over $u$ yields
\begin{equation}\label{p_of_e}
   P(E)=\int \mathrm{d}u \ P(E,u)=\rho(E)\frac{1}{2\alpha}\int_{-u_c}^{u_c}\mathrm{d}u_0 \ R(u_0) (u_c^2-u_0^2) 
   =\frac{1}{2\alpha} \ \rho(E)\int_0^{1}\mathrm{d}\tau \ y(\tau) {\left((\sqrt{2E}-\gamma_0)^2-\gamma_0^2\sin(2\pi\tau)^2\right)}_{+}
\end{equation}
where ${f}_{+}={f}\theta{(f)}$. The term on the right hand side may be interpreted as the product of $\rho(E)$ and the mean first escape time, with diffusive noise of strength $\alpha \overline{Y}$, of a fictitious ``particle" from a box of half-width $\sqrt{2E}-\gamma_0$, having started at $-\gamma(\tau)=-\gamma_0 \sin (2 \pi \tau)$. This is averaged over the possible previous yield times. Physically, it is the interplay between these two objects (the energy disorder and the timescale set by the mechanical noise) which sets the location of the transition. A further integration of (\ref{p_of_e}) over $E$ fixes the critical coupling $\alpha^{*}$. For this we separate again the two contributions to the yield rate, finding 
\begin{equation}\label{final_alpha_c}
    \alpha^{*}[\gamma_0,c,\ytwo(\tau)]=c \alphand(\gamma_0)
    +\int_0^{1}\mathrm{d}\tau \ \ytwo(\tau) \ \left(\Frhoav(\Estar(\tau))-\sqrt{2}\gamma_0 \Frhosqr(\Estar(\tau))+\frac{\gamma_0^2}{2}(\cos(2\pi\tau))^2 \Frho^0(\Estar(\tau))\right)
\end{equation}
where
\begin{equation}
    \Estar(\tau)=\frac{\gamma_0^2}{2}\left(1+|\sin(2\pi \tau)|\right)^2
\end{equation}
and 
\begin{equation}\label{alphand}
    \alphand(\gamma_0)=\Frhoav(2\gamma_0^2)-\sqrt{2}\gamma_0 \Frhosqr(2\gamma_0^2)
\end{equation}
is the critical coupling if one neglects direct yields (i.e.\ setting $c=1$). This provides an approximation for $\alpha^*$ that becomes exact for $\gamma_0 \rightarrow 0$. As stated in the main text, from this approximation one may derive an exact bound proving in general that in the presence of disorder the phase boundary lies above the original HL model. Indeed, from (\ref{alphand}) one may find analytically that the initial tangent is given by
\begin{equation}\label{bound}
    \frac{\mathrm{d}\tgamma}{\mathrm{d}\talpha}=\frac{1}{\sqrt{2}}\frac{\left(\Em\right)^{1/2}}{\langle \sqrt{E}\rangle}\geq \frac{1}{\sqrt{2}}
\end{equation}
In the original HL model, the slope is $1/\sqrt{2}$ (and in fact stays constant at this value), while the inclusion of disorder always tends to extend the size of the frozen region \footnote{Strictly speaking (\ref{bound}) only proves that the transition line for the model with disorder starts above the original 
model as $\talpha$ is decreased from $1$, but our numerics (for both exponential and Gaussian $\rho(E)$) confirm that this holds down to $\talpha=0$.}.

\begin{figure}[H]\center{\includegraphics[width=0.3\textwidth]{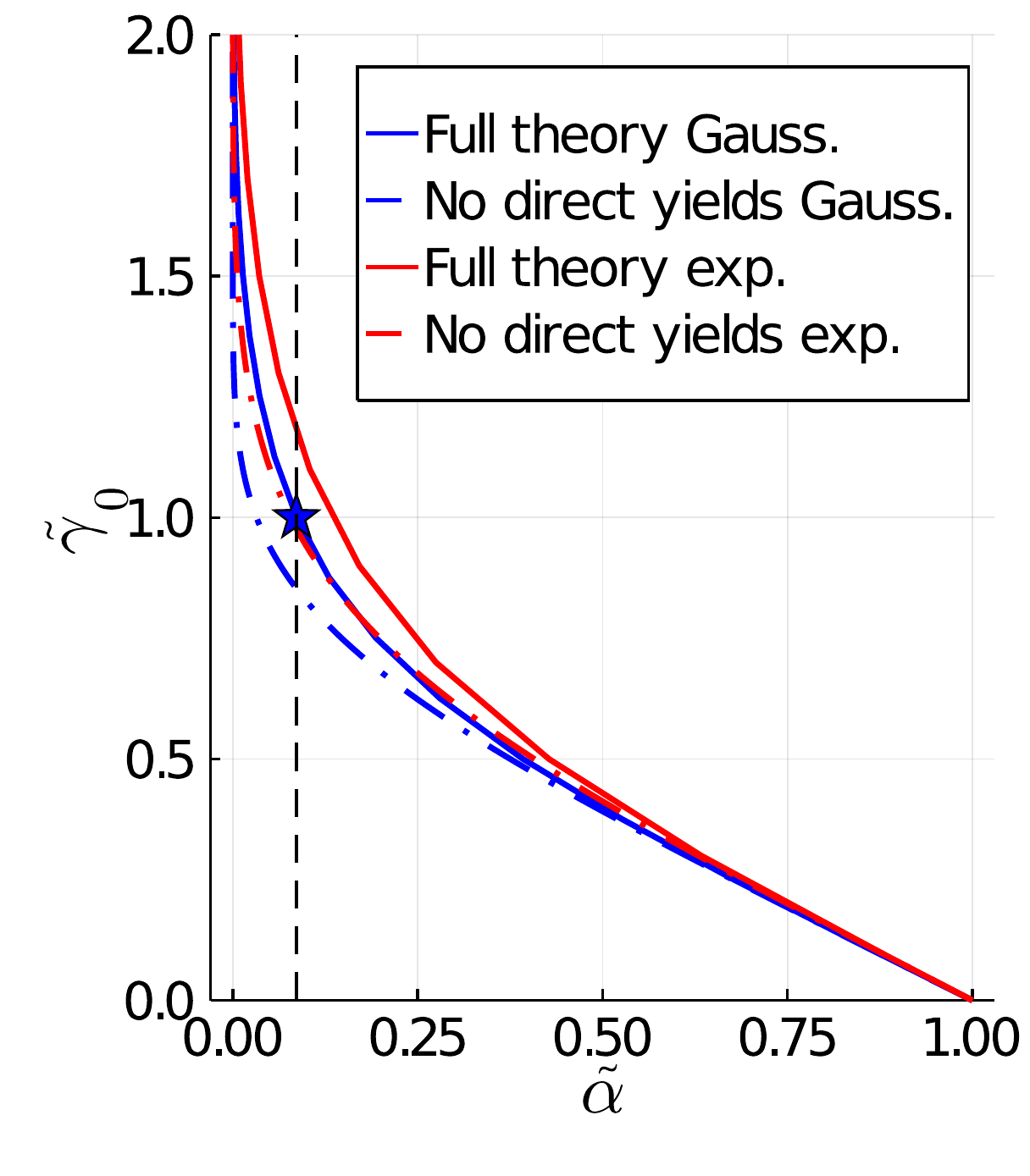}}
\caption{\label{fig:expo_gauss}Transition line in the $\talpha-\tgamma$ plane, for Gaussian (blue) and exponential (red) $\rho(E)$. We show in each case both the numerically computed result from the full theory (\ref{final_alpha_c}), full lines, and the approximate no-direct-yields analytical form (\ref{alphand}), dash-dotted lines. As in Fig. 1 of the main text, vertical dashed lines indicate the fixed coupling value $\talpha=0.086$, where in the Gaussian case ${\tgamma}^{*}=1$ (see star), chosen for studying the initial annealing dependence.}
\end{figure}

Returning to Eq. (\ref{final_alpha_c}), we now have all the ingredients necessary to compute the transition line for any given distribution $\rho(E)$. For a fixed $\gamma_0$, one must first compute the limiting yield rate, which is then inserted into (\ref{final_alpha_c}) to obtain the critical coupling. Note that as can easily be checked from Eqs. (\ref{final_alpha_c}) and (\ref{alphand}), the transition line can indeed be expressed in the rescaled form $\talpha^{*}(\tgamma)$, with $\talpha=\alpha/\Em$ and $\tgamma=\gamma_0/\sqrt{\Em}$, as given in the main text. In Fig.~\ref{fig:expo_gauss}, we show (together with the Gaussian case already displayed in Fig. 1 of the main text), the transition line for an exponential $\rho(E)$ expressed again in rescaled variables $\talpha$ and $\tgamma$. The exponential case is qualitatively very similar to the Gaussian. These two cases differ mainly in the asymptotic behaviour for $\talpha\rightarrow 0$, which is controlled by the tail of the $\rho(E)$ distribution, but this is in any case not the interesting parameter regime of the model. For any unbounded $\rho(E)$, $\tgamma$ must diverge in this limit. This divergence is most easily found from the no direct yield form (\ref{alphand}), from which one may derive that $\alphand(\gamma_0)\sim e^{-2\tilde{\gamma}_0^2}$ for the exponential, while $\alphand(\gamma_0)\sim e^{-4\tilde{\gamma}_0^4/\pi}/\tilde{\gamma}_0^4$ for the Gaussian. One can obtain the same asymptotic behaviours, up to power-law corrections, from an expansion of the full theory (\ref{final_alpha_c}).

We argue in general that, besides the transition line, we expect also the other main results of this paper, including the initial annealing dependence and the fatigue behaviour (Figs. 2 and 3 in the main text), to be qualitatively robust to the choice of $\rho(E)$ (the prior distribution whose form is difficult to ascertain), as long as it fulfils some basic properties. Indeed, there presumably needs to be an “entropic” drive towards shallow energy levels, while rare deeper local minima also need to be present, in order to be able to model the initial annealing. The Gaussian distribution is a simple ansatz fulfilling these properties, besides having been measured in numerical studies of glasses~\cite{sastry_relationship_2001}, as argued in \cite{sastry_models_2021}.

We finally address the calculation of the energy at the transition point, as denoted by the star in Fig.~2 of the main text. We recall that the energy is defined within the model as
\begin{equation}
    U=\int \mathrm{d}E \int \mathrm{d}l \ \left(l^2/2-E\right) P(E,l)
\end{equation}
so that one may consider separately the elastic energy (first term) and the potential energy
(relative to our chosen reference energy of 0, second term).

Starting with the potential energy $E_{\rm pot}$, this is simply (minus) the average over the marginal distribution $P(E)$, given by (\ref{p_of_e}) with $\alpha$ replaced by the corresponding $\alpha^{*}(\gamma_0)$. As one may easily check, $P(E)$ vanishes for energy levels with $E<\gamma_0^2/2$, where all yields are direct. Examples of the distribution $P(E)$ are shown in Fig.~\ref{fig:approach} below.

To calculate the elastic energy, it is convenient to work again with $\tilde{P}(E,u)$ ((\ref{p_e_u}), with $\alpha$ replaced by $\alpha^{*}$), giving
\begin{equation}
    E_{\rm el}=\int \mathrm{d}E \  \rho(E)\int \mathrm{d}u \ \frac{l^2}{2}\tilde{P}(E,u)
\end{equation}
Using then that $l=u+\gamma(t)$, and the symmetry $\tilde{P}(E,u)=\tilde{P}(E,-u)$, we have that
\begin{equation}
    E_{\rm el}=\int \mathrm{d}E \  \rho(E)\int \mathrm{d}u \ \frac{u^2}{2}\tilde{P}(E,u)+\frac{\gamma(t)^2}{2}
\end{equation}
which shows that the system behaves purely elastically at the transition. If we look at the energy stroboscopically, at $\gamma=0$, then we are left only with the first integral. We introduce
\begin{equation}
    \tilde{E}_{\rm el}(E)=\int \mathrm{d}u \ \frac{u^2}{2}\tilde{P}(E,u)
\end{equation}
which using the expression (\ref{p_e_u}) for $\tilde{P}(E,u)$ simplifies to
\begin{equation}
    \tilde{E}_{\rm el}(E)=\frac{1}{4 \alpha} \int_{-u_c(E)}^{u_c(E)} \mathrm{d}u_0 \ R(u_0) \ \frac{u_c^4-u_0^4}{6}
\end{equation}
We can then finally integrate over the energy spectrum to obtain the total elastic energy:
\begin{equation}
    E_{\rm el}=\int \mathrm{d}E \ \rho(E)  \tilde{E}_{\rm el}(E) 
\end{equation}

\section{Numerical methods}

As noted in the main text, the numerical solution of the model entails choosing a discrete set of energy levels $\{E_i\}$ and solving a PDE in the strain variable for each. We recall that we consider throughout the Gaussian distribution of energy levels
\begin{equation}
    \rho(E)=\sqrt{\frac{2}{\pi \sigma^2}} e^{-\frac{E^2}{2 \sigma^2}}
\end{equation}
with $\sigma=0.05$, and we fix the coupling to $\alpha=0.086\,\Em$ with $\Em=\sqrt{2/\pi}\,\sigma\approx 0.03989$. We recall from the main text that this value of the coupling corresponds to $\gamma_0^{*}\approx \sqrt{\Em}\approx 0.1997$, and that we use the fixed finite frequency $\omega=0.1$ throughout. 
We describe here firstly how the energy levels are chosen, before we turn to the numerical solution of the PDE, and finally to the question of how the levels are coupled to each other through the yield rate.

For the numerical results shown in the paper, we consider $15$ $\{E_i\}$ points. The spacings between those are chosen to ensure that we obtain a reasonable sampling of the Gaussian $\rho(E)$, including both the range of shallow minima with $E<2\gamma_0^*{}^2$ and the tail of deep minima.
Explicitly, the first five $E_i$ are chosen uniformly spaced within $(0,2 {\gamma^{*}_0}^2\approx 0.08)$, in order 
to always keep the same fixed resolution for the shallow elements.
Beyond this value, we consider another $10$ points, the spacing of which can be tuned in order to better resolve the well-annealed initial conditions. We introduce a ``Boltzmann" factor $\hat{\beta}$ (note that this is independent of and not in general equal to the $\beta$ that sets the initial condition), which defines a distribution
\begin{equation}
    \hat{\rho}(E)=N(\hat{\beta})^{-1} \sqrt{\frac{2}{\pi \sigma^2}} e^{-\frac{E^2}{2 \sigma^2}} e^{\hat{\beta}E}
\end{equation}
with the normalization factor
\begin{equation}
    N(\hat{\beta})=e^{\hat{\beta}^2\sigma^2/2}\left(1+\erf(\hat{\beta} \sigma/\sqrt{2})\right)
\end{equation}
We then set the $10$ remaining energy levels by requiring them to capture the same amounts of (weighted) probability $\hat\rho(E)$. Explicitly this means we choose 
them as on a uniform grid in $q$ within $(q(2 {\gamma_0^{*}}^2),1)$, where $q(E)$ is defined as the cumulative distribution
\begin{equation}
    \frac{\mathrm{d} q}{\mathrm{d} E}=\hat{\rho}(E)
\end{equation}
The above method therefore gives us a parameter $\hat{\beta}$ that we can tune in order to resolve better the tail of the distribution in the well-annealed cases. 
For the poorly annealed samples, $\beta=0$ to $30$, we use the intermediate value $\hat{\beta}=20$. For $\beta=50$ and $70$, we use $\hat{\beta}=50$. For the yielded state, we use $\hat{\beta}=0$. The different values of $\hat{\beta}$ lead to slight differences in macroscopic stress and energy values measured in the yielded state, as may be noticed in Fig.~2 of the paper.

Once the $\{E_i\}$ are chosen, one needs to solve the PDE in strain to evolve $P(E,l,t)$. We work with a rescaled strain $\tilde{l}=l/l_c(E)$, so that the strain thresholds are always at $\tilde{l}=\pm 1$. We discretize the $\tilde{l}$-axis by considering $N=4096$ points in the interval $[-4,4)$. We then update the distribution in time using a pseudospectral method as described in \cite{parley_aging_2020}, implementing the diffusive term in Fourier space. As stated in the main text, we initialise the strains in a narrow distribution, with standard deviation $l_c(E)/6$. In particular we use the power law form $\sim (1-|\tilde{l}|)^{\delta}$ with $\delta=6$ and $|\tilde{l}|<1$. As long as this distribution is sufficiently narrow, we do not expect its precise form to make any qualitative difference to the results. 

We do the updates of $P(E,l,t)$ in parallel for each energy level, but of course the different levels are coupled by the yield rate $Y(t)$, which must be computed from the whole system. For this we exploit the natural separation of timescales within the model. On the one hand, we notice that $Y(t)$ varies smoothly within the period on timescales of the order of $1/\omega$. We therefore introduce a fixed ``macroscopic'' timestep ${\mathrm{d}t}_{\rm macro}=10^{-2}$. After each step of ${\mathrm{d}t}_{\rm macro}$, $Y(t)$ is calculated by performing the discretised integral of $Y(E,t)$ over the $\{E_i\}$. Within each ${\mathrm{d}t}_{\rm macro}$, we update each level in parallel, with an energy dependent ${\mathrm{d}t}_{\rm micro}(E)=CE \ll {\mathrm{d}t}_{\rm macro}$. 
The reason for the linear energy dependence is that, within rescaled units $\tilde{l}=l/l_c$, one sees that the diffusive term (which is the hardest to resolve) scales as $1/E$, so that the timestep has to be scaled down accordingly. Finally, we update the prefactor $C$ adaptively during the dynamics depending on the period-average $\overline{Y}$ after each cycle, which is especially important to efficiently simulate the long transient behaviour of the well-annealed samples.


\section{Dependence on initial annealing}
\subsection{Stress amplitude}
\begin{figure}\center{\includegraphics[width=0.5\textwidth]{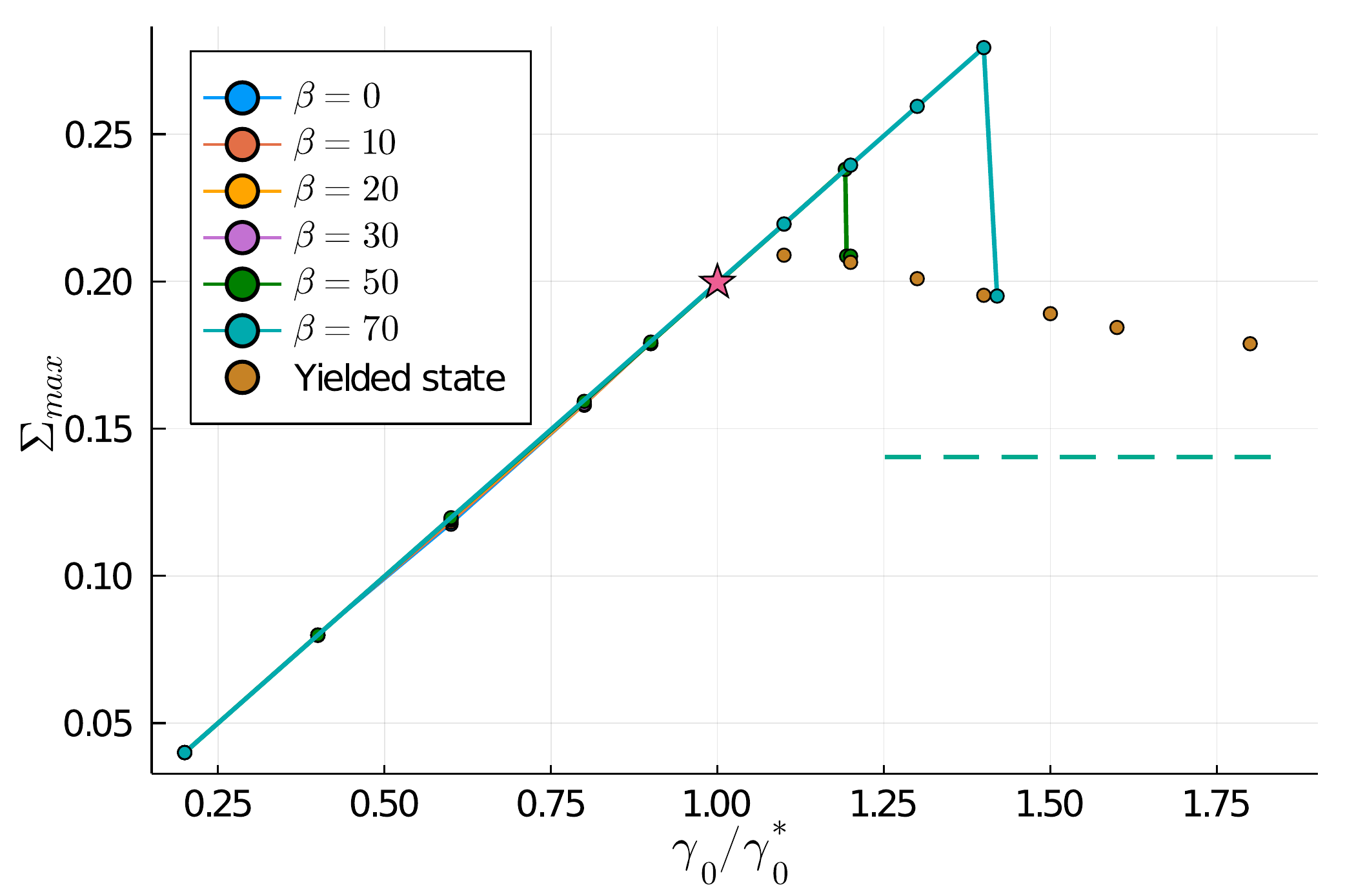}}
\caption{\label{fig:stress}Stress amplitude in steady state after many cycles of shear. Star indicates analytical value at threshold, which is simply $\Sigma=\gamma_0^{*}$ given that the system in steady state is elastic up to that strain. Dashed lines indicate the exact macroscopic yield stress $\Sigma_y$ (\ref{macro_yield}). 
}
\end{figure}

We show in Fig.~\ref{fig:stress} the equivalent of Fig.~2 in the main text, but this time for the maximum stress over the cycle once the system has reached its steady state, whether   frozen elastic or yielded. Within the precision and range of our numerics, the data are consistent with poorly annealed samples (up to $\beta=30$) showing a slight stress increase 
at or just above \footnote{As commented on below, we expect the actual yield point for this frequency to be slightly above $\gamma^{*}_0$, although we are not able to determine this precisely within our numerics. It is possible therefore that the stress amplitude in the yielded state monotonically decreases above the yield point. Another possibility is that the non-monotonicity disappears for $\omega \rightarrow 0$, but we have not studied this in detail.} the common yield point $\gamma^{*}_0$. Highly annealed samples (up to $\beta=50$, $70$) behave elastically beyond this point, until at $\gamma_c(\beta)$ the stress amplitude drops to that of the ergodic steady state.

We note that in MD studies (see e.g. \cite{bhaumik_role_2021}) the initial increase is not completely elastic; in our model it is, due to the simplification of considering the same uniform elastic modulus $k=1$ for all elements, irrespective of their energy level $E$. One could obtain non-elastic behaviour by considering a distribution of $k$-values, which should presumably be correlated with $E$.


\subsection{Limit of steady shear}
In Fig.~\ref{fig:stress} we have included the dynamical yield stress of the model, which should be attained as $\gamma_0 \rightarrow \infty$. The expression for this was derived by Agoritsas et al. \cite{agoritsas_relevance_2015}. We reproduce it here for completeness; we also give the derivation of the limiting $\gamma_0 \rightarrow \infty$ value of the energy, shown in Fig.~2 of the main text. We note that the total energy in our numerical evaluations has been measured stroboscopically at $\gamma=0$ (where the shear rate is far from being reversed), so that for large $\gamma_0$ one indeed expects it to coincide (for small $\omega)$ with the steady state energy attained under uniform shear as $\dot{\gamma}\rightarrow 0$.

In the $\dot{\gamma}\rightarrow 0$ limit of steady shear, it was found that \cite{agoritsas_relevance_2015}
\begin{equation}\label{p_of_e_ag}
    P(E)=\frac{\sqrt{\Ez}}{\alpha}\sqrt{E}\tanh{\left(\sqrt{\frac{E}{\Ez}}\right)}\rho(E)
\end{equation}
where the characteristic energy scale $\Ez$
is determined by the normalization condition:
\begin{equation}\label{norm_ag}
    1=\frac{\sqrt{\Ez}}{\alpha}\bigg\langle \sqrt{E}\tanh{\left(\sqrt{\frac{E}{\Ez}}\right)}\bigg\rangle
\end{equation}
where the average is over $\rho(E)$. An example of the distribution (\ref{p_of_e_ag}) is shown in Fig.~\ref{fig:steady_shear} later in Sec. VI where we discuss results for uniform shear.
The corresponding macroscopic yield stress was found to be \cite{agoritsas_relevance_2015} 
\begin{equation}\label{macro_yield}
   \Sigma_y=\frac{1}{\sqrt{2}}\left(\frac{\langle E \rangle}{\bigg\langle \sqrt{E}\tanh{\left(\sqrt{\frac{E}{\Ez}}\right)}\bigg\rangle}-\sqrt{\Ez}\right)
\end{equation}

To find the limiting value of the total energy, we proceed again by separating the potential and elastic contributions. For the potential energy, one simply needs to take the average over (\ref{p_of_e_ag}). To compute the elastic contribution, we proceed in the following way, which is exact for $\talpha=0$, and should be a good approximation for $\talpha \ll 1$. In particular we expect this to be fairly accurate for the particular case $\talpha=0.086$ studied in the paper, given that the macroscopic yield stress is already quite close to the limiting value at $\talpha=0$ (for the Gaussian considered in the paper $\Sigma_y (\talpha=0.086)=0.140328$, to be compared with $\Sigma_y (\talpha=0)=0.15345$).

The approximation consists in assuming that the distribution in strain is flat across positive strains in all energy wells, with an energy-dependent height fixed by the distribution $P(E)$, that is


\begin{equation}
    P(E,l)=A(E) \ \ {\rm{for}} \ \ l \in (0,l_c(E))
\end{equation}
with the height $A(E)$ fixed by
\begin{equation}
    \int_0^{l_c(E)}\mathrm{d}l\, P(E,l)=P(E)
\end{equation}
which yields
\begin{equation}
    A(E)=\sqrt{\frac{\Ez}{2}}\frac{1}{\alpha}\tanh{\left(\sqrt{\frac{E}{\Ez}}\right)}\rho(E)
\end{equation}
Now the elastic energy is simply
\begin{equation}
    \int_0^{\infty}\mathrm{d}E\int_0^{l_c(E)}\mathrm{d}l \ \frac{l^2}{2}P(E,l)
\end{equation}
Carrying out the $l$-integral explicitly one finds
\begin{equation}
   E_{\rm el}=\frac{\sqrt{\Ez}}{3\alpha}\int_0^{\infty}\mathrm{d}E\, \rho(E) \tanh{\left(\sqrt{\frac{E}{\Ez}}\right)} E^{3/2}
\end{equation}
In the limit $\alpha\rightarrow 0$, where from (\ref{norm_ag})  $\sqrt{\Ez}\sim \alpha/\langle\sqrt{E}\rangle$, one has the limiting (exact) value
\begin{equation}
    E_{\rm el}(\alpha=0)=\frac{\langle E^{3/2}\rangle}{3 \langle\sqrt{E}\rangle}
\end{equation}

\subsection{Critical behaviour}
\begin{figure}\center{\includegraphics[width=0.5\textwidth]{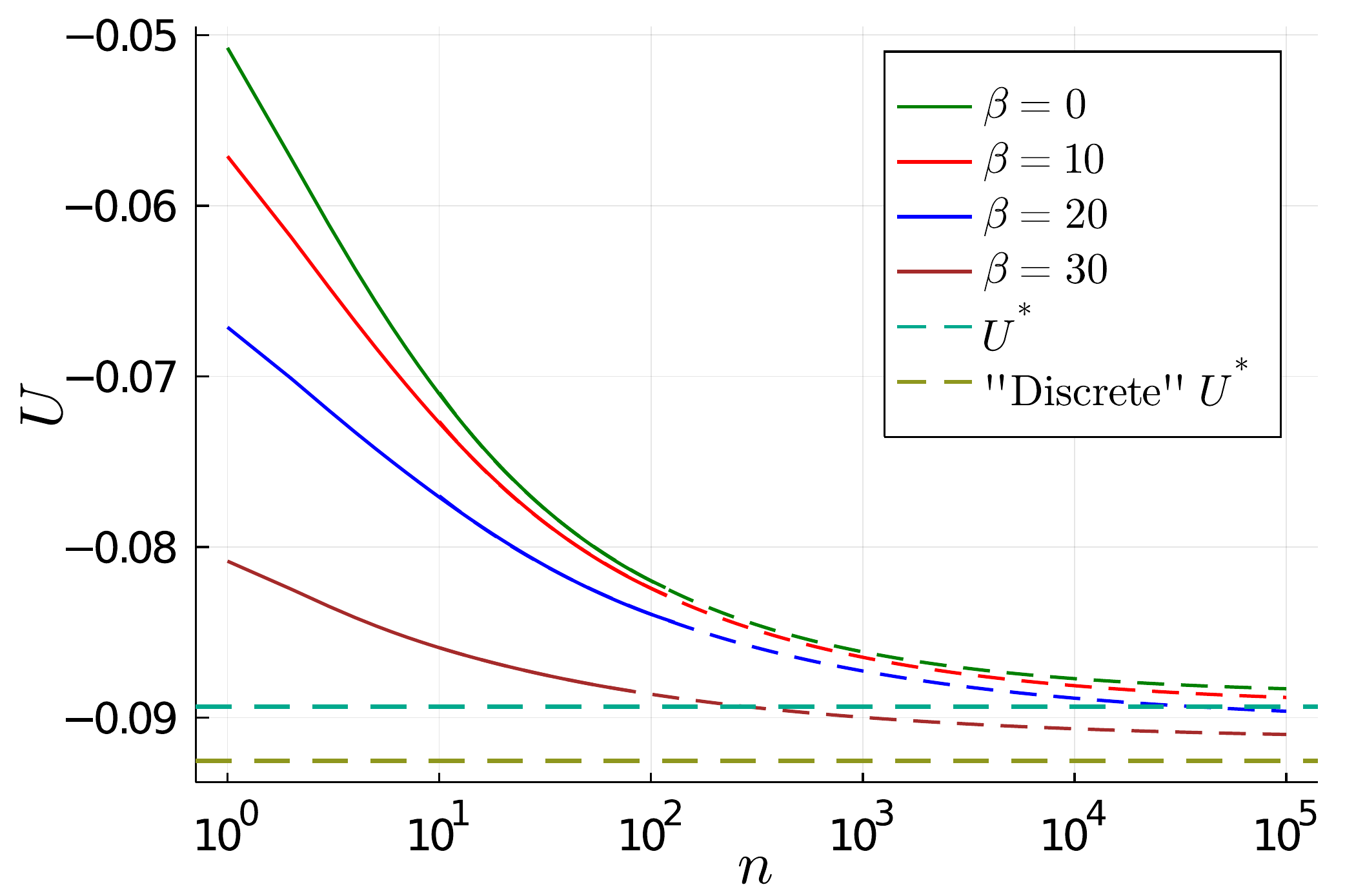}}
\caption{\label{fig:extrapolation}Decay of the stroboscopic total energy $U$ against number of cycles $n$ at strain amplitude $\gamma_0=\gamma^{*}_0$. 
Dashed lines show the fitted power-law extrapolation.}
\end{figure}

We address here details regarding the critical behaviour on the solid (frozen) 
side as the common yield point $\gamma_0^{*}$ is approached. In Fig.~\ref{fig:extrapolation} we show the evolution of the total energy, measured stroboscopically ($\gamma=0$) at each cycle, with an applied oscillatory shear amplitude $\gamma_0=\gamma^{*}_0$. Due to the finite discretisation of the strain 
axis, our numerical setup is unable to explore the dynamics reliably once the yield rate becomes very small; with our numerical parameters this range extends roughly down to $\overline{Y}\approx 10^{-3.5}$). This can be seen from the full lines in Fig.~\ref{fig:extrapolation}, which stop after around a hundred cycles. It is clear from the figure, however, that the energy is still decaying. Therefore, as noted in the main text, to obtain the steady state values of the energy for $\gamma_0=\gamma^{*}_0$ (and $0.9\gamma^{*}_0$) in Fig.~2 of the paper we extrapolate the $U(n)$ curve, using a power-law fit
\begin{equation}
    U(n)=U_{\infty}+A n^{-\delta}
\end{equation}
where we fit the three parameters $U_{\infty}$, $A$ and $\delta$. These fits are shown in Fig.~\ref{fig:extrapolation} 
for $\beta=0,10,20,30$, where we fit $\delta=0.423,0.384,0.319,0.304$ respectively.

Also shown in Fig.~\ref{fig:extrapolation} is the ``discretised'' value of $U^{*}$, which is simply the total energy (both potential and elastic) calculated at the transition using the same discrete energy points as in the numerical solutions. We note that it is slightly lower than the true $U^{*}$ calculated for a continuous spectrum of energy levels. Fig. \ref{fig:extrapolation} therefore suggests that the $U^*$ corresponding to the energy discretisation used 
has still not quite been reached at $\gamma^{*}_0$, so that the real yield point is in fact slightly higher than the one calculated analytically in the $\omega \rightarrow 0 $ limit. This is consistent with MD studies \cite{yeh_glass_2020}, where the yield point was found to increase slightly with the shear rate (related to the frequency as $\dot{\gamma}_0=\gamma_0 \omega$). We do not characterise this here in detail, as determining the precise yield point is made difficult by the fact that we are unable to numerically resolve steady states with very low $\overline{Y}$ just above yielding.

\begin{figure}\center{\includegraphics[width=0.5\textwidth]{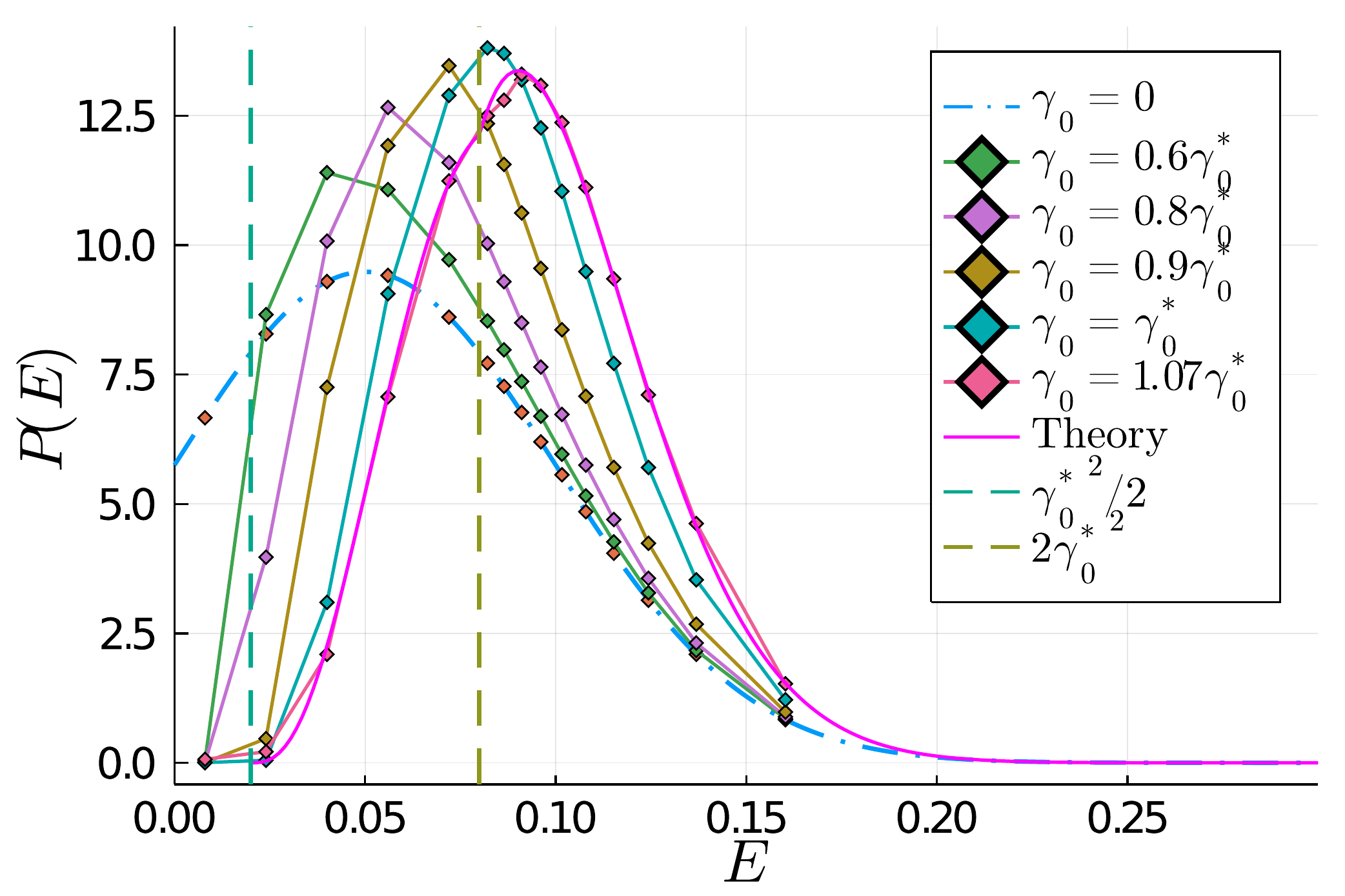}}
\caption{\label{fig:approach}Final stroboscopic $P(E)$ distribution measured from the numerical solution at different strain amplitudes, for the initial condition $\beta=20$. We also show by vertical dashed lines the two special values ${\gamma_0^{*}}^2/2$ and $2 {\gamma_0^{*}}^2$. The theoretical distribution is zero below the first value, whereas it shows a small kink at the second one, where direct yields cease to be present.}
\end{figure}

In Fig.~\ref{fig:approach}, we show the last (stroboscopic) energy distribution $P(E)$ measured from our 
numerical solutions at a given shear amplitude $\gamma_0$, for the initial condition $\beta=20$. We see that as $\gamma_0$ increases towards $\gamma^{*}_0$, the system mechanically anneals
towards the analytically calculated distribution. At $\gamma_0=\gamma^{*}_0$, the last $P(E)$ we measure has not quite reached this theoretical prediction for $\omega\to 0$, mirroring the incomplete decay of the energy in Fig. \ref{fig:extrapolation}. Here it is difficult to perform an extrapolation; instead we look at a slightly larger $\gamma_0=1.07 \gamma ^{*}_0$, where the system is expected from theory to fluidize but is still in the regime where we cannot resolve numerically the steady state. We can then extract the last $P(E)$ before our discretization limit is reached, and we find almost perfect agreement with the theory.

We next study the critical behaviour of the (period-averaged) yield rate $\overline{Y}$. In order to find clearly the power-law regime and minimise transient effects \footnote{We note that these transient effects due to the energy redistribution could slightly modify the numerically measured value of the exponent $b$.}, we use here an initial distribution already ``close" to the limiting distribution at the transition, at least in the energy variable. The ensuing relaxation dynamics therefore involves primarily the strain variable, presumably controlled by boundary layers as studied in \cite{sollich_aging_2017,parley_aging_2020} for the aging dynamics. In particular, we consider
\begin{equation}\label{ic_crit}
    P(l,E,t=0)=P^{*}(E)\frac{1}{\sqrt{2\pi}l_c \sigma_l}e^{-\frac{l^2}{2 l_c^2 \sigma_l^2}}
\end{equation}
with $\sigma_l=0.4$. The Gaussian strain distribution ensures a small but finite fraction of unstable elements in each energy level, so that the initial condition is not ``frozen''. 

\begin{figure}\center{\includegraphics[width=0.5\textwidth]{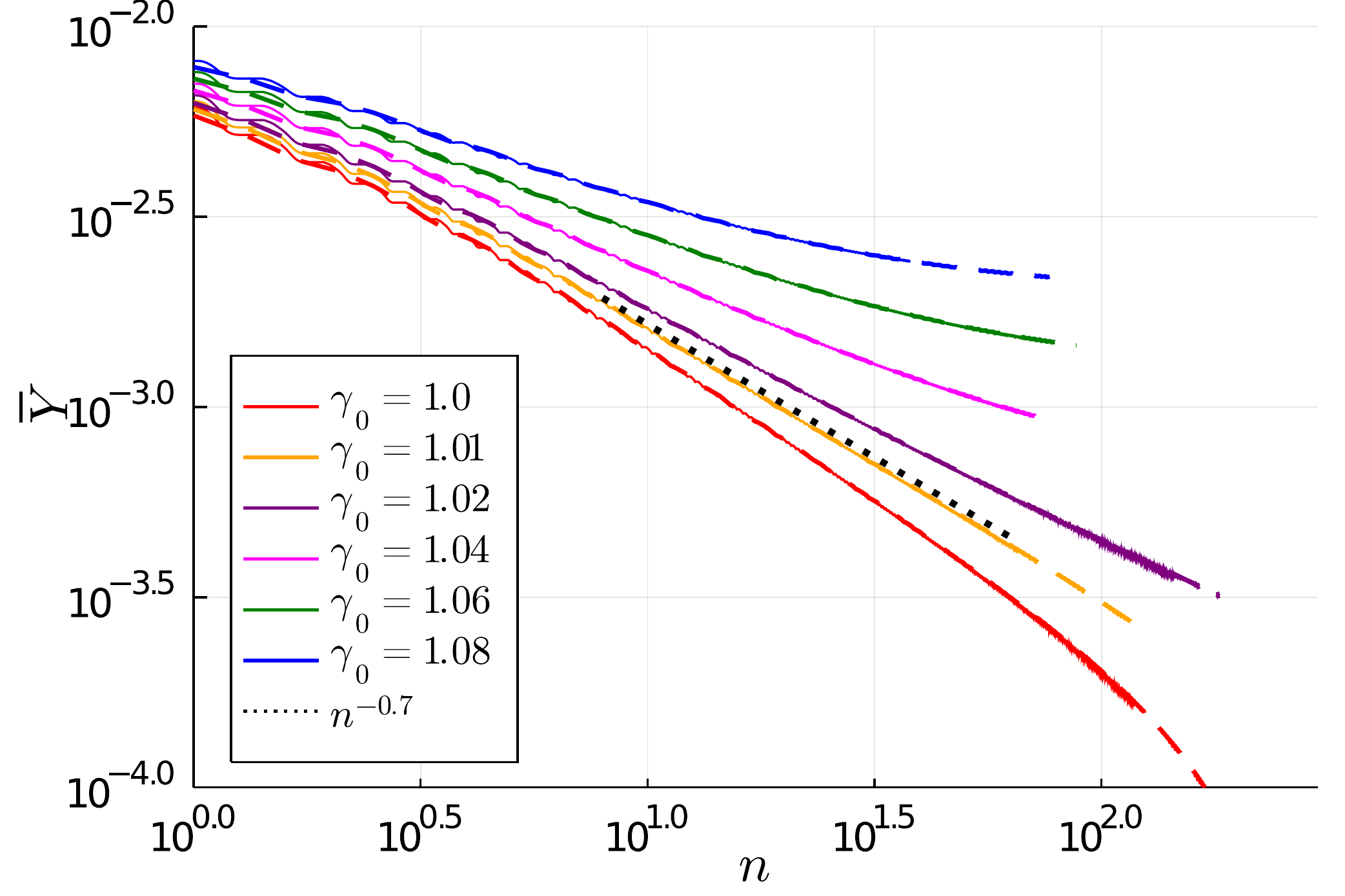}}
\caption{\label{fig:critical}Critical behaviour at $\tilde{\alpha}=0.086$, where we measure $b \approx 0.7$. Dashed lines show runs with a coarser discretisation in stress ($N=2048$), which agree within the time window shown. $\gamma_0$ values in legend are in units of $\gamma^{*}_0$.}
\end{figure}

We show results for $\talpha=0.086$ (the value considered throughout the paper) in Fig.~\ref{fig:critical}, for a range of $\gamma_0$ near $\gamma^{*}_0$. We measure a power law $n^{-b}$ with exponent $b\approx0.7$, clearly below $1$. This fact is important, as it implies a diverging number of events in the long time limit, allowing the system to lose memory of the initial condition.

Although we do not have an analytical understanding of the exponent $b$, we conjecture that is non-universal and dependent on $\talpha$. In the limiting case of $\talpha=1$ (where $\tgamma^{*}=0$), we have checked that $b=1$ for the critical dynamics in the absence of shear. This is the value found for the original HL model \cite{sollich_aging_2017,parley_aging_2020}, and can be understood from a boundary layer analysis. 



\subsection{Yielded state}

\begin{figure}\center{\includegraphics[width=0.5\textwidth]{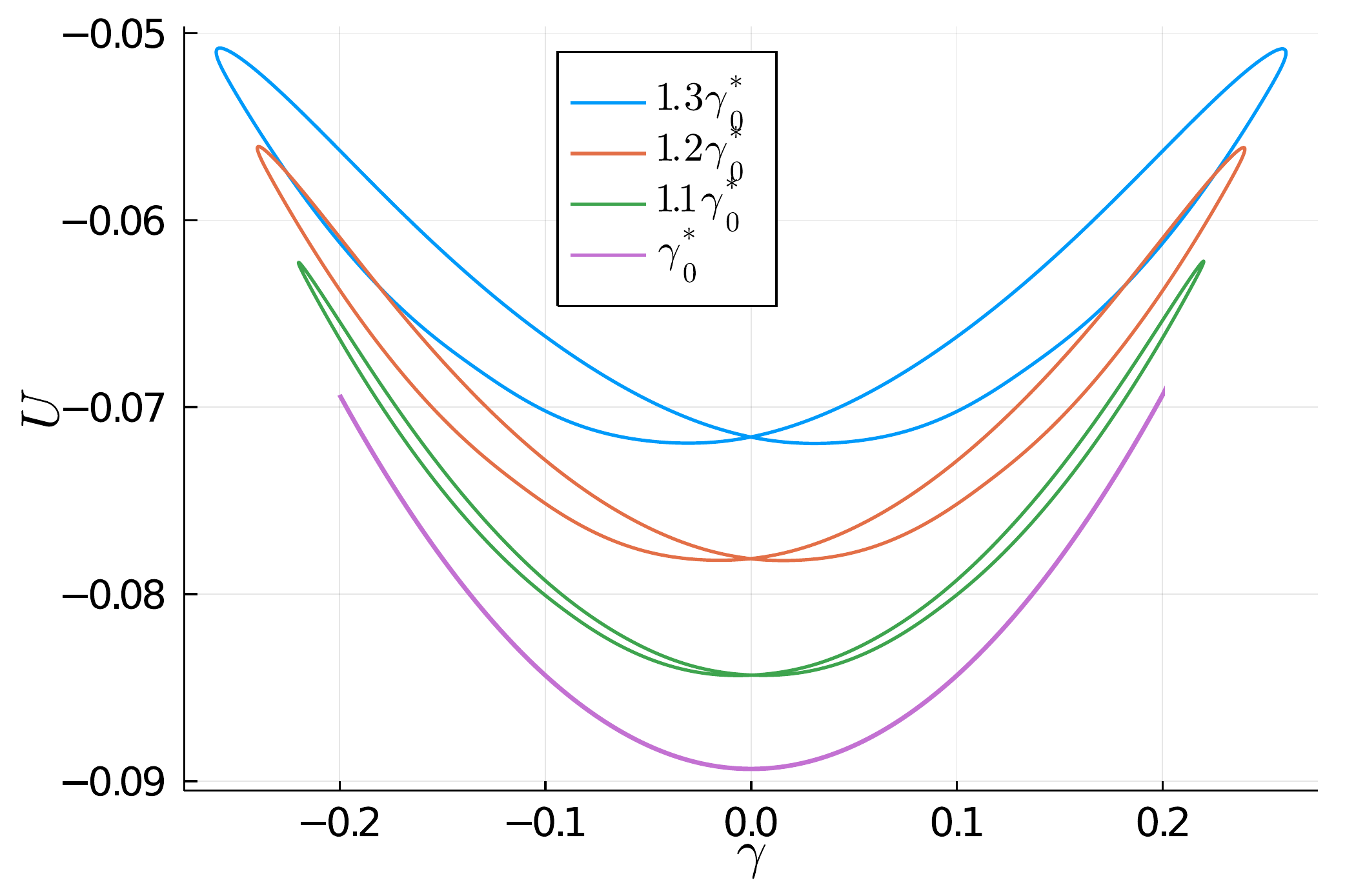}}
\caption{\label{fig:intra_energy}Energy within one cycle in the yielded steady state, plotted parametrically against the current strain $\gamma(t)=\gamma_0 \sin (\omega t)$.}
\end{figure}

\begin{figure}\center{\includegraphics[width=0.5\textwidth]{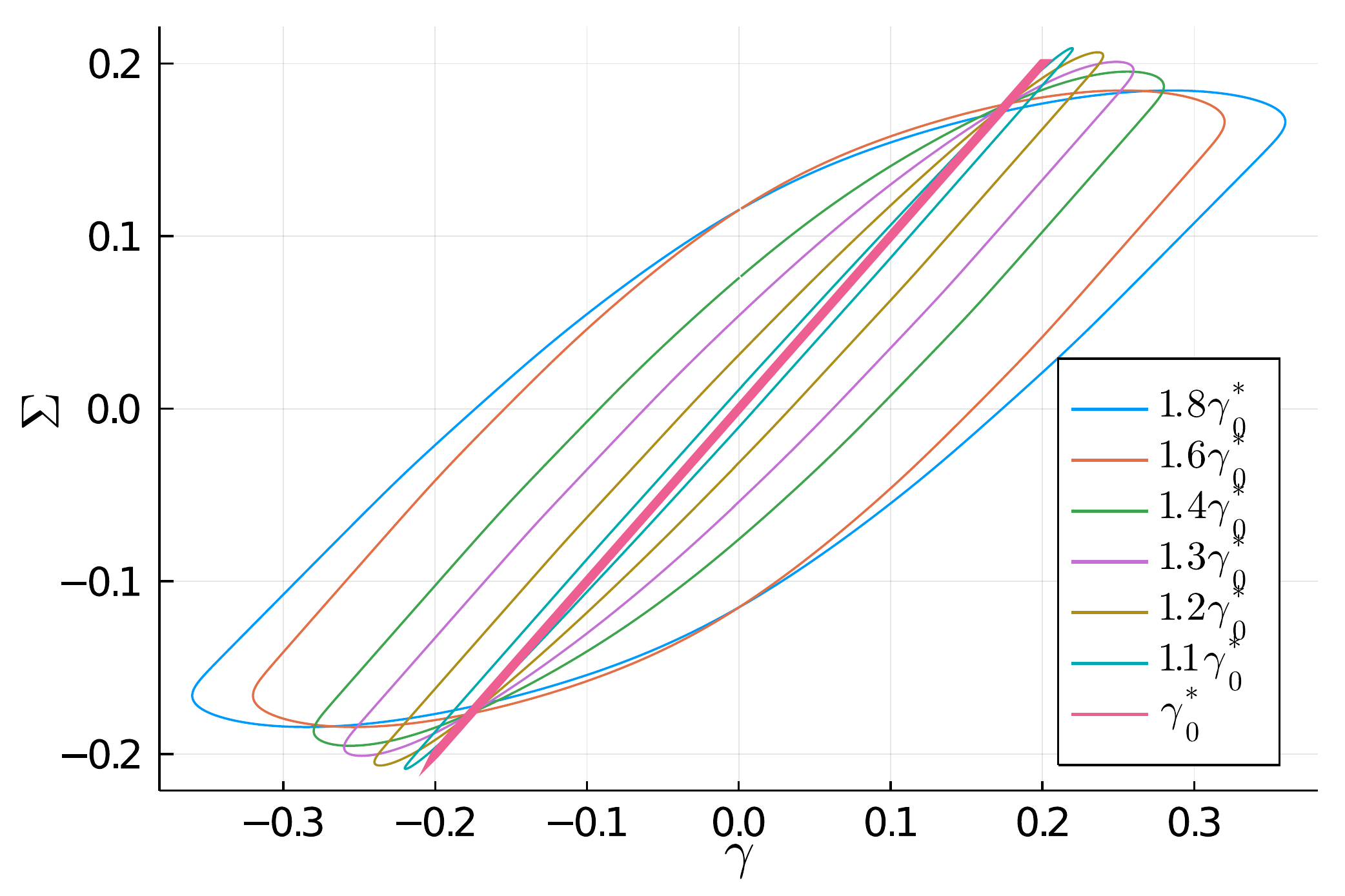}}
\caption{\label{fig:intra_stress}Macroscopic stress within one cycle in the yielded steady state, plotted parametrically against the current strain $\gamma(t)=\gamma_0 \sin (\omega t)$.}
\end{figure}

We next show two figures concerning the yielded state, for the energy (Fig.~\ref{fig:intra_energy}) and the stress (Fig.~\ref{fig:intra_stress}), within one cycle of shear. In both cases we show additionally the theoretical prediction at $\gamma_0=\gamma_0^{*}$: we recall that as $\overline{Y}$ vanishes continuously at the transition, this is simply purely elastic behaviour. As $\gamma_0$ is increased, we see that (Fig.~\ref{fig:annealed_energy}) the energy describes a ``butterfly'' shape, while the stress acquires an ever wider hysteresis loop. These are both qualitatively consistent with the behaviour in particle simulations \cite{leishangthem_yielding_2017,yeh_glass_2020,bhaumik_role_2021}.

\section{Fatigue}

\begin{figure}\center{\includegraphics[width=0.6\textwidth]{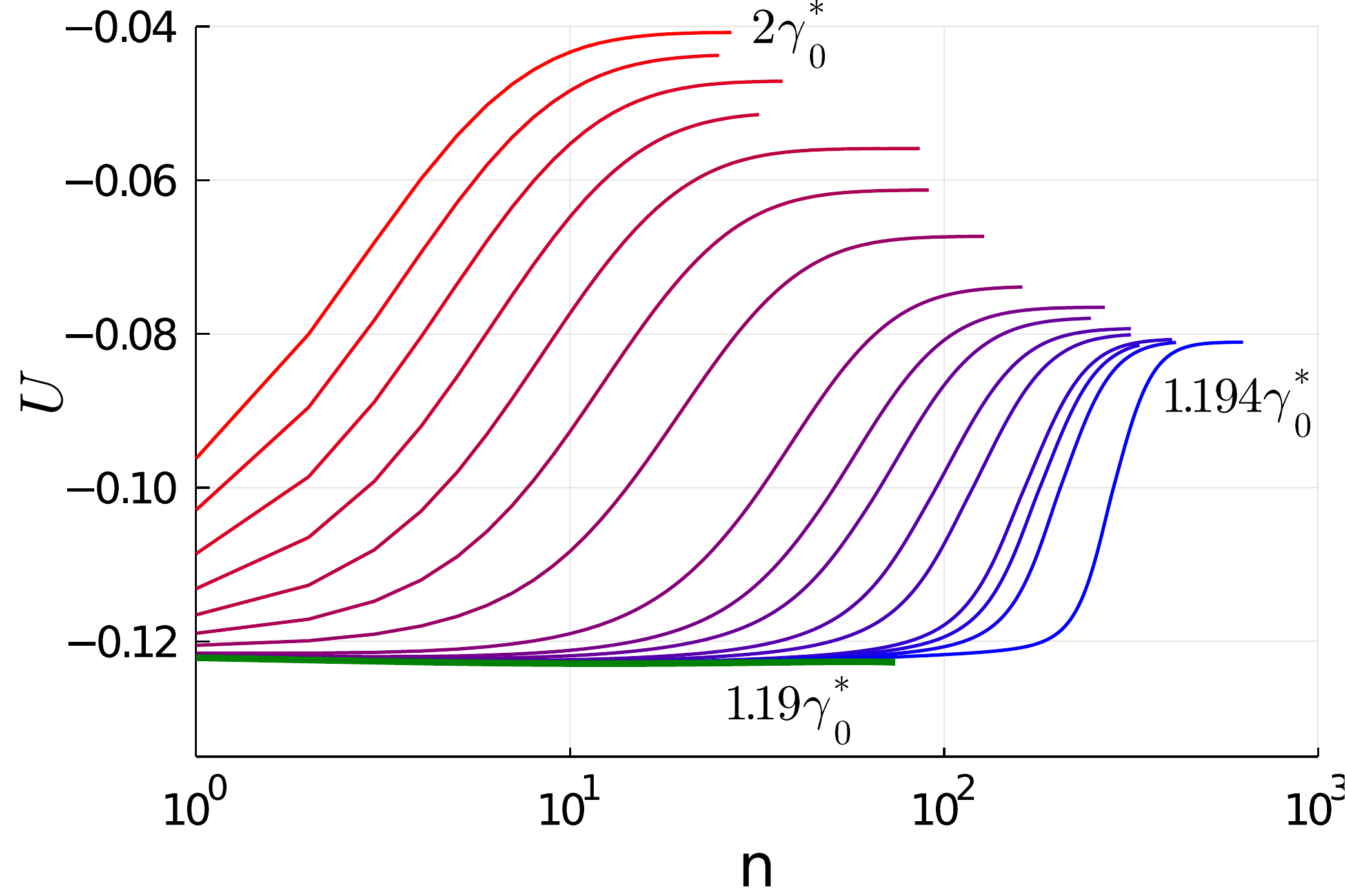}}
\caption{\label{fig:annealed_energy}Energy $U$ against number of cycles, for $\beta=50$ and a range of strain amplitudes around the fatigue limit $\gamma_c$ as given in the text.}
\end{figure}

We provide supplementary information here regarding the fatigue behaviour of the well-annealed sample ($\beta=50$). In Fig.~\ref{fig:annealed_energy}, we show the same runs as in Fig.~3 of the main text, but this time plotting the energy $U$. The $16$ strain amplitudes above yield are, from blue to red (as in Fig. 3 of the main text): 1.194, 1.196, 1.198, 1.2, 1.21, 1.22, 1.24, 1.26, 1.3, 1.4, 1.5, 1.6, 1.7, 1.8, 1.9 and 2 $\gamma_0^{*}$. Also as in Fig. 3 of the main text, we show an additional run below the fatigue limit, for 1.19 $\gamma_0^{*}$ (green). The near-constancy of this shows clearly that, approaching $\gamma_c$ from below, the energy remains very largely insensitive to the shear, due to the very low yield rates involved. 

The evolution of the energy above the fatigue limit may be compared to results shown in the SM of~\cite{yeh_glass_2020} (Fig.~S4a). There, a well-annealed sample is considered at a higher shear rate, where both initial and persistent shear banding are absent. The behaviour of the energy is qualitatively very similar to our numerical results, lending credence to the expectation that the transient behaviour described in mean field should become more accurate in this regime.


\begin{figure}\center{\includegraphics[width=0.5\textwidth]{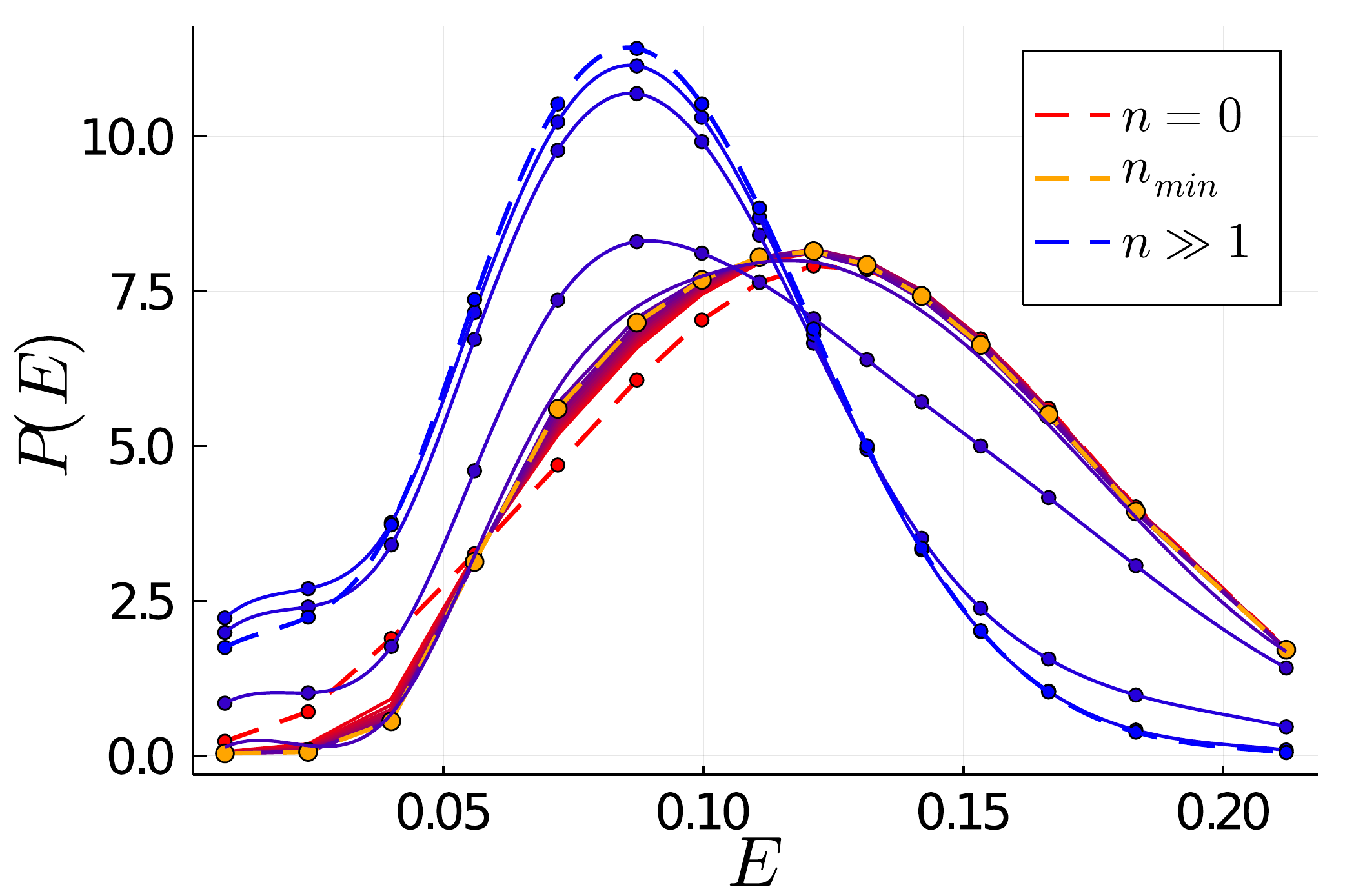}}
\caption{\label{fig:occupancies}Evolution of $P(E)$ for $\gamma_0=1.194 \gamma_0^{*}$. Log-times are uniformly spaced, from red to blue. Shown explicitly are the initial distribution (red dashed line), the almost frozen distribution at $n_{\rm min}$ (orange dashed line), and the distribution in the yielded steady state (blue dashed line). The last curves are smooth fits as a guide to the eye. Note that in general we expect $P(E)$ in the yielded state to vanish as $\sim\sqrt{E}$ as $E \rightarrow 0$ (see Fig.~\ref{fig:steady_shear}). We cannot resolve this within our energy discretisation, but the region of very small $E$ in any case does not contribute significantly to macroscopic quantities.
}
\end{figure}

In Fig.~\ref{fig:occupancies} we show the evolution of the energy distribution $P(E)$ for the run closest to $\gamma_c$ (from above), $\gamma_0=1.194 \gamma_0^{*}$. As stated in the main text, the initial energy distribution (red dashed line) initially approaches a fixed frozen state, which it comes closest to at $n_{\rm min}$ (defined as the point where $\overline{Y}$ reaches its minimum value). Beyond $n_{\rm min}$, yield events across the system drive the distribution towards the yielded state. The intriguing approach to and escape from a frozen stationary state leads to the interesting non-monotonic behaviour we observe, and associate with fatigue behaviour.

\begin{figure}\center{\includegraphics[width=0.5\textwidth]{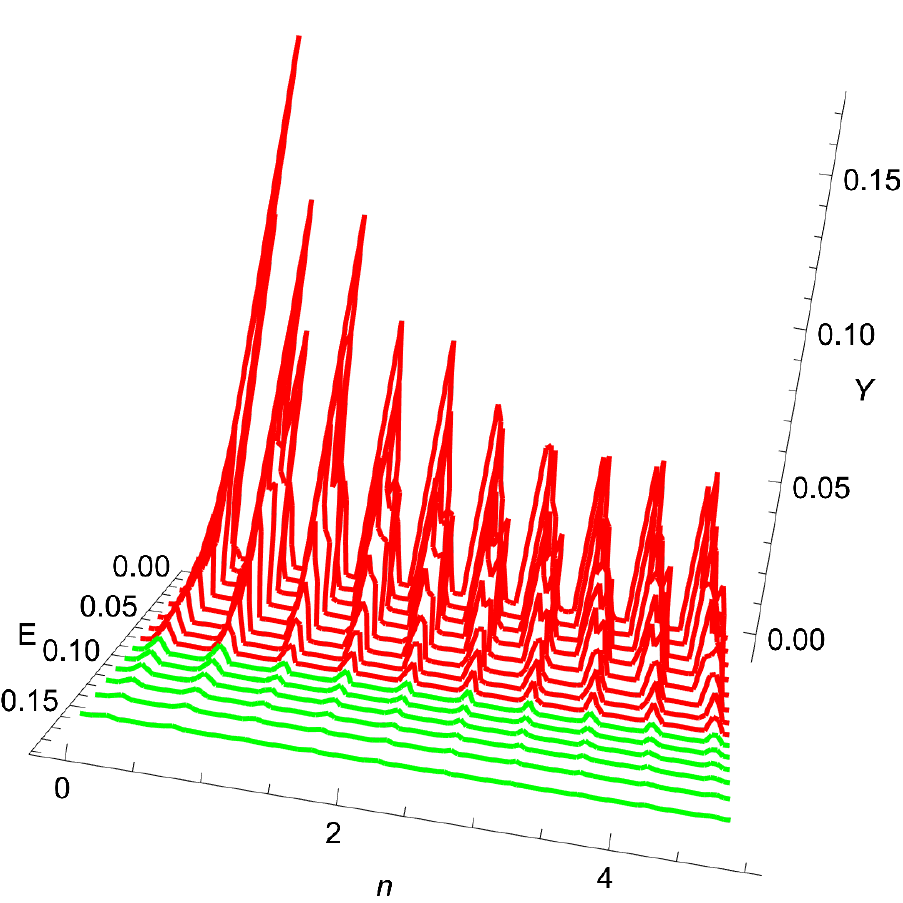}}
\caption{\label{fig:first_cycles}Yield rate $Y(E)$ of energy level $E$ in the first 5 cycles, for $\gamma_0=1.194 \gamma_0^{*}$. Shown in red are the first eight energy levels $\{E_i\}$ within our discrete set, which satisfy $E<2\gamma_0^2$, while levels with $E>2\gamma_0^2$ are shown in green. These are levels where no direct yields are possible, so that their contribution to the plastic activity during the initial cycles is vanishingly small. 
}
\end{figure}

In Fig.~\ref{fig:first_cycles}, we substantiate our claim in the main text that the plastic activity during the initial cycles (shown are the first five) is dominated by direct yielding of elements in shallow levels. Significant direct yields only occur for $E<2 \gamma_0^2$; for the strain amplitude $\gamma_0=1.194 \gamma_0^{*}$, and our discrete set of levels, this condition is fulfilled by the first eight energy levels.

\section{Effect of closure relation}

We consider here the effect of a general closure relation 
\begin{equation}\label{gen_clos}
    D(t)=\int\mathrm{d}E\, \alpha(E)Y(E,t)
\end{equation}
where
\begin{equation}
    Y(E,t)=\int \mathrm{d}l\, \theta \left(|l|-l_c(E)\right)P(E,l,t)
\end{equation}
The case considered in the main paper is the one of constant $\alpha(E)$, 
which leads back to the simple proportionality $D(t)\propto Y(t)$. Once an $E$-dependence 
is allowed for, a natural choice is $\alpha(E)=\hat{\alpha}E$, with $\hat{\alpha}$ a now dimensionless coupling parameter. We will refer to this as linear closure, and consider it for the numerics below. Physically, the linear closure corresponds to yield events at different energy levels contributing to the mechanical noise proportionally to $E$, which in turn is proportional to $\sigma_c^2$ ($\sigma_c$ being the local yield stress). This is the form one would expect if the model is derived e.g. from the KEP perspective \cite{bocquet_kinetic_2009,agoritsas_relevance_2015}. We firstly show that the general form (\ref{gen_clos}) does not change qualitatively the phase diagram of the model, i.e. the transition line. This is due to the following result. 
Starting from the master equation
\begin{equation}
    \partial_t P(E,l,t)=-\dot{\gamma}(t)\partial_l P(E,l,t)+D(t)\partial^2_l P(E,l,t)
    -\theta\left(|l|-l_c(E)\right) P(E,l,t)+Y(t)\rho(E)\delta (\sigma)
\end{equation}
and integrating over strain, one has
\begin{equation}\label{balance_Ys}
    \partial_t P(E,t)=-Y(E,t)+Y(t)\rho(E)
\end{equation}
If we then average (\ref{balance_Ys}) over a period in steady state it follows that
\begin{equation}\label{Y_rho}
    \overline{Y}(E)=\overline{Y}\rho(E)
\end{equation}
Eqs.~(\ref{gen_clos}) and (\ref{Y_rho}) imply that the following holds in any periodic steady state:
\begin{equation}\label{D_av}
    \overline{D}=\int\mathrm{d}E\, \overline{Y}(E)\alpha(E) =\overline{Y}\int\mathrm{d}E \,\rho(E)\alpha(E)\equiv \alpha_{\rm eff} \overline{Y}
\end{equation}
where we have defined 
\begin{equation}
    \alpha_{\rm eff}=\int \mathrm{d}E\, \rho(E) \alpha(E) 
\end{equation}
so that in the linear closure case $\alpha_{\rm eff}=\hat{\alpha}\langle E \rangle$.

We now recall that the equations for the limiting yield rate depend \textit{only} on $\gamma_0$. The closure relation comes in at the next step, where the normalization condition is enforced; however, one is there dealing with period-averaged quantities, so that one can simply replace $\alpha$ by $\alpha_{\rm eff}$. In the linear closure case, we then have precisely the same transition line as in the main text, provided we consider the rescaled quantity, i.e. $\tgamma^{*}(\hat{\alpha})=\tgamma^{*}(\talpha)$.

Although the transition line remains unchanged, one expects differences both in the transient and in the steady state above yield. Note that even \textit{at} the transition the actual form of $D(t)$ (\ref{gen_clos}) within the period will in general change, even though $y(t)$ is independent of the closure relation. 

\begin{figure}\center{\includegraphics[width=0.5\textwidth]{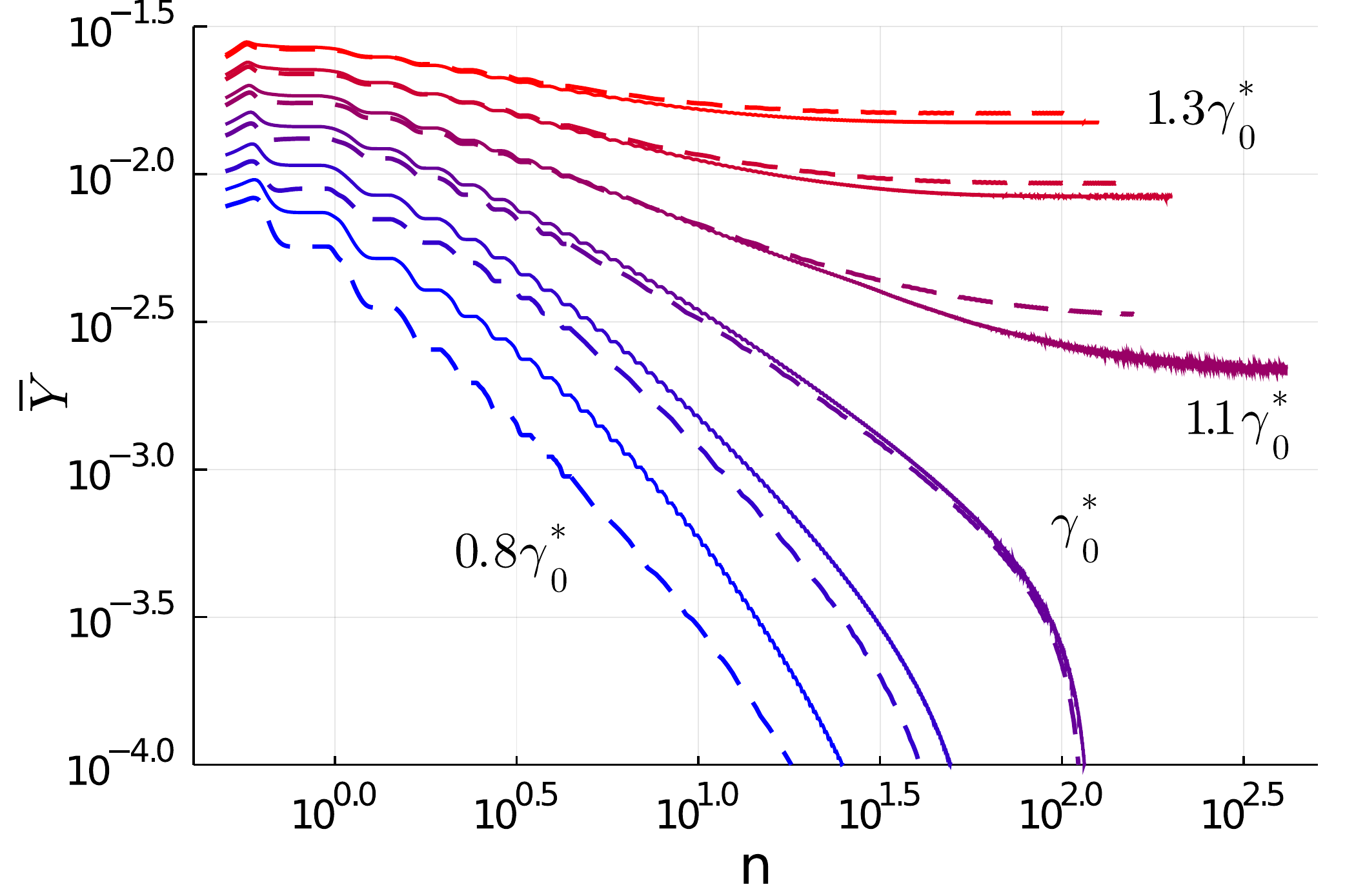}}\caption{\label{fig:beta_0_clos_simp}Dynamics at a range of strain amplitudes around the transition for $\beta=0$. These are (from blue to red): 0.8, 0.9, 1.0, 1.1, 1.2 and 1.3 $\gamma_0^{*}$. Full lines correspond to simple closure, dashed to $E$-dependent.}
\end{figure}

\begin{figure}\center{\includegraphics[width=0.5\textwidth]{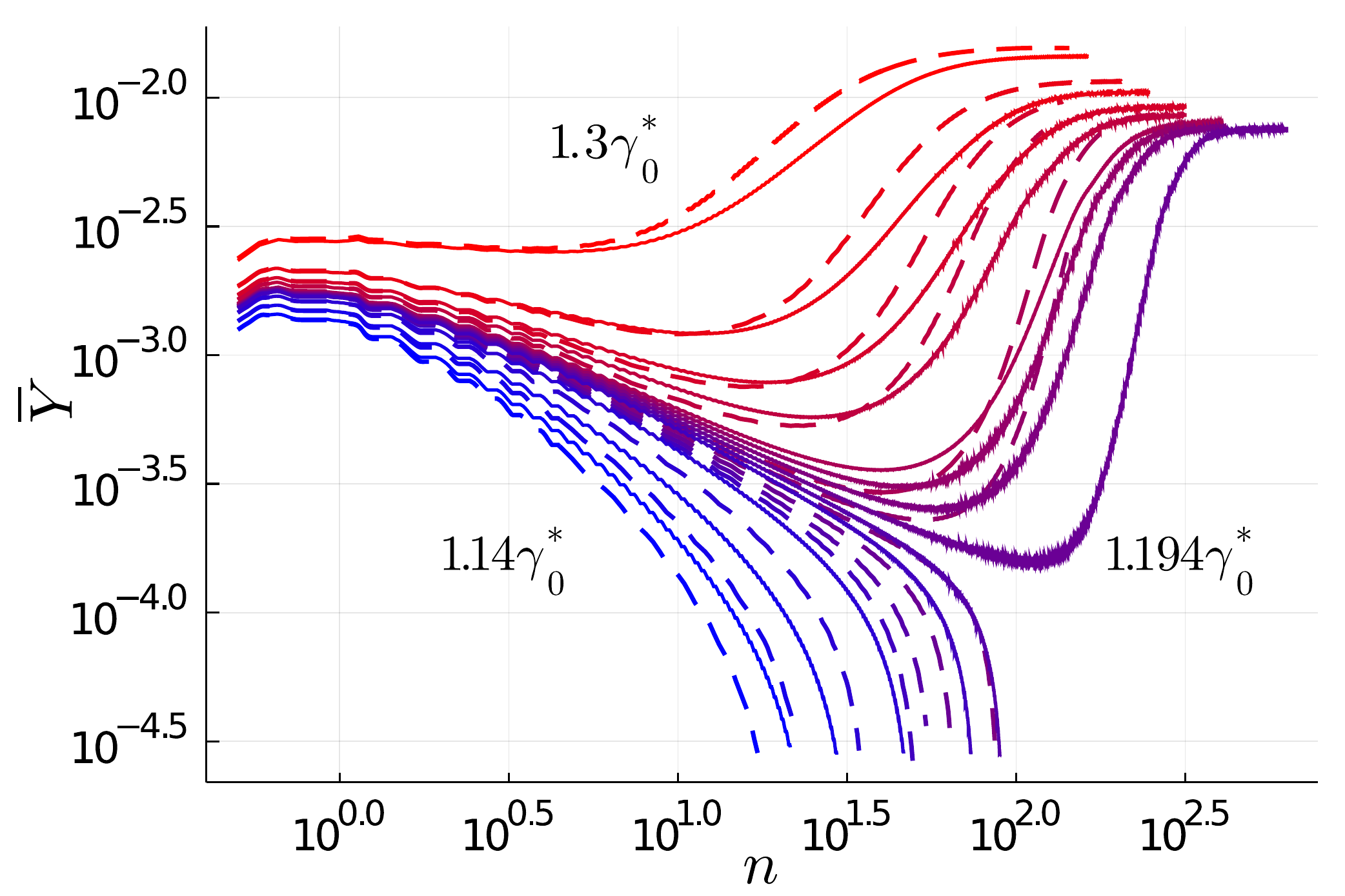}}\caption{\label{fig:beta_50_clos_simp}Dynamics at a range of strain amplitudes around the fatigue limit for $\beta=50$. These are (from blue to red): 1.14, 1.16, 1.18, 1.19, 1.192, 1.194, 1.196, 1.198, 1.2, 1.21, 1.22, 1.24 and 1.3 $\gamma_0^{*}$. Full lines correspond to the simple closure, dashed to the linear $E$-dependent one. Note that with the linear closure relation, the runs with $1.196$ and $1.194$ $\gamma_0^{*}$ end up freezing, while in the simple case these re-fluidise. This suggests that the fatigue limit is slightly higher in the linear closure case. }
\end{figure}

In Figures~\ref{fig:beta_0_clos_simp} and \ref{fig:beta_50_clos_simp} we show numerical solutions with the linear closure relation with $\hat{\alpha}=0.086$, alongside the simple version with $\alpha/\langle E\rangle=0.086$. Firstly, in Fig.~\ref{fig:beta_0_clos_simp} we show runs for the initial condition $\beta=0$: we confirm that the transition is indeed located roughly at the same strain amplitude, and overall the dynamics is largely unaffected.

In Fig.~\ref{fig:beta_50_clos_simp}, on the other hand, where we show instead runs for $\beta=50$ around the fatigue limit, we do see appreciable differences. With the $E$-dependent closure relation the system ``freezes'' faster, but also re-fluidizes at a faster rate. Also, although in this case we have not performed a thorough analysis of the yield times, the fatigue limit $\gamma_c$ appears to be slightly larger than in the simple closure relation.

\section{Uniform shear}
\begin{figure}\center{\includegraphics[width=0.5\textwidth]{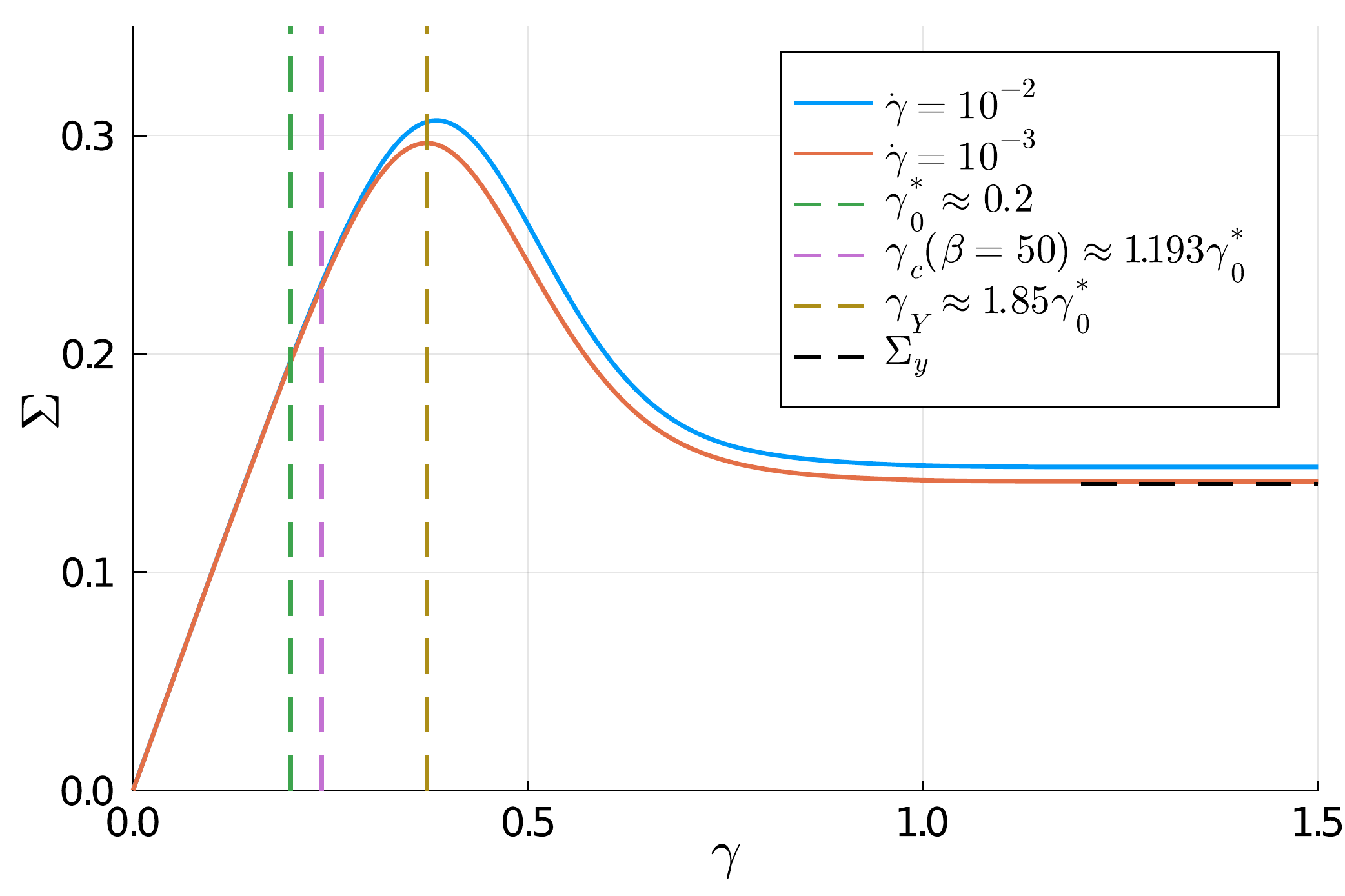}}\caption{\label{fig:beta_50_uni}Stress-strain curves for $\beta=50$, obtained with two different shear rates. Also shown is the exact macroscopic yield stress (\ref{macro_yield}).}
\end{figure}

In Fig.~\ref{fig:beta_50_uni}, we show the stress-strain curve under uniform shear for the intial condition $\beta=50$, from which we extract the yield point $\gamma_{Y}\approx 1.85 \gamma^{*}_0$ (reported in Fig.~3 of the main text) as the strain at which the overshoot is reached.

\begin{figure}\center{\includegraphics[width=0.5\textwidth]{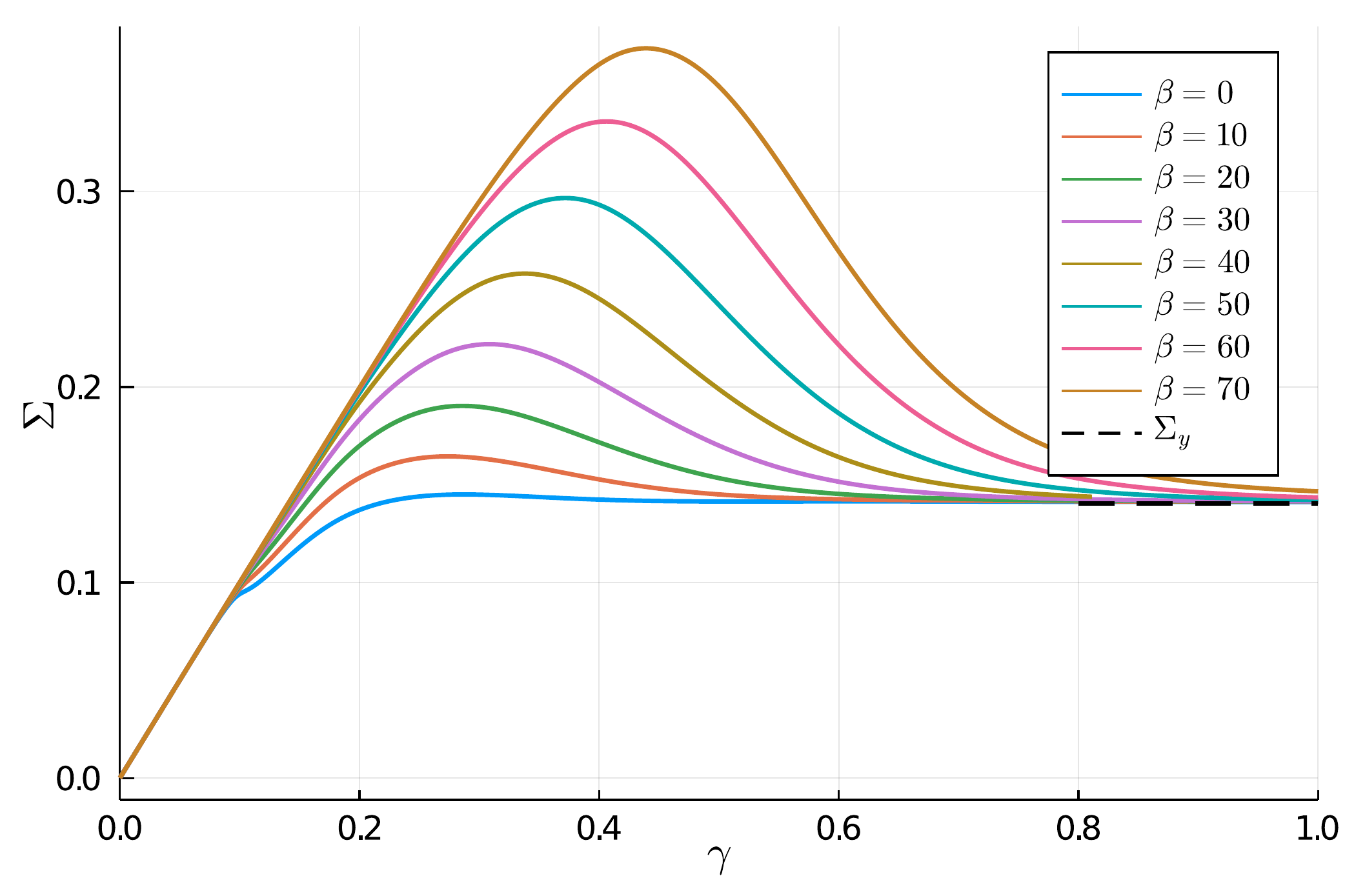}}\caption{\label{fig:all_beta}Stress-strain curves uniformed shear obtained with $\dot{\gamma}=10^{-3}$ for a range of $\beta$. Also shown is the exact macroscopic yield stress (\ref{macro_yield}). }
\end{figure}

In Fig.~\ref{fig:all_beta}, we show stress-strain curves under uniform shear with the small shear rate $\dot{\gamma}=10^{-3}$, for a range of $\beta$ values up to $\beta=70$. We see that the stress overshoot grows with $\beta$, but remains smooth and shows no sign of developing a discontinuity. 

\begin{figure}\center{\includegraphics[width=0.5\textwidth]{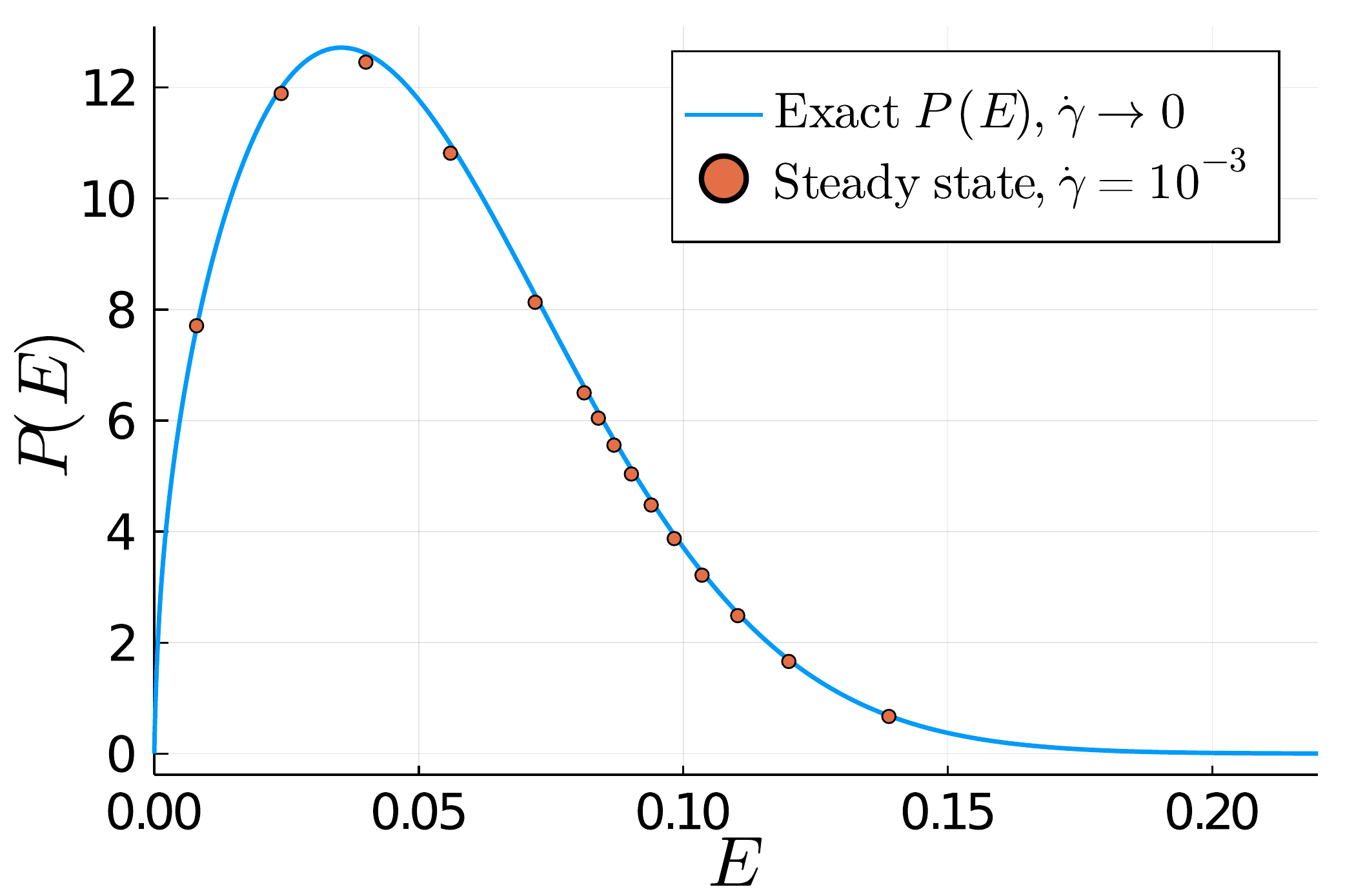}}\caption{\label{fig:steady_shear}Comparison of numerically measured $P(E)$ in the steady state with the exact form (\ref{p_of_e_ag}). Energy discretisation shown corresponds to $\hat{\beta}=0$ (higher values of $\hat{\beta}$ show similar agreement).}
\end{figure}

In Fig.~\ref{fig:steady_shear} we check that in the steady state limit under slow shear ($\dot{\gamma}=10^{-3}$) our numerics reproduces very well the exact form (\ref{p_of_e_ag}) of $P(E)$  derived by Agoritsas et al. \cite{agoritsas_relevance_2015} in the limit $\dot{\gamma}\rightarrow 0$.

We contrast our results in Fig.~\ref{fig:beta_50_uni} and \ref{fig:all_beta} with two claims in the literature,  in \cite{ozawa_random_2018} and \cite{popovic_elastoplastic_2018}, where spinodal-type discontinuous yielding was claimed to be found in mean field. We note, firstly, that \cite{ozawa_random_2018} neglects the alternating signs of the elastic kernel, in stark contrast to the fully symmetric mechanical noise considered here. To clarify the relation with~\cite{popovic_elastoplastic_2018}, where an overhanging stress-strain curve under quasistatic loading was reported in the original HL model, we now rewrite our mean field model in the same form considered there, in terms of a quasistatic evolution in plastic strain. Starting with the original master equation
\begin{equation}\label{ME}
    \partial_t P(E,l,t)=-\dot{\gamma}\partial_l P(E,l,t)+\alpha Y(t)\partial_l^2 P(E,l,t)+Y(t)\rho(E)\delta(l)-\theta\left(|l|-l_c(E)\right)P(E,l,t)
\end{equation}
we divide this by the yield rate $Y(t)$. Noting that $1/Y \partial_t =\partial_{\epsilon_p}$, with $\epsilon_p$ the plastic strain, we have that
\begin{equation}\label{plastic_aqs}
    \partial_{\epsilon_p} P(E,l,\epsilon_p)=-\frac{\dot{\gamma}}{Y}\partial_{\epsilon_p}P+\alpha \partial_l^2 P+\rho(E)\delta(l)
\end{equation}
where in the $Y \ll 1$ limit one can replace the yielding term by absorbing boundary conditions at $l_c(E)$ \footnote{Note that Eq. (\ref{plastic_aqs}) with absorbing boundary conditions only strictly conserves normalisation if the derivatives at the boundaries satisfy $\partial_l P(E,l_c)-\partial_l P(E,-l_c)=-\rho(E)/\alpha$ (valid in steady state). This severely restricts the class of initial conditions that can be studied.}. On the other hand, from (\ref{ME}) the evolution of the macroscopic stress is
\begin{equation}
    \partial_t \Sigma=\dot{\gamma}-\langle l_u \rangle Y
\end{equation}
where $\langle l_u \rangle$ is the average unstable strain, which in the quasistatic limit (where one can neglect yield events at the negative strain threshold) can be approximated by $\int \mathrm{d}E \ P(E,t) l_c(E)$. Note that in the disordered HL model this quantity is in principle time-dependent, while without disorder it is a constant. This implies that the advective term in (\ref{plastic_aqs}), denoted by $v$ in \cite{popovic_elastoplastic_2018}, is given by
\begin{equation}
    v=\frac{\dot{\gamma}}{Y}=\langle l_u \rangle+\frac{\dot{\Sigma}}{Y}=\langle l_u \rangle+\frac{\partial \Sigma}{\partial\epsilon_p}
\end{equation}
Now, as argued in \cite{popovic_elastoplastic_2018}, ``failure'' would occur if $\dot{\Sigma}/Y\rightarrow$ approaches $-\langle l_u \rangle$ (from above), where $v$ would vanish. Once this bound is surpassed, one sees that $\dot{\gamma}$ effectively has to become negative to be able to continue the dynamics, leading to an overhang in a plot of $\Sigma$ vs $\gamma$.

In the original time-dependent formulation (\ref{ME}), this implies that at some $\gamma$ during startup $Y$ has to diverge, given that one cannot relax more than $\langle l_u \rangle$ per plastic yield event. Although we do not have an exact argument, we find it implausible on physical grounds that in the disordered case $Y$ could grow without bound at some point, given that it is now coupled to the dynamics of a whole range of energy levels: due to the spectrum of different yield barriers present, the strain distributions in different energy levels will inevitably begin to approach their respective yield threshold at continuously distributed time (or strain) points. In any case it would be interesting to perform a thorough analysis of Eq. (\ref{plastic_aqs}) in the disordered HL model. 

\bibliography{supp_material.bib}